\documentclass[course]{houches}
\usepackage{epsfig}
\usepackage{axodraw}
\def\Preprint{\vspace*{-8.4cm}\mbox{}\hfill FTUV/98-46 \\ \mbox{}\hfill 
  IFIC/98-47 \\ \mbox{}\hfill June 1998 \\  \vspace*{7cm}}
\def\refjl#1#2#3#4#5#6{\bibitem{#1} #2, {\it #3} {\bf #4} (#5) #6.}
\def\refbk#1#2#3#4{\bibitem{#1} #2, {\it #3}, #4.}
\def\etal{{\it et al}}
\def\APP{Acta Phys. Polon.}

\def\NP{Nucl. Phys.}
\def\NPPS{Nucl. Phys. B (Proc. Suppl.)}
\def\PL{Phys. Lett.}
\def\PRL{Phys. Rev. Lett.}
\def\PR{Phys. Rev.}
\def\PRep{Phys. Rep.}
\def\ZP{Z. Phys.}

\def\MP{Int. J. Mod. Phys.}
\def\AP{Ann. Phys., Lpz.}
\def\APNY{Ann. Phys., NY}
\def\NC{Nuovo Cimento}
\def\RMP{Rev. Mod. Phys.}
\def\RPP{Rep. Prog. Phys.}
\def\ARNPS{Ann. Rev. Nucl. Part. Sci.}
\def\PTP{Prog. Theor. Phys.}
\def\PPNP{Prog. Part. Nucl. Phys.}

\def\SJNP{Sov. J. Nucl. Phys.}
\def\slashchar#1{\setbox0=\hbox{$#1$}\dimen0=\wd0%
\setbox1=\hbox{/}\dimen1=\wd1%
\ifdim\dimen0>\dimen1%
\rlap{\hbox to
\dimen0{\hfil/\hfil}}#1\else                                        
\rlap{\hbox to \dimen1{\hfil$#1$\hfil}}/\fi}
\newcommand{\eqn}[1]{(\ref{#1})}
\newcommand{\be}{\begin{equation}}
\newcommand{\ee}{\end{equation}}
\newcommand{\no}{\nonumber}
\newcommand{\bel}[1]{\be\label{#1}}
\newcommand{\ba}{\begin{array}{c}}
\newcommand{\bat}{\begin{array}{cc}}
\newcommand{\ea}{\end{array}}
\newcommand{\beqn}{\begin{eqnarray}}
\newcommand{\eeqn}{\end{eqnarray}}

\newcommand{\bi}{\begin{itemize}}
\newcommand{\ei}{\end{itemize}}

\newcommand{\gsim}{~{}_{\textstyle\sim}^{\textstyle >}~}
\newcommand{\lsim}{~{}_{\textstyle\sim}^{\textstyle <}~}
\newcommand{\lrder}{\stackrel{\leftrightarrow}{\partial}}
\newcommand{\toLow}{\stackrel{q^2 \ll M_W^2}{\,\longrightarrow\,}}
\newcommand{\toG}{\stackrel{G}{\,\longrightarrow\,}}
\newcommand{\dg}{\dagger}
\newcommand{\dfrac}{\displaystyle \frac}

\def\gap{\;\lower3pt\hbox{$\buildrel > \over \sim$}\;}
\def\lap{\;\lower3pt\hbox{$\buildrel < \over \sim$}\;}

\newcommand{\cL}{{\cal L}}
\newcommand{\cD}{{\cal D}}
\newcommand{\cM}{{\cal M}}
\newcommand{\cO}{{\cal O}}
\newcommand{\cP}{{\cal P}}
\newcommand{\cQ}{{\cal Q}}
\newcommand{\cH}{{\cal H}}
\newcommand{\cA}{{\cal A}}

\newcommand{\cF}{{\cal F}}
\newcommand{\cJ}{{\cal J}}

\begin{document}
\setcounter{chapter}{0}
\setcounter{page}{1}
\author[A. Pich]{Antonio Pich}
\address{Departament de F\'{\i}sica Te\`orica, IFIC, 
  Universitat de  Val\`encia --- CSIC\\
  Dr. Moliner 50, E-46100 Burjassot, Val\`encia, 
  Spain   \\ \Preprint }   
\editor{}   
\title{}    
\chapter{Effective Field Theory}

%
%

\section{Introduction}
\label{sec:introduction}

The dream of modern physics is to achieve a simple understanding 
of all observed phenomena in terms of some fundamental dynamics among
the basic constituents of nature, which would unify the different
kinds of interactions: the so-called {\it theory of everything}. 
However, even if such a marvelous theory is found at some point,
a quantitative analysis at the most elementary level
is going to be of little use for providing a
comprehensive description of nature at all physical scales.

The complicated laws of chemistry  have their origin in
the well--known electromagnetic interaction;
however, it does not seem very appropriate to attempt a quantitative
analysis starting from the fundamental Quantum Electrodynamics (QED) among
quarks and leptons.
A simplified description in terms of non-relativistic electrons
orbiting around the nuclear Coulomb potential turns out to be more
suitable to understand in a simple way the most relevant physics
at the atomic scale.
Thus, to a first approximation, the rules governing the chemical bond
among atoms can be understood in terms of
the electron mass $m_e$ and the fine structure constant
$\alpha\approx 1/137$, while only the proton mass $m_p$ is needed to
estimate the dominant corrections.
But, even this simplified description becomes too cumbersome 
to provide a useful understanding of condensed matter phenomena or
biological systems.

In order to analyze a particular physical system amid the impressive
richness of the surrounding world, it is necessary to isolate the
most relevant ingredients from the rest, so that one can obtain
a simple  description without having to understand everything.
The crucial point is to make an appropriate choice of variables,
able to capture the physics which is most important for the problem at hand.

Usually, a physics problem involves widely separated energy scales;
this allows us to study the low-energy dynamics, independently of the
details of the high-energy interactions. 
The basic idea is to identify those parameters which are very large (small)
compared with the relevant energy scale of the physical system
and to put them to infinity (zero).
This provides a sensible approximation to the problem, which can always be
improved by taking into account the corrections induced by the neglected
energy scales as small perturbations.

Effective field theories
are the appropriate theoretical tool to describe
low-energy physics, where {\it low} is defined with respect to some
energy scale $\Lambda$.
They only take explicitly into account
the relevant
degrees of freedom, i.e. those states with $m\ll\Lambda$, while the
heavier excitations with $M\gg\Lambda$ are integrated out from 
the action.
One gets in this way a string of non-renormalizable interactions
among the light states, which can be organized as an expansion
in powers of energy/$\Lambda$.
The information on  the heavier degrees of freedom is then
contained in the couplings of the resulting low-energy Lagrangian.
Although effective field theories contain an 
infinite number of terms, renormalizability
is not an issue since, at a given order in the energy expansion,
the low-energy theory is specified by a finite number of couplings;
this allows for an order-by-order renormalization.

The theoretical basis of effective field theory (EFT) can be formulated
as a theorem \cite{WE:79,LE:94a}: 
\begin{quote}
{\it 
For a given set of asymptotic states, perturbation theory with the
most general Lagrangian containing all terms allowed by the assumed
symmetries will yield the most general S-matrix elements consistent
with analyticity, perturbative unitarity, cluster decomposition
and the assumed symmetries.}
\end{quote} 

These lectures provide an introduction to the basic ideas and methods
of EFT, and a description of a few interesting
phenomenological applications in particle physics.
The main conceptual foundations are discussed in sections 
\ref{sec:expansion} and \ref{sec:loops},
which cover the momentum expansion and the most important issues
associated with the renormalization process.
Section \ref{sec:chpt} presents an overview of Chiral Perturbation
Theory (ChPT), the low--energy realization of 
Quantum Chromodynamics (QCD) in the light quark sector.
The ChPT framework is applied to weak transitions in 
section \ref{sec:kaon}, where the physics of non-leptonic
kaon decays is analyzed.
The so-called Heavy Quark Effective Theory (HQET)
is briefly discussed in
section \ref{sec:hqet}; further details on this EFT can be found in the 
lectures of M.B. Wise \cite{Wise}.
The electroweak chiral EFT is described in section \ref{sec:EWChEFT},
which contains a brief overview of the effective Lagrangian associated with 
the spontaneous electroweak symmetry breaking; this subject is analyzed
in much more detail in the lectures of R.S. Chivukula \cite{Chivukula}.
Some summarizing comments are finally given in section \ref{sec:summary}.

To prepare these lectures, I have made extensive use of several
reviews and lecture notes
\cite{GE:93,GE:84,KA:95,MA:96,EC:95,Sorrento,PI:95,Orsay,dR:95,GE:91,GR:92,NE:94,WI:94,FE:93}
already existing in the literature.
Further details on particular subjects can be found in those references.

%
%

\section{Momentum Expansion}
\label{sec:expansion}

To build an EFT describing physics at a given energy scale $E$, one
makes an expansion in powers of $E/\Lambda_i$,
where $\Lambda_i$ are the various scales involved in the problem
which are larger than $E$.
One writes the most general effective Lagrangian involving the relevant light
degrees of freedom, which is consistent with the underlying symmetries.
This Lagrangian can be organized in powers of momentum or, equivalently,
in terms of an increasing number of derivatives.
In the low-energy domain we are interested in, the terms with lower 
dimension will dominate.

\subsection{The Euler--Heisenberg Lagrangian}

A simple example of EFT
is provided by QED at very low energies,
$E_\gamma \ll m_e$.
In this limit, one can describe the light-by-light scattering
using an  effective Lagrangian in terms of the electromagnetic 
field only.
Gauge, Lorentz, Charge Conjugation and Parity invariance
constrain the possible structures present
in the effective Lagrangian:
\beqn\label{eq:L_EH}
\cL_{\mathrm{eff}} & = & -{1\over 4} F^{\mu\nu} F_{\mu\nu}
   + {a\over m_e^4} \, (F^{\mu\nu} F_{\mu\nu})^2
   + {b\over m_e^4} \, F^{\mu\nu} F_{\nu\sigma} 
     F^{\sigma\rho} F_{\rho\mu}
\no\\ &&\mbox{}
   + \cO (F^6/m_e^8) \, . 
\eeqn
In the low-energy regime, 
all the information on the original QED dynamics
is embodied in the values of the two low-energy couplings $a$ and $b$.
The values of these constants can be computed,
by explicitly integrating out the electron field from the original
QED generating
functional (or equivalently, by computing the relevant light-by-light
box diagrams). One then gets the well-known result
\cite{EU:36,EH:36}:
\bel{eq:EH}
a = -{\alpha^2\over 36}\, , \qquad\qquad\qquad
b = {7 \alpha^2\over 90} \, .
\ee
The important point to realize is that, even in the absence of an
explicit computation of the couplings $a$ and $b$, the Lagrangian
\eqn{eq:L_EH} contains non-trivial information, 
which is a consequence
of the imposed symmetries. The dominant contributions to the
amplitudes for different
low--energy photon reactions 
can be directly obtained from
$\cL_{\mathrm{eff}}$.
Moreover, the order of magnitude of the constants $a$, $b$ can also be
easily estimated through a na\"{\i}ve counting of powers of the
electromagnetic coupling and combinatorial and loop
[$1/(16 \pi^2)$] factors.

A simple dimensional analysis allows us to derive the scaling behaviour
of a given process. For instance, the $\gamma\gamma\to \gamma\gamma$
scattering amplitude should be proportional to
$\alpha^2 E^4/m_e^4$ since each photon carries a factor $e$ and each
gradient produces a power of energy.
The corresponding cross-section must have dimension $-2$, so the phase
space is proportional to $1/E^2$. Therefore,
\bel{eq:2gamma}
\sigma(\gamma\gamma\to \gamma\gamma) \propto
{\alpha^4 E^6\over m_e^8} \, .
\ee
Higher-order corrections will induce a relative uncertainty of
$\cO(E^2/m_e^2)$.

\subsection{Rayleigh Scattering}

Let us consider the low--energy scattering of photons with neutral atoms
in their ground state.
Here, low energy means that the photon energy is small enough
not to excite the internal states of the atom, i.e.
\bel{eq:small_E}
E_\gamma \ll \Delta E \ll a_0^{-1} \ll M_A \, ,
\ee
where $\Delta E \sim \alpha^2 m_e$ is the atom excitation energy,
$a_0^{-1}\sim \alpha m_e$ the inverse Bohr radius
and $M_A$ the atom mass.
Thus, the scattering is necessarily elastic.
Moreover, since $E_\gamma/M_A \ll 1$,
a non-relativistic description of the atomic field is 
appropriate.\footnote{
A Lorentz--covariant description of this process, using the
velocity--dependent formalism for heavy fields 
(see section~\protect\ref{sec:hqet})
can be found in ref.~\protect\cite{KA:95}.
}

Denoting by $\psi(x)$ the field operator that creates an atom at the 
point $x$, the effective Lagrangian for the atom has the form
\bel{eq:free_atom}
\cL \, =\,\psi^\dagger\left(i \,\partial_t - {p^2\over 2 M_A}\right)\psi 
  \, + \, \cL_{\mathrm{int}} \, .
\ee
Since the atom is neutral, the interaction term
$\cL_{\mathrm{int}}$
will involve the field
strength $F_{\mu\nu} = (\vec{E},\vec{B})$
(gauge invariance forbids a direct dependence on the vector potential
$A_\mu$).
The lowest--dimensional interaction Lagrangian contains two
possible terms \cite{MA:96}:
\bel{eq:L_Rayleigh}
\cL_{\mathrm int} \, =\, 
   a_0^3 \,\psi^\dagger\psi\,\left( c_1 \vec{E}^2 + c_2 \vec{B}^2\right)
  + \ldots
\ee
We have put an explicit factor $a_0^3$, so that the couplings $c_i$
are dimensionless ($\psi$ has dimension $3/2$ and the electromagnetic
field strength tensor has dimension 2).
Extremely low-energy photons cannot probe the internal structure of the
atom; therefore, the cross-section ought to be classical and
the typical momentum scale of the elastic scattering is set by the atom size
$a_0$.
The couplings $c_i$ are then expected to be of $\cO(1)$.

The interaction \eqn{eq:L_Rayleigh} produces a scattering amplitude
$\cA\sim c_i a_0^3 E_\gamma^2$. The corresponding cross-section,
\bel{eq:sigma_R}
\sigma \propto a_0^6 E_\gamma^4 \, ,
\ee
scales as the fourth power of the photon energy. Thus, the blue light
is scattered more strongly than the red one, which explains why the
sky looks blue.

Note that we have obtained the correct energy dependence of the
Rayleigh scattering cross-section, without doing any calculation.
Once the correct degrees of freedom have been identified,
dimensional analysis is good enough to understand qualitatively the
main properties of the process.

Higher--dimension operators induce corrections to \eqn{eq:sigma_R}
of $\cO(E_\gamma/\Lambda)$, with 
$\Lambda\sim \Delta E, a_0^{-1}, M_A$.
Since $\Delta E$ is the smallest scale, one expects our approximations to
break down as $E_\gamma$ approaches $\Delta E$. 

\subsection{The Fermi Theory of Weak Interactions}

In the Standard Model, weak decays proceed at lowest order through
the exchange of a $W^\pm$ boson between two fermionic left--handed
currents (except for the heavy quark {\it top} which decays into a real $W^+$).
The momentum transfer carried by the intermediate $W$ is very small 
compared to $M_W$. Therefore, the vector--boson propagator reduces to a 
contact interaction:
\bel{eq:toLow}
{- g_{\mu\nu} + q_\mu q_\nu /M_W^2\over q^2-M_W^2} \quad\toLow\quad
{g_{\mu\nu}\over M_W^2} \, .
\ee
These flavour--changing transitions can then be described through an
effective local 4--fermion Hamiltonian,
\bel{eq:Fermi_H}
\cH_{\mathrm{eff}} = {G_F\over\sqrt{2}} \, \cJ_\mu \, \cJ^{\mu\dagger} \, ,
\ee
where
\bel{eq:current}
\cJ_\mu = \sum_{ij} \bar u_i\gamma_\mu (1-\gamma_5) V_{ij} d_j
\, + \, \sum_l \bar\nu_l \gamma_\mu (1-\gamma_5) l \, ,
\ee
with $V_{ij}$ the Cabibbo--Kobayashi--Maskawa mixing matrix,
and
\bel{eq:Fermi_G}
{G_F\over\sqrt{2}} = {g^2\over 8 M_W^2} 
\ee
the so-called Fermi coupling constant.

At low energies ($E \ll M_W$), there is no reason to include the $W$ field
in the theory, because there is not enough energy to produce a physical
$W$ boson. The transition amplitudes corresponding to the different
weak decays of leptons and quarks are well described by the effective
4--fermion Hamiltonian \eqn{eq:Fermi_H}, which contains operators of
dimension 6 and, therefore, a coupling with dimension $-2$.
Equation~\eqn{eq:Fermi_G} establishes the relation between the effective
coupling and the parameters ($g$, $M_W$) of the underlying electroweak
theory (this is technically called a {\it matching condition}).

Expanding further the $W$ propagator in powers of $q^2/M_W^2$, one would
get fermionic operators of higher dimensions, which generate corrections
to \eqn{eq:Fermi_H}. We can neglect those contributions, provided we are
satisfied with an accuracy not better than $m_f^2/M_W^2$, where $m_f$ is 
the mass of the decaying fermion.

Let us consider the leptonic decay $l\to\nu_l l'\bar\nu_{l'}$. The
decay width can be easily computed, with the result:
\bel{eq:decay_width}
\Gamma(l\to\nu_l l'\bar\nu_{l'}) = {G_F^2 m_l^5\over 192\pi^3}\,
f(m_{l'}^2/m_l^2) \, ,
\ee
where $f(x) = 1 - 8x + 8 x^3 - x^4 - 12 x^2 \ln x$.
The global mass dependence, $\Gamma\sim G_F^2 m_l^5$, results from the known 
dimension of the Fermi coupling ($\Gamma$ must have dimension 1);
it is then a universal property of all
weak decays of fermions (except the top) and could have been fixed just by 
dimensional analysis. The three--body phase space generates a factor
$1/(4\pi)^3$; thus, the explicit calculation is only needed to fix the
remaining factor of $1/3$  and the dependence with the final lepton mass
contained in $f(m_{l'}^2/m_l^2)$.

The Fermi coupling is usually determined in $\mu$ decay; 
eq.~\eqn{eq:decay_width}
provides then a parameter--free prediction for the leptonic $\tau$ decays.
Equivalently, the $m_l^5$ dependence of the decay width, implies the
relation
\bel{eq:mu_tau}
\mbox{\rm Br}(\tau^-\to\nu_\tau e^-\bar\nu_e) =
\Gamma(\tau^-\to\nu_\tau e^-\bar\nu_e) \;\tau_\tau = 
{m_\tau^5\over m_\mu^5}\, {\tau_\tau\over \tau_\mu} = 17.77 \% ,
\ee
to be compared with the experimental value $(17.786\pm 0.072)\% $
\cite{PI:97}.

Including the additional 4--fermion operators induced by $Z$ exchange,
the effective Hamiltonian can also be used to describe
the low--energy neutrino scattering with either quarks or leptons. 
The same dimensional argument forces the cross-section to scale with energy as
\bel{eq:nu_scattering}
\sigma_\nu \sim G_F^2\, s \, ,
\ee
where $s$ is the square of the total energy in the centre-of-mass frame.

\subsection{Relevant, Irrelevant and Marginal}

An EFT is characterized by some effective Lagrangian,
\bel{eq:eff_L}
\cL = \sum_i c_i\, O_i \, ,
\ee
where $O_i$ are operators constructed with the light fields, and the
information on any heavy degrees of freedom is hidden in the couplings
$c_i$. The operators $O_i$ are usually organized according to their dimension,
$d_i$, which fixes the dimension of their coefficients:
\bel{eq:ci_scale}
[O_i] = {d_i} \qquad\longrightarrow\qquad  
c_i\sim {1\over \Lambda^{d_i-4}} \, ,
\ee
with $\Lambda$ some characteristic heavy scale of the system.

At energies below $\Lambda$, the behaviour of the different operators
is determined by their dimension.
We can distinguish three types of operators:

-- {\it Relevant} ($d_i < 4$) 

-- {\it Marginal} ($d_i = 4$) 

-- {\it Irrelevant} ($d_i > 4$)

All the operators we have seen in the previous examples have dimension
greater than four. They are called {\it irrelevant} because their effects
are suppressed by powers of $E/\Lambda$ and are thus small
at low energies.
Of course, this does not mean that they are not important. In fact,
they usually contain the interesting information about the
underlying dynamics at higher scales. The point is that
{\it irrelevant} operators are weak at low energies.

The interactions induced by the Fermi Hamiltonian \eqn{eq:Fermi_H}
are suppressed by two powers of $M_W$, and are thus {\it irrelevant}.
In spite of being weak, the four--fermion interactions
are important because they generate
the leading contributions to flavour--changing processes or to 
low--energy neutrino scattering.
However, if the masses of the $W$ and the $Z$ bosons were
$10^{16}$ GeV  we would have never seen any signal of the weak
interaction.

In contrast a coupling of positive mass dimension gives rise to effects
which become large at energies much smaller than the scale of this
coupling. Operators of dimension less than four are therefore called
{\it relevant}, because they become more important at lower energies.

In a four--dimensional relativistic field theory, the number of possible
{\it relevant} operators is rather low:

-- $d = 0$: The unit operator

-- $d = 2$: Boson mass terms ($\phi^2$)

-- $d = 3$: Fermion mass terms ($\bar\psi \psi$) and cubic scalar
  interactions ($\phi^3$)

Finite mass effects are negligible at very high energies ($E \gg m$), 
however they become {\it relevant} when the energy scale is comparable to
the mass. The role of {\it relevant} operators at low energies can be
easily understood through a simple example. Let us consider two real
scalar fields $\phi$ and $\Phi$ described by the Lagrangian \cite{KA:95}:
\bel{eq:L_relevant}
\cL = {1\over 2} (\partial\phi)^2 + {1\over 2} (\partial\Phi)^2 
  - {1\over 2} m^2 \phi^2 - {1\over 2} M^2 \Phi^2
  - {\lambda\over 2} \phi^2 \Phi \, .
\ee
The two kinetic terms are {\it marginal} operators,
with dimensionless coefficients
(the canonical ${1\over 2}$ normalization
in this case). The mass terms and the scalar interaction are {\it relevant};
therefore, they appear multiplied by coefficients with positive dimension:
$[m^2] = [M^2] = 2$, $[\lambda] = 1$.

Let us assume that $m , \lambda \ll M$, and consider the tree--level
elastic scattering of two light scalar fields $\phi\phi\to\phi\phi$,
which proceeds through the exchange of a heavy scalar $\Phi$.
The scattering amplitude is
proportional to $\lambda^2$ divided by the appropriate $\Phi$ propagator.
The behaviour of the cross--section at very high or low energies is
given by:
\bel{eq:beh_sig}
\sigma \sim {1\over E^2} \times
\left\{
  \begin{array}{lc}
    \left( {\lambda/E} \right)^4 \; , \; &
     (E \gg M) \\
    \left( {\lambda/ M} \right)^4 \; , \; &
     (m \ll E \ll M) 
  \end{array}
\right. \, .
\ee
The factor $1/E^2$ appears because the cross--section should have dimension 
$-2$.
The different energy behaviour stems from the $\Phi$ propagator.
At energies much greater than $M$, the cross--section goes rapidly to zero
as $1/E^6$. However, when $m\ll E\ll M$ the heavy propagator can be 
contracted to a point, generating a contact $\phi^4$ interaction with an
effective coupling $\lambda^2/M^2$.

We have seen a similar situation before with the Fermi theory of weak
interactions; but now, since $\lambda$ has dimension 1,
we have got the opposite low--energy behaviour.
The $d=6$ four--fermion Hamiltonian predicts a neutrino cross--section
proportional to $E^2$, which becomes {\it irrelevant} at very low energies.
In contrast, the {\it relevant} ($d=3$)\ \  $\phi^2\Phi$ interaction generates
a $\phi\phi\to\phi\phi$ cross--section which, at low energies,
increases as $1/E^2$ when the energy decreases.

Operators of dimension 4 are equally important at all energy scales and
are called {\it marginal operators}. They lie between relevancy and irrelevancy
because quantum effects could modify their scaling behaviour on either
side. Well--known examples of {\it marginal} operators are
$\phi^4$, the QED and QCD interactions and the Yukawa $\bar\psi\psi\phi$
interactions.

In any situation where there is a large mass gap between the energy scale
being analyzed and the scale of any heavier states (i.e.
$m,E\ll M$), the effects induced by {\it irrelevant} operators are always 
suppressed by powers of $E/M$, and can usually be neglected. The resulting
EFT, which only contains {\it relevant} and {\it marginal} operators, is called
{\it renormalizable}. Its predictions are valid up to $E/M$ corrections.

Dimensional analysis offers a new perspective on the old concept
of renormalizability. QED was constructed to be the most general
renormalizable ($d\leq 4$) Lagrangian consistent with the electromagnetic
$U(1)$ gauge symmetry. However, there exist other interactions
($Z$--exchange) contributing to $e^+e^-\to e^+ e^-$, which at low energies
($E\ll M_Z$) generate additional {\it non-renormalizable} local couplings
of higher dimensions. The lowest--dimensional contribution takes the form
of a Fermi $(\bar e \Gamma e)(\bar e \Gamma e)$ operator.
The reason why QED is so successful to describe
the low--energy scattering of electrons with positrons is not 
renormalizability, but rather the fact that $M_Z$ is very heavy and
the leading non-renormalizable contributions are suppressed by  
$E^2/M_Z^2$.

\subsection{Principles of Effective Field Theory}

We can summarize the basic ingredients used to build an EFT as a set
of general principles:

\begin{enumerate}

\item Dynamics at low energies (large distances) does not depend on details
of dynamics at high energies (short distances).

\item Choose the appropriate description of the important physics
at the considered scale. If there are large energy gaps, put to zero 
(infinity) the light (heavy) scales, i.e.
$$ 0\leftarrow m \ll E \ll M \rightarrow \infty .$$
Finite corrections induced by these scales can be incorporated
as perturbations.

\item Non-local heavy--particle exchanges are replaced by a tower of local
({\it non-renormalizable}) interactions among the light particles.

\item The EFT describes the low--energy physics, to a given accuracy 
$\epsilon$, in terms of a finite set of parameters:
$$(E/M)^{(d_i-4)} \gsim \epsilon \quad\Longleftrightarrow\quad
d_i \lsim 4 + {\log{(1/\epsilon)}\over\log{(M/E)}} \, . $$

\item The EFT has the same infra-red (but different ultra-violet)
behaviour than the underlying fundamental theory.

\item The only remnants of the high--energy dynamics are in the
low--energy couplings and in the symmetries of the EFT.

\end{enumerate}

%
%

\section{Quantum Loops}      
\label{sec:loops}    

Our previous dimensional arguments are quite trivial at tree--level.
It is less obvious what happens when quantum loop corrections are
evaluated. Since the momenta flowing through the internal lines
are integrated over all scales, the behaviour of {\it irrelevant} operators
within loops appears to be problematic. In fact, in order to build
{\it well--behaved} quantum field theories, {\it irrelevant} operators are
usually discarded in many textbooks, because they are {\it non-renormalizable}:
an infinite number of counter-terms is needed to get finite predictions.
Thus, at first sight, a Lagrangian including {\it irrelevant} operators seems
to lack any predictive power.

\begin{figure}[tbh]
\centering
\begin{picture}(100,55)
\ArrowLine(0,5)(50,5)
\ArrowLine(50,5)(100,5)
\ArrowArc(50,30)(25,-90,90)
\ArrowArc(50,30)(25,90,-90)
\GCirc(50,5)20
\end{picture}
\caption{Self-energy contribution to the fermion mass.}
\label{fig:self_mf}
\end{figure}
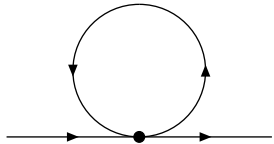

Let us try to understand the problem with a simple fermionic Lagrangian:
%
\be
\cL = \bar\psi \left( i \gamma^\mu \partial_\mu - m\right) \psi
 - {a\over\Lambda^2} (\bar\psi \psi)^2 - {b\over\Lambda^4}
 (\bar\psi \Box\psi) (\bar\psi \psi) + \cdots
\ee
The dimension--six four--fermion interaction generates a divergent contribution 
to the fermion mass, through the self-energy graph shown in 
fig.~\ref{fig:self_mf}:
\bel{eq:self_mass}
\delta m \sim 2 i \, {a \over\Lambda^2}\, m\, \int\,
{d^4k\over (2\pi)^4} \, {1\over k^2-m^2} \, .
\ee
Since the EFT is valid up to energies of order $\Lambda$, we could try
to estimate the quadratically divergent integral using $\Lambda$ as a 
natural momentum cut-off. This gives:
\bel{eq:delta_m}
\delta m \sim  {m \over\Lambda^2}\,\Lambda^2 \sim m \, .
\ee
Thus, the {\it irrelevant} four--fermion operator generates a quantum correction
to the fermion mass, which is not suppressed by any power of the scale 
$\Lambda$; i.e. it is $\cO (1)$ in the momentum expansion.
Similarly, higher--order terms  such as the dimension--eight 
operator $(\bar\psi \Box\psi) (\bar\psi \psi)$ are equally important,
and the entire expansion breaks down.

This problem can be cured if one adopts a mass--independent renormalization
scheme, such as dimensional regularization and minimal subtraction
(MS or $\overline{\mbox{\rm MS}}$). 
Performing the calculation in $D=4 + 2\epsilon$ dimensions,
the correction to the fermion mass induced
by the diagram of fig.~\ref{fig:self_mf} takes the form:
\bel{eq:delta_m2}
\delta m \sim  2 a m \; {m^2\over 16\pi^2 \Lambda^2}\, \mu^{2\epsilon}\,
\left\{ {1\over\hat{\epsilon}} + 
\log{\left({m^2\over\mu^2}\right) } -1 + \cO(\epsilon)\right\} , 
\ee
where
\bel{eq:hateps}
{1\over\hat{\epsilon}}\equiv {1\over \epsilon} + \gamma_E -\log{(4\pi)}
\, ,
\ee
and $\gamma_E= 0.577\, 215\ldots $ is the Euler's constant.
The important thing is that the arbitrary dimensional scale $\mu$ only
appears in the logarithm, and does not introduce any explicit powers
such as $\mu^2$.
The $1/\Lambda^2$ factor weighting the {\it irrelevant} operator
$(\bar\psi\psi)^2$ is then necessarily compensated by two powers of a
light physical scale, $m^2$ in this case. The integral is now of
$\cO [m^2/(16\pi^2\Lambda^2)]$, which is small provided $m\ll\Lambda$.

This is a completely general result. In a mass--independent
renormalization scheme, loop integrals do not have a power law dependence
on any big scale $\mu\sim\Lambda$. Thus, one can count powers of
$1/\Lambda$ directly from the effective Lagrangian. Operators proportional
to $1/\Lambda^n$ need only to be considered when probing effects of
$\cO(1/\Lambda^n)$ or smaller.
The EFT produces then a well--defined expansion in powers of momenta
over the heavy scale $\Lambda$.

To a given order in $E/\Lambda$ the EFT contains only a finite number
of operators. Therefore, working to a given accuracy, the EFT behaves
for all practical purposes like a renormalizable quantum field theory:
only a finite number of counter-terms are needed to reabsorb the 
divergences.

Of course, physical predictions should be independent of our 
renormalization conventions. Thus, one should get the same answers using
a mass--dependent subtraction scheme, such as our previous momentum
cut-off. The only problem is that in the cut-off scheme one needs to
consider an infinite number of contributions to each order in
$1/\Lambda$. If one was able to resum all contributions
of a given order, the net effect would be to reproduce the results
obtained in a much more simple way using a
mass--independent scheme.
Within the context of EFT, a mass--independent renormalization scheme is
very convenient, because it provides an efficient way of organizing
the $1/\Lambda$ expansion, so that only a finite number of operators
(and Feynman graphs) are needed.

Our toy--model calculation \eqn{eq:delta_m2} shows two additional 
important features. 
The first one is the logarithmic dependence on the
renormalization scale $\mu$. The physical content of this type of
logarithms will be analyzed in the next subsections, where the
concept of renormalization and the associated renormalization--group
equations will be briefly discussed.

The second interesting feature is that $\delta m \propto m$. Thus,
if $m=0$ the quantum correction also vanishes. There is a deep
symmetry reason behind this fact. The kinetic term and the
four--fermion interactions are invariant under the chiral transformation
\ $\psi\to\gamma_5\psi$, \ $\bar\psi\to -\bar\psi\gamma_5$,
which, however, is not a symmetry of the mass term.
In the $m=0$ limit, the chiral symmetry of the Lagrangian protects
the fermion from acquiring a mass through quantum corrections.
It is then {\it natural} that the fermion mass might be small, even
if there are other heavy scales in the problem such as $\Lambda$.

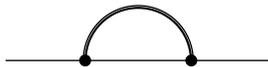
\begin{figure}[tbh]
\centering
\begin{picture}(100,40)
\Line(0,0)(100,0)
\CArc(50,0)(20,0,180)
\CArc(50,0)(19.7,0,180)\CArc(50,0)(20.7,0,180)
\GCirc(30,0)20
\GCirc(70,0)20
\end{picture}
\caption{Self-energy contribution to the light scalar mass.
The thick line denotes a heavy--scalar propagator.}
\label{fig:self_ms}
\end{figure}

The behaviour is rather different in scalar theories, because a scalar
mass term does not usually break any symmetry. Let us go back to
the toy  model in eq.~\eqn{eq:L_relevant}, and consider the 
self-energy diagram in fig.~\ref{fig:self_ms}. Even if one takes
$m=0$ at tree level, the coupling to the heavy scalar generates a
non-zero contribution to the light mass:
\bel{eq:self_ms}
\delta m^2 \sim  {\lambda^2\over 16\pi^2}\;  \mu^{2\epsilon}\,
\left\{ {1\over\hat{\epsilon}} + 
\log{\left({M^2\over\mu^2}\right) } -1 + \cO(\epsilon)\right\}
\, .
\ee
Thus, it is {\it unnatural} to have a light scalar mass much smaller
than $\lambda/(4\pi)$; that would require a fine tuning between
the bare mass and $\lambda$ such that the tree--level and loop contributions
cancel each other to all orders.
The Lagrangian has however an additional symmetry ($\delta\phi =$ constant)
when both $m$ and $\lambda$ are zero \cite{KA:95}; 
therefore $\phi$ can be light if it does not couple to the heavy scalar.

The problem of {\it naturalness} is present in the electroweak
symmetry breaking, which, in the Standard Model, is associated
with the existence of a scalar sector. While fermion masses can be
protected of becoming heavy through some kind of chiral symmetry,
the presence of a relatively light scalar Higgs (which presumably
couples to some higher new--physics scale) seems {\it unnatural}.

\subsection{Renormalization}

\begin{figure}[tbh]
\centering
\begin{picture}(100,56)
\Photon(0,20)(30,20)33
\Photon(70,20)(100,20)33
\ArrowArcn(50,20)(20,0,180)
\ArrowArcn(50,20)(20,180,360)
\GCirc(30,20)20
\GCirc(70,20)20
\Text(15,30)[c]{$\vec{q}$}
\Text(50,50)[c]{$\vec{k}$}
\Text(25,8)[c]{$\mu$}\Text(75,8)[c]{$\nu$}
\end{picture}
\caption{Vacuum--polarization diagram.}
\label{fig:QED_vp}
\end{figure}
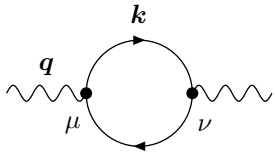

Let us consider the QED vacuum polarization induced by a fermion with
electric charge $Q_f$.
The corresponding one--loop diagram, shown in 
fig.~\ref{fig:QED_vp}, is clearly divergent
[$\sim\int d^4k\, (1/k^2)$].
We can define the loop integral through {\it dimensional regularization};
i.e. performing the calculation in $D= 4+2\epsilon$ dimensions, where
the resulting expression is well defined.
The ultraviolet divergence is then recovered through the pole of the Gamma
function $\Gamma\left({-\epsilon}\right)$ at $D=4$.

For simplicity, let us neglect the mass of the internal fermion.
Since the loop integration is going to generate logarithms of the
external momentum transfer $q^2$, it is convenient to introduce an
arbitrary mass scale $\mu$ to compensate the $q^2$ dimensions.
The result can then be written as
\bel{eq:vp_Lorentz}
\Pi^{\mu\nu}(q) = \left( -q^2 g^{\mu\nu} + q^\mu q^\nu\right)\,
\Pi(q^2) \, ,
\ee
where
\bel{eq:vp_fun}
\Pi(q^2) = - {4\over 3} \, Q_f^2\, {\alpha\over 4\pi}\;\mu^{2\epsilon}\,
\left\{ {1\over\hat{\epsilon}} + 
\log{\left({-q^2\over\mu^2}\right) } -{5\over 3} + \cO(\epsilon) \right\}
\, .
\ee
This expression does not depend on $\mu$, but written in this form
one has a dimensionless quantity inside the logarithm.

Owing to the ultraviolet divergence, eq.~\eqn{eq:vp_fun} does not determine the
wanted self-energy contribution. Nevertheless, it does show how this effect
changes with the energy scale. If one could fix the value of
$\Pi(q^2)$ at some reference momentum transfer $q_0^2$, the result
would be known at any other scale:
\bel{eq:pi_1_2}
\Pi(q^2) = \Pi(q_0^2) - {4\over 3} \, Q_f^2\,  {\alpha\over 4\pi}
\,\log{(q^2/q_0^2)} \, .
\ee

 We can split the self-energy contribution into a meaningless divergent piece
and a finite term, which includes the $q^2$ dependence,
\bel{eq:Pi_splitting}
\Pi(q^2) \,\equiv\, \Delta\Pi_\epsilon(\mu^2) + \Pi_R(q^2/\mu^2)\, .
\ee
This separation is ambiguous, because the finite $q^2$--independent
contributions can be splitted in many different ways. A given choice defines
a {\it scheme}:
\beqn 
\lefteqn{
\Delta\Pi_\epsilon(\mu^2)  = 
 -{\alpha Q_f^2\over 3\pi}  \;\mu^{2\epsilon} \, \times \, 
\left\{ 
 \begin{array}{lc}
 \left[ {\displaystyle {1\over\hat{\epsilon}}
         -{5\over 3}} \right]  \quad\qquad &
        (\mu)  \\ \\
  {\displaystyle {1\over \epsilon}}  &
	       (\mbox{\rm MS})  \\ \\
  {\displaystyle {1\over\hat{\epsilon}} \phantom{\left[ {5\over 3}\right]}} & 
       (\overline{\mbox{\rm MS}})  
  \ea\right. ; } &&
\no\\ && \label{eq:Pi_eps_R} \\
\lefteqn{
  \Pi_R(q^2/\mu^2)  = -{\alpha Q_f^2\over 3\pi}  \times \left\{ 
   \begin{array}{lc}
     \log{\left( \dfrac{-q^2}{\mu^2}\right)}  &  (\mu)  \\ \\
    \left[ \log{\left( \dfrac{-q^2}{\mu^2}\right)}
        + \gamma_E -\log(4\pi) -{\displaystyle {5\over 3}}\right]   \quad & 
        (\mbox{\rm MS}) \\ \\
    \left[ \log{\left( \dfrac{-q^2}{\mu^2}\right)} -
    {\displaystyle {5\over 3}}\right]  & 
    (\overline{\mbox{\rm MS}}) 
   \ea\right.  . 
} && \no
\eeqn
In the $\mu$--scheme, one uses the value of $\Pi(-\mu^2)$ to define the
divergent part.
MS and $\overline{\mbox{\rm MS}}$ stand for minimal subtraction 
\cite{THO:73} and
modified minimal subtraction schemes \cite{BBDM:78}; 
in the MS case, one subtracts only
the divergent $1/\epsilon$ term, while the $\overline{\mbox{\rm MS}}$
scheme puts also the $\gamma_E-\log(4\pi)$ factor into the
divergent piece.
Notice that the logarithmic $q^2$--dependence of
$\Pi_R(q^2/\mu^2)$ is always the same.

Let us now consider the corrections induced by the photon self-energy
on the electromagnetic interaction between two electrons.\footnote{
The QED Ward identity, associated with the conservation of the vector current,
guarantees that the sum of the corresponding vertex and
wave--function corrections is finite. Since we are only interested in
the divergent pieces, and their associated logarithmic dependences, we
don't need to specify those contributions. 
}
The scattering amplitude takes the form
\bel{eq:T_ee}
T(q^2) \,\sim\, -J^\mu J_\mu \, {4\pi\alpha\over q^2} \, 
\left\{ 1 -\Pi(q^2) + \ldots
\right\} \, ,
\ee
where $J^\mu$ denotes the electromagnetic fermion current.

\begin{figure}[tbh]
\centerline{\epsfig{file=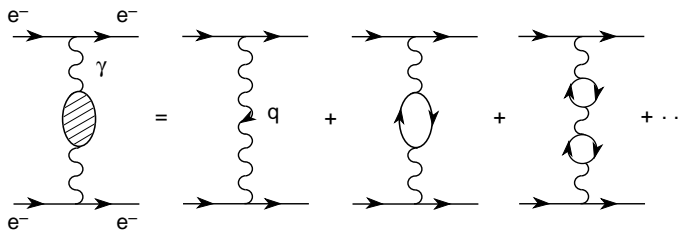, width=0.8\linewidth}}
\caption{Photon self-energy contributions to $e^-e^-$.}
\label{fig:EEint}
\end{figure}

The divergent correction generated by quantum loops can be reabsorbed into a
redefinition of the coupling:
\bel{eq:alpha_R}
\alpha_B\, \left\{ 1 -\Delta\Pi_\epsilon(\mu^2)
- \Pi_R(q^2/\mu^2)\right\} \,\equiv\,
\alpha_R(\mu^2)\, \left\{ 1 
- \Pi_R(q^2/\mu^2)\right\} \, ,
\ee
where $\alpha_B\equiv e_B^2/(4\pi)$ denotes the {\it bare} QED coupling
and
\bel{eq:alpha_Rb}
\alpha_R(\mu^2) \, = \, \alpha_B \,
\left\{ 1 + Q_f^2\, {\alpha_B \over 3\pi} \,\mu^{2\epsilon}
\left[{1\over\epsilon} + C_{\mathrm{scheme}}\right] + \ldots
\right\} \, .
\ee
The resulting scattering amplitude is finite and gives rise
to a definite prediction for the cross--section, which can be compared with
experiment. Thus, one actually measures the {\it renormalized} coupling
$\alpha_R(\mu^2)$.

The redefinition \eqn{eq:alpha_R} is meaningful provided that it can be done
in a self-consistent way: all ultraviolet divergent contributions to all
possible scattering processes should be eliminated through the same redefinition
of the coupling (and the fields). The nice thing of gauge theories, such as
QED or QCD, is that the underlying gauge symmetry guarantees the
renormalizability of the quantum field theory.

The renormalized coupling $\alpha_R(\mu^2)$ depends
on the arbitrary scale $\mu$ and on the chosen {\it renormalization scheme}
[the constant $C_{\mathrm{scheme}}$ denotes the corresponding finite terms in
eq.~\eqn{eq:Pi_eps_R}]. Quantum loops have introduced a scale dependence in a
quite subtle way. Both $\alpha_R(\mu^2)$ and the renormalized self-energy
correction $\Pi_R(q^2/\mu^2)$ depend on $\mu$, but the physical scattering
amplitude $T(q^2)$ is of course $\mu$--independent:
\beqn\label{eq:mu_dep}
T(q^2) &\sim & -  
{\alpha_R(\mu^2)\over q^2}\, \left\{ 1 + Q_f^2\,
{\alpha_R(\mu^2)\over 3\pi} \left[ \log{\left({-q^2\over\mu^2}\right)} +
C'_{\mathrm{scheme}}\right] + \ldots  \right\}\no\\
& = &   
{\alpha_R(Q^2)\over Q^2} \, \left\{ 1 +  Q_f^2\,
{\alpha_R(Q^2)\over 3\pi} C'_{\mathrm{scheme}} + \cdots \right\}\, ,
\eeqn
where $Q^2\equiv -q^2$.

The quantity $\alpha(Q^2)\equiv\alpha_R(Q^2)$ is called the QED
{\it running coupling}.
The ordinary fine structure constant $\alpha\approx1/137$ is defined through the
classical Thomson formula; therefore, it corresponds to a very low scale
$Q^2= -m_e^2$. Clearly, the value of $\alpha$ relevant for LEP experiments
is not the same.

The scale dependence of $\alpha(Q^2)$ is regulated by 
the so-called $\beta$ function:
\bel{eq:beta}
\mu \, {d\alpha\over d\mu} \,\equiv\, \alpha \,\beta(\alpha) \, ;
\qquad\qquad \beta(\alpha)\, =\, \beta_1\, {\alpha\over \pi} +
\beta_2 \left({\alpha\over\pi}\right)^2 + \cdots
\ee
Only renormalized quantities appear in \eqn{eq:beta}; thus, the $\beta$
function is non-singular in the limit $\epsilon\to 0$.

At the one--loop level, the $\beta$ function reduces to the first coefficient,
which is fixed by eq.~\eqn{eq:alpha_Rb}:
\bel{eq:beta_1}
\beta_1^{\mathrm{QED}} \, = \, {2\over 3}\, Q_f^2\, . 
\ee
The first--order differential equation \eqn{eq:beta} can then be easily solved,
with the result:
\bel{eq:alpha_run}
\alpha(Q^2) \, = \, {\alpha(Q_0^2)\over 1 - {\beta_1 \alpha(Q_0^2)\over 2\pi}
\log{(Q^2/Q_0^2)}} \, .
\ee
Since $\beta_1>0$, the QED running coupling increases with the energy scale:
$\alpha(Q^2)>\alpha(Q_0^2)$ \ if \ $Q^2>Q_0^2$;
i.e. the electromagnetic charge decreases at large distances. This can be
intuitively understood as a screening effect of the virtual 
fermion--antifermion pairs
generated, through quantum effects, around the electron charge.
The physical QED vacuum behaves as a polarized dielectric medium.

Notice that taking $\mu^2=Q^2$ in eq.~\eqn{eq:mu_dep} we have eliminated
all dependences on $\log{(Q^2/\mu^2)}$ to all orders in $\alpha$.
The running coupling \eqn{eq:alpha_run} makes a resummation of all
leading logarithmic corrections, i.e 
\bel{eq:alpha_logs}
\alpha(Q^2) \, = \, \alpha(\mu^2)\,\sum_{n=0}^\infty
\left[ {\beta_1 \alpha(\mu^2)\over 2\pi}
\log{(Q^2/\mu^2)}\right]^n  .
\ee
These higher--order logarithms correspond to the contributions from an
arbitrary number of one--loop self-energy insertions along the intermediate
photon propagator:
$1 - \Pi_R(q^2/\mu^2) + \left(\Pi_R(q^2/\mu^2)\right)^2 + \cdots $


\begin{figure}[tbh]
\centerline{\epsfig{file=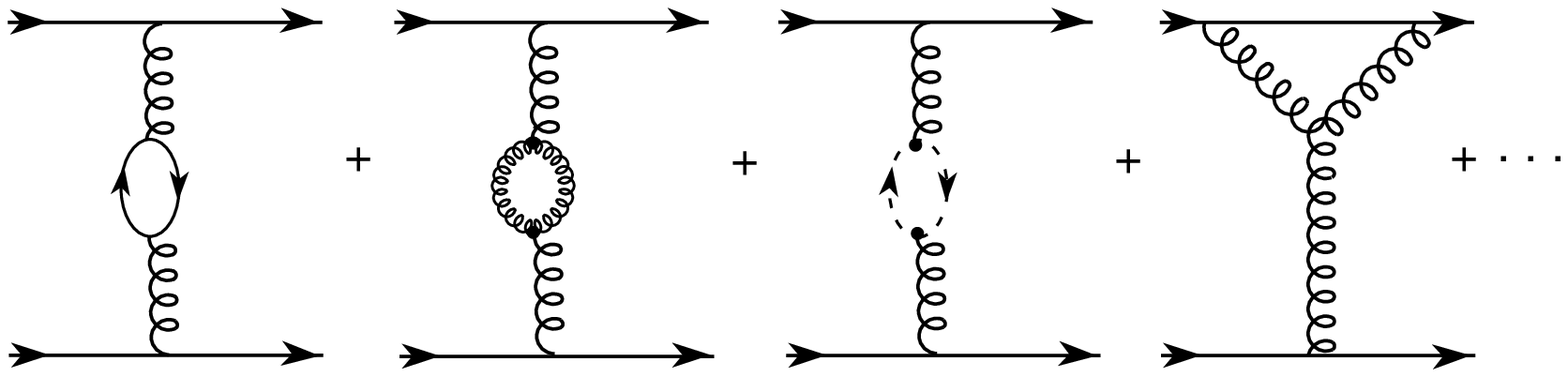, width=0.9\linewidth}}  
\caption{Feynman diagrams contributing to the renormalization of the
  strong coupling.}
\label{fig:g_si}
\end{figure}

The renormalization of the QCD coupling proceeds in a similar
way. Owing to the non-abelian character of $SU(3)_C$, there are additional
contributions involving gluon self-interactions. From the 
calculation of the relevant one-loop diagrams, shown in fig.~\ref{fig:g_si}, 
one gets the value of the first $\beta$--function 
coefficient \cite{GW:73,PO:73}:
\bel{eq:QCD_beta}
\beta_1 \, = \, {2\over 3} T_F n_f  -  {11\over 6} C_A
\, = \, {2 n_f - 11 N_C \over 6} \, .
\ee
The positive contribution proportional to the number of quark flavours
$n_f$ is generated by the
$q$-$\bar q$ loops and corresponds to the QED result 
(except for the $T_F=\frac{1}{2}$ factor).
The gluonic self-interactions introduce the additional {\it negative}
contribution proportional to $C_A=N_C$, where $N_C=3$ is the
number of QCD colours.
This second term is responsible for the
completely different behaviour of QCD: $\beta_1 < 0$ if $n_f \leq 16$.
The corresponding QCD running coupling,
decreases at short distances, i.e.
\bel{eq:af_limit}
\lim_{Q^2\to\infty} \, \alpha_s(Q^2) \, = \, 0 \, .
\ee
Thus, for $n_f\leq 16$, QCD has the required property of asymptotic 
freedom.
The gauge self-interactions of the gluons {\it spread out} the QCD charge,
generating an {\it anti-screening} effect. This could not happen in QED, because
photons do not carry electric charge. Only non-abelian gauge theories,
where the intermediate gauge bosons are self-interacting particles, have this
antiscreening property \cite{CG:73}.

Quantum effects have introduced a dependence of the coupling with the
energy, modifying the na\"{\i}ve scaling of the {\it marginal} QED and QCD
interactions.
Owing to the different sign of their associated $\beta$ functions, these
two gauge theories behave differently.
Quantum corrections make QED {\it irrelevant} at low energies
($\lim_{Q^2\to 0} \, \alpha(Q^2)  = 0$),
while the QCD interactions become highly {\it relevant}
($\lim_{Q^2\to 0} \, \alpha_s(Q^2)  = \infty$).

Notice that a dynamical scale dependence has been generated, in spite
of the fact that we are considering dimensionless interactions
among massless fermions.
An explicit reference scale can be introduced through the solution
of the $\beta$--function differential equation \eqn{eq:beta}. 
At one loop, one gets
\bel{eq:Lambda_def}
\log{\mu} + {\pi\over\beta_1\alpha(\mu^2)} \, = \, 
\log{\Lambda} \, ,
\ee
where $\log{\Lambda}$ is just an integration constant. Thus,
\bel{eq:alpha_Lambda}
\alpha(\mu^2) \, = \, 
{2\pi\over -\beta_1 \log{\left({\mu^2/\Lambda^2}\right)}} \, .
\ee
In this way, we have traded the dimensionless coupling by the
dimensionful scale $\Lambda$, which indicates when a given energy scale
can be considered large or small.
The number of free parameters is the same (1 for massless fermions).
Although, eq.~\eqn{eq:alpha_run} gives the impression that the 
scale--dependence of
$\alpha(\mu^2)$ involves two parameters, $\mu_0^2$ and
$\alpha(\mu_0^2)$, only the combination \eqn{eq:Lambda_def}
matters, as explicitly shown in \eqn{eq:alpha_Lambda}.

The renormalization of a general EFT is completely analogous to
the simpler QED and QCD cases.
The only difference is that one needs to deal with as many couplings
as operators appearing in the corresponding effective Lagrangian.
In a mass--independent subtraction scheme, the number of couplings
to be renormalized is finite because
only a finite number of operators have to be considered (to a given 
accuracy).

\subsection{Decoupling}

Let us consider again the QED vacuum--polarization diagram in
fig.~\ref{fig:QED_vp}, and let us study the effects associated
with the fermion mass:
$$   
\Pi(q^2) = - {\alpha Q_f^2\over 3 \pi}\,\left\{
{\mu^{2\epsilon}\over\hat{\epsilon}} + 6
\int_0^1 dx\, x (1-x) \,\log{\left( {m_f^2 - q^2 x (1-x)\over\mu^2}\right)}
\right\} .
$$   

In a mass--dependent renormalization scheme, such as the $\mu$--scheme,
the renormalized self-energy takes he form
\bel{eq:Pi_mu_scheme}
\Pi_R(q^2/\mu^2) = - Q_f^2\, {\alpha\over 3 \pi}\, 6
\int_0^1 dx\, x (1-x) \,\log{\left[ 
{m_f^2 - q^2 x (1-x)\over m_f^2 + \mu^2 x (1-x)}\right]}
\, ,
\ee
while the fermion contribution to the
one--loop $\beta$--function coefficient is easily found to be
\bel{eq:beta1_mu_scheme}
\beta_1 = 4\, Q_f^2\,
\int_0^1 dx\, {\mu^2 x^2 (1-x)^2 \over m_f^2 + \mu^2 x (1-x)}
\, .
\ee
%

\begin{figure}[tbh]
\centerline{\epsfig{file=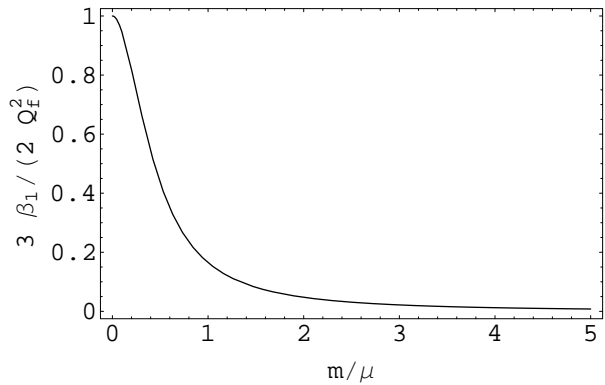,width=0.7\linewidth}}
\caption{Mass--dependence of $\beta_1$ in the $\mu$--scheme.}
\label{fig:beta1}
\end{figure}

The mass--dependence of $\beta_1$ is plotted in fig.~\ref{fig:beta1}.
In the limit $m_f^2\ll\mu^2, q^2$ we recover the massless result
$\beta_1 = 2 Q_f^2/3$; while for
high masses ($m_f^2\gg\mu^2, q^2$) the fermion contribution
to the $\beta$ function decreases as $1/m_f^2$:
\bel{eq:beta1_dec}
\beta_1 \sim {2\over 15}\, Q_f^2 \, {\mu^2\over m_f^2} \, .
\ee
The same happens with the heavy--fermion contribution to the 
renormalized self-energy:
\bel{eq:Pi_dec}
\Pi_R(q^2/\mu^2) \sim Q_f^2\, {\alpha\over 15\pi}\,
{q^2 + \mu^2\over m_f^2} \, .
\ee
Thus, at energies much smaller than $m_f$
the fermion {\it decouples} \cite{AC:75}.

In the $\overline{\mbox{\rm MS}}$ scheme, the $\beta$ function is independent
of the mass. Therefore, the fermion generates the same contribution,
$\beta_1 = 2 Q_f^2/3$,
to the running of the QED coupling at all energy scales:
a heavy fermion does not decouple as it should.
Moreover, the renormalized self-energy,
\bel{eq:Pi_MS}
\Pi_R(q^2/\mu^2) = - Q_f^2\, {\alpha\over 3 \pi}\, 6
\int_0^1 dx\, x (1-x) \,\log{\left[ 
{m_f^2 - q^2 x (1-x)\over \mu^2}\right]}
\, ,
\ee
grows as $\log{(m_f^2/\mu^2)}$.
For $\mu \ll m_f$ the logarithm becomes large and perturbation theory
breaks down.

The mass--independent subtraction gives rise to an unphysical
behaviour when $q^2,\mu^2\ll m_f^2$. 
The $\overline{\mbox{\rm MS}}$ coupling runs incorrectly at low
energies, because one is using a {\it wrong} $\beta$ function which includes
contributions from very high scales.
The large logarithm in $\Pi_R(q^2/\mu^2)$ is compensating the {\it wrong}
running, in such a way that the
low--energy ($E \ll m_f$) physical amplitudes are not affected by the
heavy--fermion contributions.

Decoupling of heavy particles is not manifest in mass--independent
subtraction schemes. This is an important drawback for schemes
such as MS or $\overline{\mbox{\rm MS}}$. However, they are much
easier to use than the mass--dependent ones.
One way out is to implement decoupling by hand,
{\it integrating out} the heavy particles \cite{WE:80,HA:81,OS:81}.
At energies above the heavy particle mass one uses the full theory
including the heavy field, while 
a different EFT without the heavy field is used below threshold.

In the previous example, for $\mu> m_f$ 
one would use the QED Lagrangian with an explicit massive fermion $f$;
the corresponding one--loop $\beta$ function would be
$\beta_1 = {2\over 3} Q_f^2 + \beta_1^{\mathrm{light}}$,
where $\beta_1^{\mathrm{light}}$ stands for the light--field contributions.
When $\mu < m_f$, one takes instead QED with the light fields only;
i.e. $\beta_1 = \beta_1^{\mathrm{light}}$.

\subsection{Matching}

The effects of a heavy particle are included in the low--energy theory
through higher--dimension operators, which are suppressed by inverse powers
of the heavy--particle mass.
Around the heavy--threshold region, the physical predictions should be
identical in the full and effective theories.
Therefore, the two descriptions are related by
a {\bf matching condition:}
{\it at $\mu = m_f$,
the two EFTs (with and without the heavy field)  should give rise to
the same $S$--matrix elements for light--particle scattering.}

Since the light--particle content is the same, the infra-red properties
of the two theories will be identical.  The EFT without the heavy
field only distorts the high--energy behaviour. The {\it matching
conditions} mock up the effects of heavy particles and high--energy
modes into the low--energy EFT.
In practice, one should {\it match} all the 
{\it one--light--particle--irreducible} diagrams
(those that cannot be disconnected by cutting a single light--particle 
line) with external light particles.

Thus, in the $\overline{\mbox{\rm MS}}$ scheme one uses a series of EFTs with
different particle content. When running from higher to lower
energies, every time a particle threshold is crossed
one {\it integrates out} the corresponding field and imposes the
appropriate {\it matching condition} on the resulting low--energy theory.
This procedure guarantees the correct decoupling properties, while
keeping at the same time the calculational simplicity of
mass--independent subtraction schemes.

The following examples illustrate how the {\it matching conditions}
are implemented.

\medskip
\noindent {\it\thesubsection.1. The $\phi^2\Phi$ interaction}
\smallskip

\noindent
Let us consider again the scalar Lagrangian in eq.~\eqn{eq:L_relevant}.
At energies below the heavy scalar mass $M$, one {\it integrates out} the
heavy field $\Phi$:
\bel{eq:S_fund_eff}
 \exp{\{iZ\}} \equiv \int \cD\phi \cD\Phi\, 
\exp{\{i S(\phi,\Phi)\}} =
\int \cD\phi\, \exp{\{i S_{\mathrm{eff}}(\phi)\}} .
\ee
The resulting EFT, which only contains the light
scalar $\phi$, is described by the effective Lagrangian:
\bel{eq:L_phi}
\cL_{\mathrm{eff}} = {1\over 2}\, a \, (\partial\phi)^2 
  - {1\over 2} \, b \, \phi^2 + c\, {\lambda^2 \over 8 M^2} \,
  \phi^4 + \cdots
\ee
%

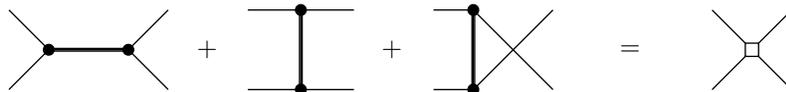
\begin{figure}[tbh]
\centering
\begin{picture}(295,30)

\Line(0,0)(15,15)
\Line(0,30)(15,15)
\Line(15,15)(45,15)\Line(15,14.5)(45,14.5)\Line(15,15.5)(45,15.5)
\Line(45,15)(60,30)
\Line(45,15)(60,0)
\GCirc(15,15)20
\GCirc(45,15)20
\Text(75,15)[c]{$+$}

\Line(90,30)(130,30)
\Line(90,0)(130,0)
\Line(110,0)(110,30)\Line(109.5,0)(109.5,30)\Line(110.5,0)(110.5,30)
\GCirc(110,30)20
\GCirc(110,0)20
\Text(145,15)[c]{$+$}

\Line(160,30)(175,30)
\Line(160,0)(175,0)
\Line(175,0)(175,30)\Line(174.5,0)(174.5,30)
  \Line(175.5,0)(175.5,30)
\Line(175,30)(205,0)
\Line(175,0)(205,30)
\GCirc(175,30)20
\GCirc(175,0)20
\Text(235,15)[c]{$=$}

\Line(265,30)(295,0)
\Line(265,0)(295,30)
\BBoxc(280,15)(5,5)
\end{picture}

\caption{Tree--level matching condition.
The thick lines denote heavy--scalar propagators in the full scalar 
theory.
The rhs diagram corresponds to the low--energy EFT.}
\label{fig:tree_matching}
\end{figure}

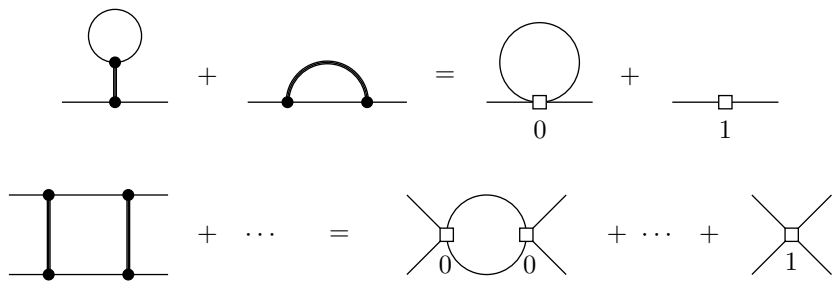
\begin{figure}[tbh]
\centering
\begin{picture}(310,100)
\Line(90,65)(150,65)
\CArc(120,65)(15,0,180)
\CArc(120,65)(14.5,0,180)\CArc(120,65)(15.5,0,180)
\GCirc(105,65)20
\GCirc(135,65)20
\Text(165,75)[c]{$=$}
\Text(235,75)[c]{$+$}

\Line(180,65)(220,65)
\Line(250,65)(290,65)
\CArc(200,80)(15,0,360)
\BBoxc(200,65)(5,5)
\BBoxc(270,65)(5,5)
\Text(201,55)[c]{$0$}
\Text(271,55)[c]{$1$}

\Line(20,65)(60,65)
\Line(40,65)(40,80)\Line(39.5,65)(39.5,80)\Line(40.5,65)(40.5,80)
\GCirc(40,65)20\GCirc(40,80)20
\CArc(40,90)(10,0,360)
\Text(75,75)[c]{$+$}


\Line(0,0)(60,0)
\Line(0,30)(60,30)
\Line(15,0)(15,30)\Line(14.5,0)(14.5,30)\Line(15.5,0)(15.5,30)
\Line(45,0)(45,30)\Line(44.5,0)(44.5,30)\Line(45.5,0)(45.5,30)
\GCirc(15,0)20
\GCirc(15,30)20
\GCirc(45,0)20
\GCirc(45,30)20
\Text(75,15)[c]{$+$}
\Text(95,15)[c]{$\cdots$}
\Text(125,15)[c]{$=$}
\Text(230,15)[c]{$+$}
\Text(245,15)[c]{$\cdots$}
\Text(265,15)[c]{$+$}
\Line(150,0)(165,15)
\Line(150,30)(165,15)
\Line(195,15)(210,0)
\Line(195,15)(210,30)
\CArc(180,15)(15,0,360)
\BBoxc(165,15)(5,5)
\BBoxc(195,15)(5,5)

\Line(280,30)(310,0)
\Line(280,0)(310,30)
\BBoxc(295,15)(5,5)
\Text(165,4)[c]{$0$}
\Text(197,4)[c]{$0$}
\Text(296,5)[c]{$1$}
\end{picture}

\caption{One--loop matching condition for the two and four--point vertices.
The numbers beneath the rhs vertices indicate the corresponding loop order.}
\label{fig:loop_matching}
\end{figure}

The couplings $a$, $b$, $c$, \ldots\ are fixed by {\it matching}
the effective Lagrangian with the full underlying scalar theory.
At tree level, $a=1$, $b=m^2$, and the $\phi^4$ interaction is
generated through $\Phi$--exchange.
The {\it matching condition},
shown in fig.~\ref{fig:tree_matching}, implies $c=1$.

At the quantum level, the {\it matching} is slightly more involved:
\beqn\label{eq:loop_matching}
a & = & 1 + a_1 \, {\lambda^2\over 16\pi^2 M^2} + \cdots
\no\\
b & = & m^2 + b_1 \, {\lambda^2\over 16\pi^2} + \cdots
\\
c & = & 1 + c_1 \, {\lambda^2\over 16\pi^2 M^2} + \cdots
\no
\eeqn
The one--loop matching conditions \cite{KA:95} with both 2 and 4
external $\phi$ fields, shown in fig.~\ref{fig:loop_matching},
determine the coefficients $a_1$, $b_1$ and $c_1$.
This calculation is left as an exercise.
Nevertheless, it is worth while to stress some general features:

--- The ultra-violet divergences are dealt with in $\overline{\mbox{\rm MS}}$;
the {\it matching conditions} relate then well--defined finite quantities.

--- The effective couplings are $\mu$--dependent, where $\mu$ is the
renormalization scale.
{\it Matching} is imposed at the scale $\mu=M$ in order to avoid
large $\log{(M^2/\mu^2)}$ corrections.

--- The two theories have the same infra-red properties, therefore all
infra-red divergences cancel out in the {\it matching conditions}.
Non-analytic dependences on light--particle masses and momenta
(e.g. $\log{m^2}$ or $\log{p^2}$) also cancel out.
$\cL_{\mathrm{eff}}$ has then a local expansion in powers of $1/M$.

\medskip
\noindent {\it\thesubsection.2. QCD matching}
\smallskip

\noindent
Let us consider the QCD Lagrangian with $n_f-1$ light--quark flavours
plus one heavy quark of mass $M$. 
At $\mu < M$, one {\it integrates out} the heavy quark; the resulting
EFT is $\cL^{(n_f-1)}_{\mathrm{QCD}}$
plus a tower of higher--dimensional operators
suppressed by powers of $1/M$.
The {\it matching conditions} relate this EFT to
the original QCD Lagrangian with $n_f$ flavours:
\bel{eq:mat_QCD}
\cL^{(n_f)}_{\mathrm{QCD}} \quad\Longleftrightarrow\quad
\cL^{(n_f-1)}_{\mathrm{QCD}} + \sum_{d_i>4} {\tilde{c}_i\over M^{d_i-4}}
\; O_i \, .
\ee

At low energies, one usually neglects the small effect of the
{\it irrelevant} ($d_i>4$) operators. The EFT reduces then to the normal
QCD Lagrangian with $(n_f-1)$ quark flavours, which contains
all the {\it marginal} ($d=4$) 
and {\it relevant} (light--quark mass terms)
operators allowed by gauge invariance.
Remember that, owing to the quantum corrections, the {\it marginal}
QCD interaction becomes highly {\it relevant} at low scales.

The two QCD theories have different $\beta$ functions (the $\beta_i$
coefficients depend explicitly on the number of quark flavours).
Thus, the running of the corresponding couplings $\alpha_s^{(n_f)}(\mu^2)$ and
$\alpha_s^{(n_f-1)}(\mu^2)$ is different.
The two effective couplings are related through a matching condition:
\bel{eq:QCDmatch}
 \alpha_s^{(n_f)}(\mu^2) = \alpha_s^{(n_f-1)}(\mu^2) \,\left\{
 1 + \sum_{k=1} C_k(L) \,\left( {\alpha_s^{(n_f-1)}(\mu^2)\over\pi}\right)^k
 \right\}  ,
\ee
where $L\equiv \log{(\mu/M)}$.
Since we use a mass--independent subtraction scheme ($\overline{\mbox{\rm MS}}$),
the neglected higher--dimensional operators $O_i$ cannot affect this
matching condition.

The logarithmic dependence of the $C_k(L)$ coefficients on the scale $\mu$
can be easily obtained, by taking the derivative of eq.~\eqn{eq:QCDmatch}
with respect to $\log{\mu}$ and using the corresponding $\beta$--function 
equation obeyed by each coupling.
At one loop, this gives:
\bel{eq:dC1}
{d C_1(L)\over dL} = \beta_1^{(n_f)} - \beta_1^{(n_f-1)} =
{1\over 3} \, ,
\ee
i.e.
\bel{eq:C1}
C_1(L) = c_{1,0} + {1\over 3}\, L \, ,
\qquad\qquad c_{1,0} = 0 \, .
\ee
The value of the integration constant $c_{1,0}$ can only be fixed by 
matching the explicit calculation of some Green function in both effective 
theories. One easily gets $c_{1,0}=0$, which corresponds to
$\alpha_s^{(n_f)}(M^2) = \alpha_s^{(n_f-1)}(M^2)$.

Similarly, using the calculated value of the two--loop $\beta$--function
coefficient in the $\overline{\mbox{\rm MS}}$ scheme
\cite{CA:74,JO:74},
\bel{eq:beta2}
\beta_2 = {19\over 12} n_f - {51\over 4} \, ,
\ee
one obtains \cite{BW:82}:
\bel{eq:C2}
C_2(L) = c_{2,0} + {19\over 12}\, L + {1\over 9}\, L^2 \, .
\ee
The value of the two--loop integration constant is no longer zero.
Moreover, it depends on the adopted definition for the heavy quark mass 
$M$ \cite{LRV:95}:
\bel{eq:c20}
 c_{2,0} = \left\{
   \begin{array}{cc}
    -11/72 \, ,\qquad & [ M\equiv M(M^2)] \\
    \phantom{-1}7/24 \, , \qquad & [ M\equiv M_{\mathrm{pole}}]
   \ea \, . \right.
\ee
In the first case, the quark mass is defined to be the 
$\overline{\mbox{\rm MS}}$
running mass, while in the second line $M$ refers to the pole of the
perturbative quark propagator.
Notice, that the running coupling constant has now a discontinuity
at the matching point:
\bel{eq:dis_alpha}
\alpha_s^{(n_f)}(M^2) = \alpha_s^{(n_f-1)}(M^2) \, \left\{
1 + c_{2,0} \,\left( {\alpha_s^{(n_f-1)}(M^2)\over \pi}\right)^2 + \cdots
\right\} \, .
\ee
Thus, at the two--loop (or higher) level the $\overline{\mbox{\rm MS}}$
QCD coupling is not continuous when crossing a heavy--quark threshold.
There is nothing wrong with that. The running QCD coupling is not
a physical observable; it is just a  parameter which depends on our
renormalization conditions. Moreover, the couplings $\alpha_s^{(n_f)}(\mu^2)$
and $\alpha_s^{(n_f-1)}(\mu^2)$ are defined in different EFTs;
they are different parameters and there is no reason why they should be
equal at the matching point.
Of course, physical observables should be the same independently of
which conventions (or EFT) have been used to compute them. But this is precisely
the content of the {\it matching} conditions we have imposed, which
require a discontinuous coupling.

Analogously, the running masses of the light quarks are defined differently
in the two EFTs. The so-called $\gamma$ function,
which governs their evolution,
\bel{eq:gamma}
\mu\, {d m\over d\mu} \equiv - m \,\gamma(\alpha_s) \, ;
\qquad\qquad
\gamma(\alpha_s) = \gamma_1\, {\alpha_s\over\pi} + \gamma_2
\left( {\alpha_s\over\pi}\right)^2 + \cdots
\ee
starts to depend on the flavour number at the two--loop level \cite{TA:81}:
\bel{eq:gamma_coef}
\gamma_1 = 2 \, , \qquad\qquad
\gamma_2 = {101\over 12} - {5\over 18} n_f \, .
\ee
The quark--mass matching conditions can be easily implemented in the same
way as for the strong coupling.
Since the $\beta$ and $\gamma$ QCD functions are already known to
four loops \cite{RVS:97,CH:97} 
the logarithmic scale dependence of the $\alpha_s^{(n_f)}(\mu^2)$
and $m^{(n_f)}(\mu^2)$ matching conditions can be worked out at
this level in a quite straightforward way \cite{RPS:98}.
The corresponding non-logarithmic contributions are however only known to
three loops \cite{ChKS:97}.

\subsection{Scaling}

We have seen already that quantum corrections can change the scaling
dimension of operators from their classical value. This is
specially important for marginal operators because they become either 
relevant (like QCD) or irrelevant (like QED). Although the effect is less
dramatic for operators with dimension different from four, the modified
quantum scaling generates sizeable corrections 
whenever two widely separated physical scales are involved.

The change of the scaling properties is associated with the introduction
of the new scale $\mu$, in the renormalization process. 
The statement that {\it physical
observables should be independent of our renormalization conventions},
provides a powerful tool to analyze the quantum scaling, which is called
{\it renormalization group}
\cite{SP:53,GL:54,CA:70,SY:70}.

Let us consider some Green function
$\Gamma(p_i;\alpha,m)$, where $p_i$ are physical momenta.
To simplify the discussion we assume that $\Gamma$ depends on a single
coupling $\alpha$ and mass $m$, but the following arguments are completely
general and can be extended to several couplings and masses in
a quite straightforward way.

The {\it renormalized} ($\Gamma_R$) and {\it bare} ($\Gamma_B$) Green 
functions are related through some equation of the form
\bel{eq:zeta_rel}
\Gamma_B(p_i;\alpha_B,m_B;\epsilon) = Z_\Gamma(\epsilon,\mu) \,
\Gamma_R(p_i;\alpha,m;\mu) \, ,
\ee
where $\alpha_B$, $m_B$ ($\alpha$, $m$) denote the {\it bare} 
({\it renormalized})
coupling and mass. The appropriate product of renormalization factors
is contained in $Z_\Gamma(\epsilon,\mu)$, which reabsorbs
the divergences of the {\it bare} Green function $\Gamma_B$. 
The dependence on $\epsilon$ refers to the dimensional regulator
$(D-4)/2$.
Obviously,
$Z_\Gamma(\epsilon,\mu)$ depends on our choice of renormalization scheme.
We have explicitly indicated that both $Z$ and $\Gamma_R$ depend on the
renormalization scale $\mu$.
Since the {\it bare} Green function $\Gamma_B$ does not depend on 
the arbitrary scale $\mu$, the corresponding {\it renormalized} Green 
function should obey the renormalization--group equation:
\bel{eq:rge}
\left( \mu\, {d\over d\mu} + \gamma_\Gamma(\alpha) \right)\,
\Gamma_R(p_i;\alpha,m;\mu) = 0 \, ,
\ee
where
\bel{eq:gamma_G}
\gamma_\Gamma(\alpha) \equiv {\mu\over Z_\Gamma}\,
{d Z_\Gamma\over d \mu} .
\ee
The function $\gamma_\Gamma(\alpha)$ is necessarily non-singular,
because only renormalized quantities appear in eq.~\eqn{eq:rge};
moreover, in a mass--independent renormalization scheme
\cite{THO:73,WE:73}, it only depends on the coupling.

The dependence on $\mu$ can be made more explicit,
using the $\beta$-- and $\gamma$--function equations \eqn{eq:beta} and 
\eqn{eq:gamma}:
\bel{eq:rge2}
\left( \mu\, {\partial\over \partial\mu} + 
\beta(\alpha) \,\alpha\, {\partial\over \partial\alpha}
- \gamma(\alpha)\, m\, {\partial\over \partial m} +
\gamma_\Gamma(\alpha) \right)
\Gamma(p_i;\alpha,m;\mu) = 0 \, .
\ee
Since it is  no-longer necessary, we have dropped the subscript $R$.

Using the $\beta$ function to trade the dependence on $\mu$ by $\alpha$,
the solution of the renormalization--group equation \eqn{eq:rge} is easily
found to be
\bel{eq:rge_sol}
\Gamma(p_i;\alpha,m;\mu) = \Gamma(p_i;\alpha_0,m_0;\mu_0)\;
\exp{\left\{-\int_{\alpha_0}^\alpha\, {d\alpha\over\alpha}\,
{\gamma_\Gamma (\alpha)\over\beta(\alpha)} \right\}} .
\ee
This equation  relates the Green functions obtained at two
different renormalization points $\mu$ and $\mu_0$.

The information provided by the 
renormalization--group equations allows us to relate the
values of the Green function at different physical scales.
A global scale transformation of all external momenta by a factor
$\xi$ will induce the change
\bel{eq:scale_ch}
\Gamma(\xi p_i;\alpha,m;\mu) = \xi^{d_\Gamma}\,
\Gamma(p_i;\alpha,m/\xi;\mu/\xi) \, ,
\ee
where $d_\Gamma$ is the classical dimension of $\Gamma$.
Taking the derivative of this equation with respect to $\log{\xi}$
and using eq.~\eqn{eq:rge2}, one gets
\beqn\label{eq:rge_sc}
\lefteqn{
\left\{   \xi\, {\partial\over\partial\xi}  - 
\beta(\alpha)\, \alpha\, {\partial\over \partial\alpha}
+ \left[ 1 + \gamma(\alpha)\right]\, m\, {\partial\over \partial m} 
\right. } && \no\\ && \left. \phantom{{\partial\over\partial\xi}}
\qquad\qquad\qquad\qquad\quad
- \left[ d_\Gamma + \gamma_\Gamma(\alpha) \,\right]\right\}
\,\Gamma(\xi p_i;\alpha,m;\mu) = 0  \, .
\eeqn
The general solution of this equation can be obtained with the
standard method of characteristics to solve linear partial
differential equations.
One obtains the relation
\beqn\label{eq:rge_res}
\lefteqn{
\Gamma\left(\xi p_i;\alpha(\mu^2),m(\mu^2);\mu\right) = 
} && \\ && \qquad\quad
\xi^{d_\Gamma}\,
\exp{ \left\{\int_{\alpha(\mu^2)}^{\alpha(\xi^2\mu^2)} {d\alpha\over\alpha}
 {\gamma_\Gamma(\alpha)\over\beta(\alpha)}\right\} } \;
\Gamma\left(p_i;\alpha(\xi^2\mu^2),m(\xi^2\mu^2)/\xi;\mu\right)  ,   \no
\eeqn
which is the fundamental result of the renormalization group.
For a fixed value of the renormalization scale $\mu$, the behaviour of the
Green function under the scaling of all external momenta is given by the
corresponding running of the parameters of the theory (couplings and masses) as
functions of the scale factor.
Moreover, the global scale factor $\xi^{d_\Gamma}$ is modified by the
exponential term.
The function $\gamma_\Gamma(\alpha)$ is called the {\it anomalous dimension}
of the Green function $\Gamma$, since it modifies its classical
dimension. The usual $\gamma(\alpha)$ function is the
{\it anomalous dimension} of the mass.
The role of the anomalous dimensions is rather transparent in 
eq.~\eqn{eq:rge_sc} where  we see the explicit factors
$\left[ d_\Gamma + \gamma_\Gamma(\alpha) \right]$
and 
$\left[ 1 + \gamma(\alpha)\right]$.

\subsection{Wilson Coefficients}
\label{subsec:wilson}

Let us consider a low--energy EFT with Lagrangian
\bel{eq:EFT_lag}
\cL = \sum_i\, {c_i\over\Lambda^{d_i-4}}\; O_i .
\ee
We have written explicit factors of $1/\Lambda$, 
in order to have dimensionless coefficients $c_i$.
Using the renormalization group, we can learn how these
coefficients change with the scale.

To simplify the discussion, let us assume that the operators $O_i$
do not mix under renormalization
(this would be the case if, for instance,
there is a single operator for each dimension),
i.e.
\bel{eq:O_ren}
\langle O_i\rangle_B = Z_i(\epsilon,\mu) \, 
\langle O_i(\mu)\rangle_R ,
\ee
where $\langle O_i\rangle$ denotes the matrix element of the operator $O_i$
between asymptotic states of the theory.
The renormalized operators satisfy then the renormalization--group
equation
\bel{rge_op}
\left( \mu\, {d\over d\mu} + \gamma_{O_i}\right) \langle O_i\rangle_R = 0 \, ,
\ee
with
\bel{eq:gamma_i}
\gamma_{O_i} \equiv  {\mu\over Z_i}\,
{d Z_i\over d \mu} \, = \,
\gamma_{O_i}^{(1)}\, {\alpha\over\pi} + 
\gamma_{O_i}^{(2)} \left({\alpha\over\pi}\right)^2 + \cdots 
\ee
the corresponding anomalous dimension of the operator $O_i$.
Since the product $c_i \langle O_i\rangle$ is scale independent, this
implies an analogous equation for the so-called {\it Wilson coefficients}
\cite{WI:69} $c_i$,
\bel{rge_coeff}
\left( \mu\, {d\over d\mu} - \gamma_{O_i}\right) c_i = 0 \, ,
\ee
which has the solution:
\beqn\label{rge_coeff_sol}
c_i(\mu) &=& c_i(\mu_0) \; \exp{\left\{
  \int_{\alpha_0}^\alpha\, {d\alpha\over\alpha}\,
{\gamma_{O_i}(\alpha)\over\beta(\alpha)} \right\}} 
\no\\ & = & 
c_i(\mu_0) \;\left[ 
{\alpha(\mu^2)\over\alpha(\mu_0^2)}\right]^{\gamma_{O_i}^{(1)}/\beta_1}
\,\biggl\{ 1 \, + \,\cdots \biggr\} .
\eeqn

\medskip
\noindent {\it\thesubsection.1. Operator mixing}
\smallskip

\noindent
In general, there are several operators of the same dimension which
mix under renormalization,
\bel{eq:O_mix}
\langle O_i\rangle_B = \sum_j\, \mat{Z}_{ij}(\epsilon,\mu)\, 
\langle O_j(\mu)\rangle_R .
\ee
This complicates slightly the previous derivation, because one has
to consider a set of coupled renormalization--group equations.

The anomalous dimensions of the mixed operators are now given by
the matrix
\bel{eq:g_mat}
\mat{\gamma}_O \equiv \mat{Z}^{-1}\mu\, {d\over d\mu}\, \mat{Z} \, .
\ee
The renormalization--group equations obeyed by the operators and
the Wilson coefficients are easily found to be:
\bel{eq:rge_vec}
\left( \mu\, {d\over d\mu} + \mat{\gamma}_O\right) \langle \vect{O}\rangle_R 
= 0 \, ,
\qquad\qquad
\left( \mu\, {d\over d\mu} - \mat{\gamma}^T_O\right) \langle 
\vect{c}\,\rangle_R = 0 \, ,
\ee
where $\vect{O}$ and $\vect{c}$ are 1--column vectors containing the
operators $O_i$ and the coefficients $c_i$ respectively.

With this compact matrix notation, the equations have the same form than
in the simpler unmixed case. They can be solved in a straightforward way,
diagonalizing the anomalous--dimension matrix:
\bel{eq:diag}
\left( \mat{U}^{-1} \mat{\gamma}^T \mat{U}\right)_{ij} = 
\tilde{\gamma}_{O_i} \,\delta_{ij}
\, ; \qquad\qquad
\tilde{c}_i = \mat{U}^{-1}_{ij}\, c_j \, .
\ee
The diagonal coefficients $\tilde{c}_i$ obey the unmixed renormalization
group equations \eqn{rge_coeff}, but with the diagonal anomalous
dimensions $\tilde{\gamma}_{O_i}$.
Therefore,
\bel{eq:sol_mix}
c_i(\mu) = \sum_{j,k} \,\mat{U}_{ij}\,
\exp{\left\{
  \int_{\alpha_0}^\alpha\, {d\alpha\over\alpha}\,
{\tilde{\gamma}_{O_j}(\alpha)\over\beta(\alpha)} \right\}}\,
\mat{U}^{-1}_{jk}\,  c_k(\mu_0)\, .
\ee

\medskip
\noindent {\it\thesubsection.2. Wilson coefficients in the Fermi Theory}
\smallskip

\noindent
We can illustrate how the previous formulae work in practice, with a
simple but important example. Let us consider the usual $W$--exchange
between two quark lines, which is responsible for the weak decays of
hadrons. At energies much lower than the $W$ mass, the interaction is
described by the four--quark Fermi coupling
\bel{eq:Fe_coup}
\cL_{\mathrm{eff}} = {G_F\over\sqrt{2}}\, V^{\phantom{*}}_{12} 
V^*_{43}\; O(1,2;3,4) , 
\ee
with
\bel{eq:O_def}
O(1,2;3,4) \equiv \left[\bar q_1\gamma^\mu (1-\gamma_5) q_2\right]\,
\left[\bar q_3\gamma_\mu (1-\gamma_5) q_4\right] .
\ee

Gluon exchanges between the quark legs induce important QCD corrections,
which are responsible for the very different behaviour observed in
strange, charm and beauty decays [all of them governed by an underlying
weak interaction of the form \eqn{eq:Fe_coup}].
The main qualitative effect generated by the exchanged gluons can
be simply understood, if one remembers the following colour
[$\lambda^a$ are Gell-Mann's matrices with Tr$(\lambda^a\lambda^b) =
2 \delta^{ab}$],
\bel{eq:colour}
\sum_a\, \lambda^a_{ij} \lambda^a_{kl} = -{2\over N_C} \,\delta_{ij}\,
\delta_{kl}
+ 2 \,\delta_{il}\,\delta_{kj} ,
\ee
and Fierz,
\bel{eq:fierz}
\left[ \gamma^\mu (1-\gamma_5) \right]_{\alpha\beta} \,
\left[ \gamma_\mu (1-\gamma_5) \right]_{\gamma\delta} =
- \left[ \gamma^\mu (1-\gamma_5) \right]_{\alpha\delta} \,
\left[ \gamma_\mu (1-\gamma_5) \right]_{\gamma\beta} ,
\ee
algebraic relations.
Thus, owing to the colour matrices introduced by the gluonic vertices,
a new four--quark operator with a permutation of two quark legs appears.
Therefore,
\bel{eq:Fe_corr}
\cL_{\mathrm{eff}} = {G_F\over 2\sqrt{2}}\, V^{\phantom{*}}_{12} 
V^*_{43}\, \left\{
c_+(\mu)\, Q_+ + c_-(\mu)\, Q_- \right\}, 
\ee
where
\bel{eq:Qpm}
Q_\pm \equiv O(1,2;3,4) \pm O(1,4;3,2) .
\ee
In the absence of gluons, $c_+ = c_- = 1$, and we recover the effective
Lagrangian \eqn{eq:Fe_coup}.
The QCD interaction modifies the values of these coefficients, which,
moreover,
will depend on the chosen renormalization scale (and scheme).
We have written the Lagrangian in terms of the operators $Q_\pm$,
because they form a diagonal basis under renormalization.

In order to describe hadronic decays we also need to compute the 
corresponding matrix elements of the four--quark operators between the
asymptotic hadronic states, 
$\langle Q_\pm(\mu)\rangle$,
which is a difficult non-perturbative problem.
At the scale $M_W$, where the underlying electroweak Lagrangian applies,
the short--distance correction induced by
the exchanged gluons is small and can be rigorously calculated
in perturbation theory; however, it is very difficult to  compute
the four--quark hadronic matrix elements at such high scale.
It seems more feasible to estimate those matrix elements at a typical
hadronic scale, where approximate non-perturbative hadronic tools are
available.
The final result for the physical amplitude
$\langle\cL_{\mathrm{eff}}\rangle $ should not depend on the chosen
renormalization scale. 
Changing the value of $\mu$ we are just shifting corrections between
the hadronic matrix elements and their Wilson coefficients.
The idea is to put all calculable short--distance ($k>\mu$) contributions
into the coefficients $c_i(\mu)$ and leave the remaining
long--distance ($k<\mu$) pieces in the matrix elements, for which a
non-perturbative calculation is required.

The calculational procedure goes as follows:

\begin{enumerate}

\item One computes the QCD corrections perturbatively at the scale $M_W$,
 using the full Standard Model.
 
\item One performs a matching with the four--quark operator description
  \eqn{eq:Fe_corr}. This gives the coefficients $c_\pm(M_W)$.
  
\item The renormalization group tells us how the short--distance coefficients
  change with the scale, which allows us to compute $c_\pm(\mu)$
  at low energies.
  
\item Finally, we choose any available non-perturbative tools to calculate
  the hadronic matrix elements at the scale $\mu$.
  
\end{enumerate}

The scale $\mu$ should be chosen low enough that we can apply hadronic
methods to estimate matrix elements, but high enough that our perturbative
approach can  still be trusted.

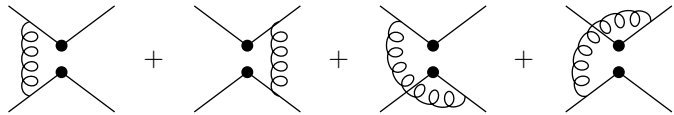
\begin{figure}[tbh]
\centering
\begin{picture}(250,40)

\Line(0,0)(8,6)\Line(8,6)(20,15)
\Line(20,15)(40,0)\GCirc(20,15)20
\Line(0,40)(8,34)\Line(8,34)(20,25)
\Line(20,25)(40,40)\GCirc(20,25)20
\Gluon(8,6)(8,34)34
\Text(55,20)[c]{$+$}

\Line(70,0)(90,15)
\Line(90,15)(102,6)\Line(102,6)(110,0)
\GCirc(90,15)20
\Line(70,40)(90,25)
\Line(90,25)(102,34)\Line(102,34)(110,40)
\GCirc(90,25)20
\Gluon(102,6)(102,34)34
\Text(125,20)[c]{$+$}

\Line(140,0)(160,15)\Line(180,0)(160,15)\GCirc(160,15)20
\Line(140,40)(160,25)\Line(180,40)(160,25)\GCirc(160,25)20
\GlueArc(165.005,24.29)(19.582,150.273,290.929)37
\Text(195,20)[c]{$+$}

\Line(210,0)(230,15)\Line(250,0)(230,15)\GCirc(230,15)20
\Line(210,40)(230,25)\Line(250,40)(230,25)\GCirc(230,25)20
\GlueArc(235.005,15.71)(19.582,69.071,209.727)37
\end{picture}
\caption{Gluon exchanges generating the one--loop anomalous dimensions.}
\label{fig:anomalous_dim}
\end{figure}

At lowest order, the calculation is very simple. 
Since the coupling $\alpha_s(M_W^2)$ is small, the uncorrected Wilson
coefficients provide a very good approximation at the $M_W$ scale,
i.e.
$c_\pm(M_W)\approx 1$.
To evolve these values to lower scales, we need to know the 
corresponding one--loop anomalous dimensions;
this only requires to compute the divergent gluonic
contributions. Moreover, owing to the conservation (for massless quarks)
of the quark currents, the vertex and 
wave--function contributions cancel among them. Thus, we only need to
consider gluonic exchanges between the two currents.
The explicit calculation of the diagrams in fig.~\ref{fig:anomalous_dim}
gives:
\bel{eq:Z_pm}
Z_+ = 1 + {\alpha_s\over 2\pi\hat{\epsilon}} + \cdots \; ;
\qquad\qquad\qquad
Z_- = 1 - {\alpha_s\over \pi\hat{\epsilon}} + \cdots 
\ee
i.e.
\bel{eq:g_pm}
\gamma_+ = {\alpha_s\over\pi} + \cdots \; ;
\qquad\qquad\qquad\qquad
\gamma_- = -2 {\alpha_s\over\pi} + \cdots
\ee
Therefore,
\beqn\label{eq:c_pm_mu}
c_\pm(\mu) & = & c_\pm(M_W) \;
\exp{ \left\{\int_{\alpha_s(M_W^2)}^{\alpha_s(\mu^2)} 
{d\alpha_s\over\alpha_s}\, 
{\gamma_\pm(\alpha_s)\over\beta(\alpha_s)}\right\} }
\no\\ & \approx &
\left( {\alpha_s(M_W^2) \over \alpha_s(\mu^2)}\right)^{a_\pm} ,
\eeqn
where $a_\pm \equiv - \gamma_\pm^{(1)}/\beta_1$.
From eqs~\eqn{eq:g_pm} and \eqn{eq:QCD_beta}, one gets:
\bel{eq:a_pm}
a_+ = {6\over 33 -2 n_f} \; ;
\qquad\qquad\qquad
a_- = {-12\over 33 -2 n_f} .
\ee

Thus, when running to lower energies,
the QCD interaction enhances the coefficient $c_-(\mu)$ and
suppresses $c_+(\mu)$ \cite{GL:74,AM:74}.
Taking $\mu=1$~GeV and $N_f=4$, we get
$c_-\approx  1.8$ and $c_+\approx 0.7$.
This is one of the crucial ingredients in the
understanding of the famous $\Delta I=1/2$ rule observed in
non-leptonic kaon decays.

A much more detailed analysis of the QCD interplay in weak transitions
(including higher--order corrections, quark--mass effects, 
additional operators, \ldots)
is given in the lectures of A.J. Buras \cite{buras}.

\subsection{Evolving from High to Low Energies}

\begin{figure}[tbh]      
\setlength{\unitlength}{0.6mm} \centering
\begin{picture}(160,120)
\thicklines

\put(65,110){\makebox(30,10){\large Large $\mu$}}
\put(5,95){\makebox(60,10){\large $\cL(\phi_i) + \cL(\phi_i,\Phi)$}}
\put(105,95){\makebox(20,10){\large $\phi_i,\Phi$}}

\put(95,80){\makebox(60,10){\large Renormalization Group}}

\put(65,55){\makebox(30,10){\large $\mu = M$}}
\multiput(10,60)(5,0){11}{\line(1,0){2.5}}
\multiput(97.5,60)(5,0){6}{\line(1,0){2.5}}

\put(125,55){\makebox(40,10){\large Matching}}

\put(5,40){\makebox(60,10){\large $\cL(\phi_i) + \delta\cL(\phi_i)$}}
\put(105,40){\makebox(20,10){\large $\phi_i$}}

\put(95,25){\makebox(60,10){\large Renormalization Group}}
\put(65,0){\makebox(30,10){\large Low $\mu$}}

\linethickness{0.3mm}
\put(80,108){\vector(0,-1){41}}
\put(80,53){\vector(0,-1){41}}

\end{picture}
\caption{Evolution from high to low scales.  \label{fig:evolution}}
\end{figure}
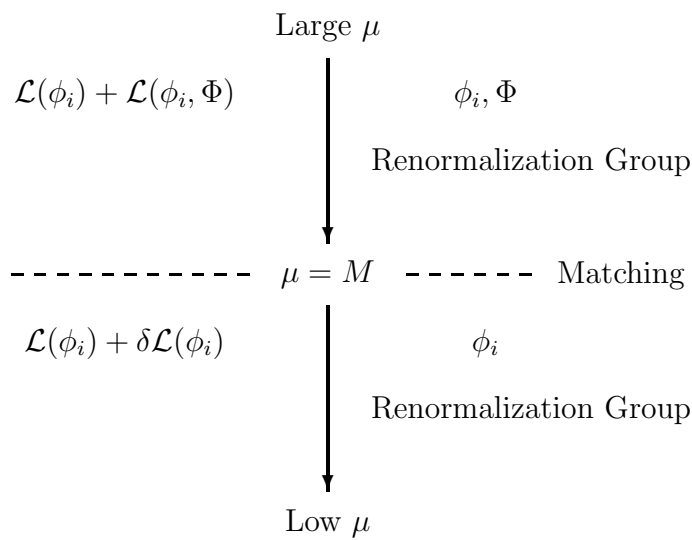

Figure~\ref{fig:evolution} shows schematically the general
procedure to evolve down in energy.
At some high scale, the physics is described by a field
(or set of fields) $\Phi$, with the heaviest mass $M$,
and a set of light--particle fields $\phi_i$. The Lagrangian,
\bel{eq:L_high}
\cL(\phi_i) + \cL(\phi_i,\Phi) ,
\ee
has a piece $\cL(\phi_i)$, which only contains light fields,
while $\cL(\phi_i,\Phi)$ encodes the dependences on the heavy field
$\Phi$ and its interactions with the lighter particles.

Using the renormalization group, one can evolve down to lower energies
up to scales of the order of the heavy mass $M$.
To proceed further down in energy, one should {\it integrate out} the
field $\Phi$; i.e. one should change to a different EFT which only
contains the light fields $\phi_i$. The Lagrangian of this new EFT takes
the form
\bel{eq:L_low}
\cL(\phi_i) + \delta\cL(\phi_i) ,
\ee
where $\delta\cL(\phi_i)$ contains a tower of operators constructed with the
light fields only, with coefficients which scale as powers of $1/M$.
Matching the high-- and low--energy theories at the scale $\mu=M$,
determines the coefficients of the new interactions.
Thus, $\delta\cL(\phi_i)$ encodes the information on the heavy field $\Phi$.
The parameters of $\cL(\phi_i)$ are not the same in the
high-- and low--energy theories; the differences are also given by
the matching conditions.

Once the matching has been performed, one can continue the evolution down
to lower scales, using the renormalization--group equations associated with
the EFT \eqn{eq:L_low}. This evolution will follow until a new particle
threshold is encountered. Then the whole procedure of integrating the
new heavy scale and matching to another EFT starts again.

In this picture, the physics is described by a chain of different
EFTs, with different particle content, which match each other at the
corresponding boundary (heavy threshold).
Each theory is the {\it low--energy} EFT of the previous underlying theory.
Going backwards in this evolution, one goes from an {\it effective} to a more
{\it fundamental} theory containing heavier scales.
One could wonder whether going up in energy should bring us at some point
to the ultimate fundamental theory of everything.
Clearly, we would stop the process at the highest physical scale we are
aware of. Thus, the word {\it fundamental} would only apply within
the context of our limited knowledge of nature.

%
%

\section{Chiral Perturbation Theory}
\label{sec:chpt}

QCD is nowadays the established theory of the strong interactions.
Owing to its asymptotic--free nature, 
perturbation theory can be applied at short distances; the resulting 
predictions have achieved a remarkable success,
explaining a wide range of phenomena where large momentum
transfers are involved.
In the low--energy domain, however, 
the growing of the running QCD coupling and the associated
confinement of quarks
and gluons make very difficult to perform a thorough analysis
of the  QCD dynamics in terms of these fundamental degrees
of freedom.
A description in terms of the hadronic asymptotic states seems
more adequate; unfortunately, given the richness of the
hadronic spectrum, this is also a formidable task.

At very low energies, a great simplification of the 
strong--interaction dynamics occurs.
Below the resonance region ($E<M_\rho$), the hadronic
spectrum only contains an octet of very light pseudoscalar
particles ($\pi$, $K$, $\eta$),
whose interactions can be easily understood with
global symmetry considerations.
This has allowed the development of a
powerful theoretical framework, Chiral Perturbation Theory
(ChPT) \cite{WE:79,GL:85}, 
to systematically analyze the low--energy implications
of the QCD symmetries.
This formalism is based on two key ingredients:
the chiral symmetry properties of QCD
and the concept of EFT.

\subsection{Chiral Symmetry}
\label{subsec:symmetry}
 
   In the absence of quark masses, the QCD Lagrangian
[$q = \hbox{\rm column}(u,d,\ldots)$]
\bel{eq:LQCD}
{\cL}_{\mathrm{QCD}}^0 = -{1\over 4}\, G^a_{\mu\nu} G^{\mu\nu}_a
 + i \bar q_L \gamma^\mu D_\mu q_L  + i \bar q_R \gamma^\mu D_\mu q_R
\ee
is invariant under
independent {\it global} $G\equiv SU(n_f)_L\otimes SU(n_f)_R$
transformations\footnote{    
Actually, the Lagrangian \eqn{eq:LQCD}
has a larger $U(n_f)_L\otimes U(n_f)_R$ global symmetry. However, the
$U(1)_A$ part is broken by quantum effects [$U(1)_A$ anomaly],
while the
quark--number symmetry $U(1)_V$ is trivially realized 
in the meson sector.}
of the left-- and right--handed quarks in flavour space:
\bel{eq:qrot}
q_L  \toG  g_L \, q_L  , \qquad
q_R  \toG  g_R \, q_R  , \qquad
g_{L,R} \in SU(n_f)_{L,R}  .
\ee
The Noether currents associated with the chiral group $G$ are:
\bel{eq:noether_currents}
J^{a\mu}_X = \bar q_X \gamma^\mu {\lambda^a\over 2} q_X ,
\qquad\qquad (X = L,R;\quad a = 1,\,\ldots,\, 8) .
\ee
The corresponding Noether charges
$Q^a_X = \int d^3x \, J^{a0}_X(x)$ satisfy the familiar 
commutation relations
\bel{eq:commutation_relations}
[Q_X^a,Q_Y^b] = i \delta_{XY} f_{abc} Q^c_X ,
\ee
which were the starting point of the Current--Algebra
methods \cite{AD:68,AL:73} of the sixties.

This chiral symmetry, which should be
approximately good in the light quark
sector ($u$,$d$,$s$), is however not seen in
the hadronic spectrum. Although hadrons can be nicely classified in
$SU(3)_V$ representations,
degenerate multiplets with opposite parity do not exist.
Moreover, the octet of pseudoscalar mesons happens to be much 
lighter than all the other hadronic states.
  To be consistent with this experimental  fact, 
the ground state of the
theory (the vacuum) should not be symmetric under the chiral group.
The $SU(3)_L \otimes SU(3)_R$ symmetry
spontaneously breaks down to
$SU(3)_{L+R}$
and, according to Goldstone's theorem \cite{GO:61},
an octet of pseudoscalar massless bosons appears in the theory.
 
More specifically,
let us consider a Noether charge $Q$,
and assume the existence of an
operator $O$ that satisfies
\bel{eq:order}
\langle 0 | [Q,O] | 0 \rangle \not= 0 \, ;
\ee
this is clearly only possible if $Q|0\rangle\not= 0$.
Goldstone's theorem then tells us that there exists a
massless state $|G\rangle$
such that
\bel{eq:Goldstone_theorem}
\langle 0|J^0|G\rangle \, \langle G|O|0\rangle\not= 0 \, .
\ee
The quantum numbers of the Goldstone boson are
dictated by those of $J^0$ and $O$.
The quantity in the left--hand side of eq.~\eqn{eq:order}
is called the order parameter of the spontaneous symmetry breakdown.
 
Since there are eight broken axial generators of the chiral
group, $Q^a_A = Q^a_R - Q^a_L$,
there should be eight pseudoscalar Goldstone states
$|G^a\rangle$, which we can identify with
the eight lightest hadronic states
($\pi^+$, $\pi^-$, $\pi^0$, $\eta$, $K^+$, $K^-$, $K^0$ 
and $\bar{K}^0$);
their small masses being generated by the quark--mass matrix,
which explicitly
breaks the global symmetry of the QCD Lagrangian.
The corresponding $O^a$ must be pseudoscalar operators. The simplest
possibility are $O^a = \bar q \gamma_5 \lambda^a q$, which satisfy
\bel{eq:vev_relation}
\langle 0|[Q^a_A, \bar q \gamma_5 \lambda^b q] |0\rangle=
-{1\over 2} \,\langle 0|\bar q \{\lambda^a,\lambda^b\} q |0\rangle =
-{2\over 3} \,\delta_{ab} \,\langle 0|\bar q q |0\rangle \, .
\ee
The quark condensate
\bel{eq:quark_condensate}
\langle 0|\bar u u |0\rangle =
\langle 0|\bar d d |0\rangle =
\langle 0|\bar s s |0\rangle \not = 0
\ee
is then
the natural order parameter of
Spontaneous Chiral Symmetry Breaking (SCSB).

\subsection{Effective Chiral Lagrangian at Lowest Order}
\label{subsec:lo}

The Goldstone nature of the pseudoscalar mesons implies strong
constraints on their interactions, which can be most easily analyzed
on the basis of an effective Lagrangian.
Since there is a mass gap separating the pseudoscalar octet from the
rest of the hadronic spectrum, we can build an EFT
containing only the Goldstone modes.
Our basic assumption is the pattern of SCSB:
\bel{eq:scsb}
G \equiv SU(3)_L\otimes SU(3)_R \;
\stackrel{\rm SCSB}{\longrightarrow} \; H \equiv SU(3)_V .
\ee

Let us denote $\phi^a$ ($a=1,\ldots,8$)  the coordinates
describing the Goldstone fields in the coset space $G/H$, and choose
a coset representative
$\bar\xi(\phi)\equiv(\xi_L(\phi),\xi_R(\phi))\in G$.
The change of the Goldstone coordinates under
a chiral transformation
$g\equiv(g_L,g_R)\in G$ is given by
\bel{eq:h_def}
\xi_L(\phi) \toG g_L\,\xi_L(\phi)\, h^\dagger(\phi,g)  ,
\qquad\quad
\xi_R(\phi) \toG g_R\,\xi_R(\phi)\, h^\dagger(\phi,g)  ,
\ee
where $h(\phi,g)\in H$ is a compensating
transformation which is needed to return to the  
given choice of coset representative $\bar\xi$;
in general, $h$ depends both on $\phi$ and $g$.
Since the same transformation $h(\phi,g)$ occurs in the
left and right sectors
(the two chiral sectors can be related by a parity transformation,
which obviously leaves $H$ invariant), 
we can get rid of it by
combining the two chiral relations in \eqn{eq:h_def}
into the simpler form
\bel{eq:u_def}
U(\phi)\, \toG\, g_R\, U(\phi)\, g_L^\dagger \, ,
\qquad\qquad    
U(\phi)\,\equiv\,\xi_R(\phi)\,\xi_L^\dagger(\phi) \, .
\ee
Moreover, without lost of generality, we can take a canonical
choice of coset representative such that
$\xi_R(\phi) = \xi_L^\dagger(\phi) \equiv u(\phi)$.
The $3\times 3$ unitary matrix
\be
U(\phi)\, = \, u(\phi)^2\, =\,
\exp{\left\{i\sqrt{2}\Phi/f\right\}}
\label{eq:u_parametrization}
\ee
gives a very convenient parametrization of the
Goldstone fields
\be
\Phi (x) \equiv {\vect{\lambda}\over\sqrt 2} \, \vect{\phi}
 = \, 
\pmatrix{{1\over\sqrt 2}\pi^0 \, + 
\, {1\over\sqrt 6}\eta_8
 & \pi^+ & K^+ \cr
\pi^- & - {1\over\sqrt 2}\pi^0 \, + \, {1\over\sqrt 6}\eta_8   
 & K^0 \cr K^- & \bar{K}^0 & - {2 \over\sqrt 6}\eta_8 }.
\label{eq:phi_matrix}
\ee
Notice that
$U(\phi)$ transforms linearly under the chiral group,
but the induced transformation on the Goldstone fields
$\vec{\phi}$ is highly non-linear.

To get a  low--energy effective Lagrangian realization of QCD, 
for the light--quark sector ($u$, $d$, $s$),
we should write the most general Lagrangian involving the matrix
$U(\phi)$, which is consistent with chiral symmetry.
The Lagrangian can be organized in terms of increasing powers of
momentum or, equivalently, in terms of an increasing number of
derivatives 
(parity conservation requires an even number of derivatives):
\be
\cL_{\mathrm{eff}}(U) \, = \, \sum_n \cL_{2n} \, .
\label{eq:l_series}
\ee
%
 
Due to the unitarity of the $U$ matrix, $U U^\dagger = I$, at least
two derivatives are required to generate a non-trivial interaction.
To lowest order, the effective chiral Lagrangian is uniquely
given by the term
\be
\cL_2 = {f^2\over 4}
\langle \partial_\mu U^\dagger \partial^\mu U \rangle \, ,
\label{eq:l2}
\ee
where $\langle A\rangle$ denotes the trace of the matrix $A$.
 
Expanding $U(\phi)$ in a power series in $\Phi$, one obtains the
Goldstone kinetic terms plus a tower of interactions involving
an increasing number of pseudoscalars.
The requirement that the kinetic terms are properly normalized
fixes the global coefficient $f^2/4$ in eq.~\eqn{eq:l2}.
All interactions among the Goldstones can then be predicted in terms
of the single coupling $f$:
\be
{\cal L}_2 \, = \, {1\over 2} \,\langle\partial_\mu\Phi
\partial^\mu\Phi\rangle
\, + \, {1\over 12 f^2} \,\langle (\Phi\lrder_\mu\Phi) \,
(\Phi\buildrel \leftrightarrow \over {\partial^\mu}\Phi)
\rangle \, + \, \cO (\Phi^6/f^4) \, .
\label{eq:l2_expanded}
\ee

To compute the $\pi\pi$ scattering amplitude, for instance, is now
a trivial perturbative exercise. One gets the well--known 
\cite{WE:66} result  [$t\equiv (p_+' - p_+)^2$]
\be
T(\pi^+\pi^0\to\pi^+\pi^0) = {t\over f^2} \, .
\label{eq:WE1}
\ee
Similar results can be obtained for 
$\pi\pi\to 4\pi, 6\pi, 8\pi, \ldots $
The non-linearity of the effective Lagrangian relates
amplitudes with different numbers of Goldstone bosons, allowing
for absolute predictions in terms of $f$.
 
The EFT
technique becomes much more powerful if one introduces couplings
to external classical fields.
Let us consider an extended QCD Lagrangian, with quark
couplings to external Hermitian matrix--valued fields
$v_\mu$, $a_\mu$, $s$, $p$\/ :
\be
\cL_{\mathrm{QCD}} = \cL^0_{\mathrm{QCD}} +
\bar q \gamma^\mu (v_\mu + \gamma_5 a_\mu ) q -
\bar q (s - i \gamma_5 p) q \, .
\label{eq:extendedqcd}
\ee
The external fields will allow us to compute the effective realization
of general Green functions of quark currents in a very straightforward
way. Moreover, they can be used to incorporate the
electromagnetic and semileptonic weak interactions, and the
explicit breaking of chiral symmetry through the quark masses:
\beqn\label{eq:breaking}
r_\mu &\equiv & v_\mu + a_\mu \, = \, e \cQ A_\mu + \ldots
\no\\
\ell_\mu & \equiv & v_\mu - a_\mu \, =
\,  e \cQ A_\mu + {e\over\sqrt{2}\sin{\theta_W}}
(W_\mu^\dagger T_+ + {\rm h.c.}) + \ldots
\\
s & = & {\cal M} + \ldots 
\no\eeqn
Here, $\cQ$ and ${\cal M}$ 
denote the quark--charge and quark--mass matrices, respectively,
\be
\cQ = {1\over 3}\, \hbox{\rm diag}(2,-1,-1)\, , \qquad\qquad
{\cal M} = \hbox{\rm diag}(m_u,m_d,m_s) \, ,
\label{eq:q_m_matrices}
\ee
and $T_+$ is a $3\times 3$ matrix containing the relevant
quark--mixing factors
\be
T_+ \, = \, \pmatrix{
0 & V_{ud} & V_{us} \cr 0 & 0 & 0 \cr 0 & 0 & 0
}.
\label{eq:t_matrix}
\ee

The Lagrangian \eqn{eq:extendedqcd}
is invariant under the following set of {\it local}
$SU(3)_L\otimes SU(3)_R$ transformations:
\beqn
q_L  &\longrightarrow &  g_L \, q_L \, , \qquad
q_R  \,\longrightarrow\,  g_R \, q_R \, , \qquad
s + i p  \,\longrightarrow\,  g_R \, (s + i p) \, g_L^\dagger \, ,
\no\\
\ell_\mu &\longrightarrow &  g_L \, \ell_\mu \, g_L^\dagger \, + \,
i g_L \partial_\mu g_L^\dagger \, ,
\qquad
r_\mu  \,\longrightarrow\,  g_R \, r_\mu \, g_R^\dagger \, + \,
i g_R \partial_\mu g_R^\dagger \, .
\label{eq:symmetry}
\eeqn
We can use this symmetry to build a generalized effective
Lagrangian for the Goldstone bosons, in the presence of external
sources. Note that to respect the local invariance, the gauge fields
$v_\mu$, $a_\mu$ can only appear through the covariant derivatives
\be
D_\mu U = \partial_\mu U - i r_\mu U + i U \ell_\mu \, ,
\quad\;\;
D_\mu U^\dagger = \partial_\mu U^\dagger  + i U^\dagger r_\mu
- i \ell_\mu U^\dagger ,
\label{eq:covariant_derivative}
\ee
and through the field strength tensors
\be
F^{\mu\nu}_L =
\partial^\mu \ell^\nu - \partial^\nu \ell^\mu 
- i [ \ell^\mu , \ell^\nu ] \, ,
\quad\;
F^{\mu\nu}_R =
\partial^\mu r^\nu - \partial^\nu r^\mu - i [ r^\mu , r^\nu ] .
\label{eq:field_strength}
\ee
At lowest order, the most general effective Lagrangian
consistent with Lorentz invariance and (local) chiral symmetry
has the form \cite{GL:85}:
\be
{\cal L}_2 = {f^2\over 4}\,
\langle D_\mu U^\dagger D^\mu U \, + \, U^\dagger\chi  \, 
+  \,\chi^\dagger U
\rangle ,
\label{eq:lowestorder}
\ee
where
\be
\chi \, = \, 2 B_0 \, (s + i p) ,
\label{eq:chi}
\ee
and $B_0$ is a constant, which, like $f$, is not fixed by
symmetry requirements alone.

Once special directions in flavour space, like the ones in
eq.~\eqn{eq:breaking}, are selected for the external fields,
chiral symmetry is of course explicitly broken.
The important point is that \eqn{eq:lowestorder} then breaks the
symmetry in exactly the same way as the fundamental short--distance
Lagrangian \eqn{eq:extendedqcd} does.

The power of the external field technique becomes obvious when
computing the chiral Noether currents.
The Green functions are obtained as functional
derivatives of the generating functional
$Z[v,a,s,p]$, defined via the path--integral formula
\beqn\label{eq:generatingfunctional}
\exp{\{i Z\}} & = &  \int  {\cal D}q \,{\cal D} \bar q
\,{\cal D}G_\mu \,
\exp{\left\{i \int d^4x\, {\cal L}_{\mathrm{QCD}}\right\}}  
\no\\ & = & 
\int  {\cal D}U \,
\exp{\left\{i \int d^4x\, {\cal L}_{\mathrm{eff}}\right\}} .
\eeqn
At lowest order in momenta, the generating functional reduces to the
classical action $S_2 = \int d^4x \,{\cal L}_2$;
therefore, the currents can be trivially
computed by taking the appropriate derivatives
with respect to the external fields:
\beqn\label{eq:l_r_currents}
J^\mu_L \doteq {\delta S_2\over \delta \ell_\mu} \, = \, & 
 {i\over 2} f^2 D_\mu U^\dagger U =
\hphantom{-}{f\over\sqrt{2}} D_\mu \Phi -
{i\over 2} \, \left(\Phi
\stackrel{\leftrightarrow}{D^\mu}\Phi\right) +
\cO (\Phi^3/f) , 
\no\\ && \\ 
J^\mu_R \doteq {\delta S_2\over \delta r_\mu} \, = \, &
 {i\over 2} f^2 D_\mu U U^\dagger =
-{f\over\sqrt{2}} D_\mu \Phi -
{i\over 2} \, \left(\Phi
\stackrel{\leftrightarrow}{D^\mu}\Phi\right) +
\cO (\Phi^3/f) . \no
\eeqn
The physical meaning of the chiral coupling $f$ is now obvious;
at $\cO (p^2)$, $f$ equals the pion decay constant,
$f = f_\pi = 92.4$ MeV, defined as
\be
\langle 0 | (J^\mu_A)^{12} | \pi^+\rangle
 \equiv i \sqrt{2} f_\pi p^\mu .
\label{f_pi}
\ee
Similarly, by taking derivatives with respect to the external scalar
and pseudoscalar
sources,
\beqn\label{eq:s_p_currents}
\bar q^j_L q^i_R
\doteq -{\delta S_2\over \delta (s-ip)^{ji}} \, = &\, 
-{f^2\over 2} B_0 \, U(\phi)^{ij} ,
\no\\
\bar q^j_R q^i_L
\doteq -{\delta S_2\over \delta (s+ip)^{ji}} \, = &\, 
-{f^2\over 2} B_0 \, U(\phi)^{\dagger ij} ,
\eeqn
we learn that the constant $B_0$ is related to the
quark condensate:
\be
\langle 0 | \bar q^j q^i|0\rangle = -f^2 B_0 \,\delta^{ij} .
\label{eq:b0}
\ee
The Goldstone bosons, parametrized by the matrix $U(\phi)$,
correspond to the zero--energy excitations over this
vacuum condensate.
 
Taking $s = {\cal M}$ and $p=0$,
the $\chi$ term in eq.~\eqn{eq:lowestorder}
gives rise to a quadratic pseudoscalar mass term plus
additional interactions proportional to the quark masses.
Expanding in powers of $\Phi$
(and dropping an irrelevant constant), one has:
\be
{f^2\over 4} 2 B_0 \,\langle {\cal M} (U + U^\dagger) \rangle
=  B_0 \left\{ - \langle {\cal M}\Phi^2\rangle
+ {1\over 6 f^2} \langle {\cal M} \Phi^4\rangle
+ \cO\left({\Phi^6\over f^4} \right) \right\} .
\label{eq:massterm}
\ee
The explicit evaluation of the trace in the quadratic 
mass term provides
the relation between the physical meson masses and the quark masses:
\beqn
M_{\pi^\pm}^2 \, & = &  2 \hat{m} B_0 \, , \qquad\qquad\qquad
M_{\pi^0}^2  =   2 \hat{m} B_0 - \varepsilon +
\cO (\varepsilon^2) \, , 
\no\\
M_{K^\pm}^2  & = &   (m_u + m_s) B_0 \, , \qquad\;\;
M_{K^0}^2  =   (m_d + m_s) B_0 \, , 
\label{eq:masses} 
\\
M_{\eta_8}^2 \; & = &   
{2\over 3} (\hat{m} + 2 m_s)  B_0 + \varepsilon +
\cO (\varepsilon^2) \, ,   \no
\eeqn
where\footnote{    
The $\cO (\varepsilon)$ corrections to $M_{\pi^0}^2$
and $M_{\eta_8}^2$
originate from a small mixing term between the
$\pi^0$ and $\eta_8$ fields:
$ \quad
- B_0 \langle {\cal M}\Phi^2\rangle \longrightarrow
- (B_0/\sqrt{3})\, (m_u - m_d)\, \pi^0\eta_8 \, .
\quad $
The diagonalization of the quadratic $\pi^0$, $\eta_8$
mass matrix, gives
the mass eigenstates,
$\pi^0 = \cos{\delta} \,\phi^3 + \sin{\delta} \,\phi^8$
and
$\eta_8 = -\sin{\delta} \,\phi^3 + \cos{\delta} \,\phi^8$,
where
$
\tan{(2\delta)} = \sqrt{3} (m_d-m_u)/\left( 2 (m_s-\hat{m})\right) .
$
}
%
\be
\hat m = {1\over 2} (m_u + m_d) \, , \qquad\qquad
\varepsilon = {B_0\over 4} {(m_u - m_d)^2\over  (m_s - \hat m)} \, .
\label{eq:mhat}
\ee

Chiral symmetry relates the magnitude of the meson and quark masses
to the size of the quark condensate.
Using the result \eqn{eq:b0}, one gets from
the first equation in \eqn{eq:masses}
the relation \cite{GOR:68}
\be
f^2_\pi M_\pi^2 = -\hat m \,\langle 0|\bar u u + \bar d d|0\rangle\, .
\label{eq:gmor}
\ee

Taking out the common $B_0$ factor, eqs.~\eqn{eq:masses} imply
the old Current--Algebra mass ratios \cite{GOR:68,WE:77},
\be
{M^2_{\pi^\pm}\over 2 \hat m} = {M^2_{K^+}\over (m_u+m_s)} =
{M_{K^0}\over (m_d+m_s)}
\approx {3 M^2_{\eta_8}\over (2 \hat m + 4 m_s)} \, ,
\label{eq:mratios}
\ee
and
(up to $\cO (m_u-m_d)$ corrections)
the Gell-Mann--Okubo \cite{GM:62,OK:62} mass relation,
\be
3 M^2_{\eta_8} = 4 M_K^2 - M_\pi^2 \, .
\label{eq:gmo}
\ee
Note that the chiral Lagrangian
automatically implies the successful quadratic
Gell-Mann--Okubo mass relation, and not a linear one.
Since $B_0 m_q\propto M^2_\phi$, the external field $\chi$
is counted as $\cO (p^2)$ in the chiral expansion.
 
Although chiral symmetry alone cannot fix the absolute values
of the quark masses, it gives information about quark--mass
ratios. Neglecting the tiny $\cO (\varepsilon)$ effects,
one gets the relations
\be
{m_d - m_u \over m_d + m_u} \, = \,
{(M_{K^0}^2 - M_{K^+}^2) - (M_{\pi^0}^2 - M_{\pi^+}^2)
\over M_{\pi^0}^2}
\, = \, 0.29 \, , 
\label{eq:ratio1}
\ee\be
{m_s -\hat m\over 2 \hat m} \, = \,
{M_{K^0}^2 - M_{\pi^0}^2\over M_{\pi^0}^2}
\, = \, 12.6 \, . 
\label{eq:ratio2}
\ee
In eq.~\eqn{eq:ratio1} we have subtracted the pion square--mass
difference, to take into account the electromagnetic contribution
to the pseudoscalar--meson self-energies;
in the chiral limit ($m_u=m_d=m_s=0$), 
this contribution is proportional
to the square of the meson charge and it is the same for 
$K^+$ and $\pi^+$ \cite{DA:69}.
The mass formulae \eqn{eq:ratio1} and \eqn{eq:ratio2}
imply the quark--mass ratios advocated by Weinberg \cite{WE:77}:
\be
m_u : m_d : m_s = 0.55 : 1 : 20.3 \, .
\label{eq:Weinbergratios}
\ee
Quark--mass corrections are therefore dominated by $m_s$, which is
large compared with $m_u$, $m_d$.
Notice that the difference $m_d-m_u$ is not small compared with
the individual up and down quark masses; in spite of that,
isospin turns out
to be a very good symmetry, because
isospin--breaking effects are governed by the small ratio
$(m_d-m_u)/m_s$.

The $\Phi^4$ interactions in eq.~\eqn{eq:massterm}
introduce mass corrections to the $\pi\pi$ scattering amplitude
\eqn{eq:WE1},
\be
T(\pi^+\pi^0\to\pi^+\pi^0) = {t - M_\pi^2\over f^2}\, ,
\label{eq:WE2}
\ee
in perfect agreement with the Current--Algebra result 
\cite{WE:66}.
Since $f=f_\pi$ is fixed from pion decay, this result
is now an absolute prediction of chiral symmetry.
 
The lowest--order chiral Lagrangian \eqn{eq:lowestorder} encodes
in a very compact way all the Current--Algebra results obtained in
the sixties \cite{AD:68,AL:73}.
The nice feature of the EFT approach is its elegant
simplicity. Moreover, 
it allows us to estimate higher--order corrections in a systematic way.

\subsection{ChPT at $\cO (p^4)$}
\label{subsec:p4}
 
At next-to-leading order in momenta, $\cO (p^4)$, the
computation of the generating functional $Z[v,a,s,p]$ involves
three different ingredients:
\begin{enumerate}

\item  The most general effective chiral Lagrangian of
$\cO (p^4)$, $\cL_4$, to be considered at tree level.

\item  One--loop graphs associated with the lowest--order
Lagrangian $\cL_2$.

\item  The Wess--Zumino--Witten \cite{WZ:71,WI:83} functional
to account for the chiral anomaly.

\end{enumerate}

\medskip
\noindent {\it\thesubsection.1. $\cO (p^4)$ Lagrangian}
\smallskip

\noindent
At $\cO (p^4)$, the most general\footnote{
Since we will only need $\cL_4$ at tree level,
the general expression of this Lagrangian has been simplified,
using the $\cO (p^2)$ equations of motion obeyed by $U$.
Moreover, a $3\times 3$ matrix relation has been used to reduce the
number of independent terms.
For the two--flavour case, not all of these terms are independent
\protect\cite{GL:85,GL:84}.}
Lagrangian, invariant under
parity, charge conjugation and
the local chiral transformations \eqn{eq:symmetry},
is given by \cite{GL:85}
\beqn\label{eq:l4}
\lefteqn{\cL_4} & & \quad =
L_1 \,\langle D_\mu U^\dagger D^\mu U\rangle^2 \, + \,
L_2 \,\langle D_\mu U^\dagger D_\nu U\rangle\,
   \langle D^\mu U^\dagger D^\nu U\rangle
\no\\ &&\mbox{}
+ L_3 \,\langle D_\mu U^\dagger D^\mu U D_\nu U^\dagger
D^\nu U\rangle\,
+ \, L_4 \,\langle D_\mu U^\dagger D^\mu U\rangle\,
   \langle U^\dagger\chi +  \chi^\dagger U \rangle
\no\\ &&\mbox{}
+ L_5 \,\langle D_\mu U^\dagger D^\mu U \left( U^\dagger\chi +
\chi^\dagger U \right)\rangle\,
+ \, L_6 \,\langle U^\dagger\chi +  \chi^\dagger U \rangle^2
\no\\ &&\mbox{}
+ L_7 \,\langle U^\dagger\chi -  \chi^\dagger U \rangle^2\,
+ \, L_8 \,\langle\chi^\dagger U \chi^\dagger U
+ U^\dagger\chi U^\dagger\chi\rangle
 \\ &&\mbox{}
- i L_9 \,\langle F_R^{\mu\nu} D_\mu U D_\nu U^\dagger +
     F_L^{\mu\nu} D_\mu U^\dagger D_\nu U\rangle\,
+ \, L_{10} \,\langle U^\dagger F_R^{\mu\nu} U F_{L\mu\nu} \rangle
\no\\ &&\mbox{}
+ H_1 \,\langle F_{R\mu\nu} F_R^{\mu\nu} +
F_{L\mu\nu} F_L^{\mu\nu}\rangle\,
+ \, H_2 \,\langle \chi^\dagger\chi\rangle \, .
\no
\eeqn

The terms proportional to $H_1$ and $H_2$ do not contain the
pseudoscalar fields and are therefore not directly measurable.
Thus, at $\cO (p^4)$ we need ten additional coupling constants
$L_i$
to determine the low--energy behaviour of the Green functions.
These constants  parametrize our
ignorance about the details of the underlying QCD dynamics.
In principle, all the chiral couplings are calculable functions
of $\Lambda_{\mathrm{QCD}}$ and the heavy--quark masses. At the present time,
however, our main source of information about these couplings
is low--energy phenomenology.

\medskip
\noindent {\it\thesubsection.2. Chiral loops}
\smallskip

\noindent
ChPT is a quantum field theory, perfectly defined through
eq.~\eqn{eq:generatingfunctional}. As such, we must take
into account quantum loops with Goldstone--boson propagators in the
internal lines. 
The chiral loops generate non-polynomial contributions,
with logarithms and threshold factors, as required by unitarity.
 
The loop integrals are homogeneous functions 
of the external momenta and
the pseudoscalar masses occurring in the propagators.
A simple dimensional counting shows that,
for a general connected diagram with $N_d$ vertices of
$\cO (p^d)$ ($d=2,4,\ldots$) and $L$ loops,
the overall chiral dimension is given by \cite{WE:79}
\bel{eq:d_counting}
D = 2 L + 2 + \sum_d N_d \, (d-2) \, .
\ee
Each loop  adds two powers of momenta;
this power suppression of loop diagrams is at the basis of low--energy
expansions, such as ChPT.
The leading $D=2$ contributions are obtained with $L=0$ and $d=2$,
i.e. only tree--level graphs with $\cL_2$ insertions.
At $\cO (p^4)$, we  have tree--level contributions from
$\cL_4$ ($L=0$, $d=4$, $N_4=1$) and one--loop graphs with the
lowest--order Lagrangian $\cL_2$ ($L=1$, $d=2$).
 
The Goldstone loops are divergent and need to be renormalized.
If we use a regularization which preserves the symmetries of
the Lagrangian, such as dimensional regularization,
the counter-terms needed to renormalize the theory will be
necessarily symmetric.
Since by construction the full effective Lagrangian
contains all terms permitted by the symmetry,
the divergences can then be absorbed in a renormalization of the
coupling constants occurring in the Lagrangian.
At one loop (in $\cL_2$), the ChPT divergences are $\cO (p^4)$
and are therefore renormalized by the low--energy couplings
in eq.~\eqn{eq:l4}:
\be
L_i = L_i^r(\mu) + \Gamma_i {\mu^{2\epsilon}\over 32 \pi^2} \left\{
{1\over \hat{\epsilon}} - 1 \right\}  , \quad
H_i = H_i^r(\mu) + \widetilde\Gamma_i {\mu^{2\epsilon}\over 32 \pi^2} 
\left\{ {1\over \hat{\epsilon}} - 1 \right\} .
\label{eq:renormalization}
\ee
The explicit calculation\footnote{
Notice that the divergent pieces are defined with the factor
${1\over\hat{\epsilon}} - 1$. This slight modification of the
$\overline{\mathrm{MS}}$ scheme is usually adopted in ChPT
calculations.
}
of the one--loop generating functional $Z_4$
\cite{GL:85} gives:
\beqn
\Gamma_1 &=& {3\over 32}\, , \quad \Gamma_2 = \frac{3}{16}\, , 
\quad  \Gamma_3 = 0 \, ,
\quad \Gamma_4 = {1\over 8} \, , \quad
\Gamma_5 = {3\over 8} \, , \quad \Gamma_6 = \frac{11}{144} \, ,
\no\\
\Gamma_7 &=& 0 \, ,
\quad \Gamma_8 = {5\over 48}\, , \quad
\Gamma_9 = {1\over 4} \, ,\quad \Gamma_{10} = -\frac{1}{4} \, , 
\quad \widetilde\Gamma_1 = -{1\over 8} \, , \quad
\widetilde\Gamma_2 = {5\over 24}\, .
\no 
\eeqn
The renormalized couplings $L_i^r(\mu)$ depend on the arbitrary
scale of dimensional regularization $\mu$.
This scale dependence is of course
cancelled by that of the loop amplitude, in any
measurable quantity.

A typical $\cO (p^4)$ amplitude will then consist of a non-polynomial
part, coming from the loop computation, plus a polynomial 
in momenta and
pseudoscalar masses, which depends on the unknown constants $L_i$.
The non-polynomial part (the so-called chiral logarithms) is
completely predicted as a function
of the lowest--order coupling $f$ and
the Goldstone masses.

This chiral structure can be easily understood in terms
of dispersion relations.
Given the lowest--order Lagrangian $\cL_2$, the non-trivial
analytic behaviour associated with  some physical intermediate state
is calculable without the introduction of new arbitrary
chiral coefficients.
Analyticity then allows us to reconstruct the full amplitude, through
a dispersive integral, up to a subtraction polynomial.
ChPT generates (perturbatively) the correct dispersion integrals and
organizes the subtraction polynomials in a derivative expansion.
 
ChPT is an expansion in powers of momenta over some typical hadronic
scale, usually called the scale of chiral symmetry breaking
$\Lambda_\chi$.
The variation of the loop contribution under a rescaling of
$\mu$, by say $e$, provides a natural order--of--magnitude 
estimate of $\Lambda_\chi$
\cite{WE:79,MG:84}:
$\Lambda_\chi\sim 4\pi f_\pi\sim 1.2\,{\rm GeV}$.
 
\medskip
\noindent {\it\thesubsection.3. The chiral anomaly}
\smallskip

\noindent
Although the QCD Lagrangian \eqn{eq:extendedqcd}
is invariant under local
chiral transformations, this is no longer true for the associated
generating functional.
The anomalies of the fermionic determinant break chiral symmetry
at the quantum level
\cite{AD:69,BA:69,BJ:69}.
The fermionic determinant can always be defined with the convention
that $Z[v,a,s,p]$ is invariant under vector transformations.
Under an infinitesimal chiral transformation
\be
g_{L,R} = 1 + i \alpha \mp i \beta + \ldots
\label{eq:inf}
\ee
the anomalous change of the generating functional
is then given by \cite{BA:69}:
\be
\delta Z[v,a,s,p]  \, = \,
-{N_C\over 16\pi^2} \, \int d^4x \,
\langle \beta(x) \,\Omega(x)\rangle \, ,
\label{eq:anomaly}
\ee
where ($\varepsilon_{0123}=1$)
\beqn\label{eq:anomaly_b}
\Omega(x) & = &\varepsilon^{\mu\nu\sigma\rho} \!
 \left[
v_{\mu\nu} v_{\sigma\rho}
+ {4\over 3} \,\nabla_\mu a_\nu \nabla_\sigma a_\rho
+ {2\over 3} i \,\{ v_{\mu\nu},a_\sigma a_\rho\}
+ {8\over 3} i \, a_\sigma v_{\mu\nu} a_\rho
\right.\no\\ &&\qquad\quad\mbox{}\left.
+ {4\over 3} \, a_\mu a_\nu a_\sigma a_\rho \right] \! ,
\eeqn
and
\be
v_{\mu\nu} \, = \,
\partial_\mu v_\nu - \partial_\nu v_\mu - i \, [v_\mu,v_\nu] \, ,
\qquad\;
\nabla_\mu a_\nu  \, = \,
\partial_\mu a_\nu - i \, [v_\mu,a_\nu] \, .
\label{eq:anomaly_c} 
\ee
%
Note that $\Omega(x)$ only depends on the external fields $v_\mu$
and $a_\mu$.
This anomalous variation of $Z$ is an $\cO (p^4)$
effect in the chiral counting.
 
So far, we have been imposing chiral symmetry to construct the
effective ChPT Lagrangian.
Since chiral symmetry is explicitly violated
by the anomaly at the fundamental QCD level,
we need to add a functional $Z_{\cA}$ with the property that its
change under a chiral gauge transformation reproduces 
\eqn{eq:anomaly}.
Such a functional was first constructed by Wess and Zumino \cite{WZ:71},
and reformulated in a nice geometrical way by Witten \cite{WI:83}.
It has the explicit form:
\beqn
\lefteqn{
S[U,\ell,r]_{\mathrm{WZW}}  =  -\,\dfrac{i N_C}{240 \pi^2}
\int d\sigma^{ijklm} \left\langle \Sigma^L_i
\Sigma^L_j \Sigma^L_k \Sigma^L_l \Sigma^L_m \right\rangle }
\no\\ &&\mbox{}
 -\,\dfrac{i N_C}{48 \pi^2} \int d^4 x\,
\varepsilon_{\mu \nu \alpha \beta}\left( W (U,\ell,r)^{\mu \nu
\alpha \beta} - W ({\bf 1},\ell,r)^{\mu \nu \alpha \beta} \right) ,
\label{eq:WZW}
\eeqn
\beqn
\lefteqn{
W (U,\ell,r)_{\mu \nu \alpha \beta}  = 
\left\langle U \ell_{\mu} \ell_{\nu} \ell_{\alpha}U^{\dg} r_{\beta}
+ \dfrac{1}{4} U \ell_{\mu} U^{\dg} r_{\nu} U \ell_\alpha U^{\dg} 
r_{\beta}
\right. }
\no\\ && \mbox{}
+ i U \partial_{\mu} \ell_{\nu} \ell_{\alpha} U^{\dg} r_{\beta}
 +  i \partial_{\mu} r_{\nu} U \ell_{\alpha} U^{\dg} r_{\beta}
- i \Sigma^L_{\mu} \ell_{\nu} U^{\dg} r_{\alpha} U \ell_{\beta}
\no\\ && \mbox{}
+ \Sigma^L_{\mu} U^{\dg} \partial_{\nu} r_{\alpha} U \ell_\beta
 -  \Sigma^L_{\mu} \Sigma^L_{\nu} U^{\dg} r_{\alpha} U \ell_{\beta}
+ \Sigma^L_{\mu} \ell_{\nu} \partial_{\alpha} \ell_{\beta}
+ \Sigma^L_{\mu} \partial_{\nu} \ell_{\alpha} \ell_{\beta}
\no\\ &&  \left.\mbox{} 
 - i \Sigma^L_{\mu} \ell_{\nu} \ell_{\alpha} \ell_{\beta}
+ \dfrac{1}{2} \Sigma^L_{\mu} \ell_{\nu} \Sigma^L_{\alpha} \ell_{\beta}
- i \Sigma^L_{\mu} \Sigma^L_{\nu} \Sigma^L_{\alpha} \ell_{\beta}
\right\rangle 
 - \left( L \leftrightarrow R \right) , 
 \label{eq:WZW2}
\eeqn
where
\be
\Sigma^L_\mu = U^{\dg} \partial_\mu U \, , \qquad\qquad
\Sigma^R_\mu = U \partial_\mu U^{\dg} \, ,
\label{eq:sima_l_r}
\ee
and
$\left( L \leftrightarrow R \right)$ stands for the interchanges
$U \leftrightarrow U^\dg $, $\ell_\mu \leftrightarrow r_\mu $ and
$\Sigma^L_\mu \leftrightarrow \Sigma^R_\mu $.
The integration in the first term of eq.~\eqn{eq:WZW} is over a
five--dimensional manifold whose boundary is four--dimensional 
Minkowski space. The integrand is a surface term; 
therefore both the first and the
second terms of $S_{\mathrm{WZW}}$ are $\cO (p^4)$, according to the chiral
counting rules.
 
Since anomalies have a short--distance origin, their effect is
completely calculable. The translation from the fundamental
quark--gluon level to the effective chiral level is unaffected by
hadronization problems.
In spite of its considerable complexity, the anomalous action
\eqn{eq:WZW} has no free parameters.
 
The anomaly functional gives rise to interactions that break
the intrinsic parity.
It is responsible for the $\pi^0\to 2\gamma$,
$\eta\to 2 \gamma$ decays, and the $\gamma 3\pi$, 
$\gamma\pi^+\pi^-\eta$
interactions;
a detailed analysis of these processes has been given in 
ref.~\cite{BI:93}.
The five--dimensional surface term generates interactions among five
or more Goldstone bosons.

\subsection{Low--Energy Phenomenology at $\cO (p^4)$}
\label{subsec:phenomenology}
 
At lowest order in momenta,
the predictive power of the chiral Lagrangian was
really impressive; with only two low--energy couplings, it was
possible to describe all Green functions associated
with the pseudoscalar--meson interactions.
The symmetry constraints become less powerful at
higher orders. Ten additional constants appear in the
$\cL_4$ Lagrangian, and many more\footnote{ 
According to a recent analysis \protect\cite{FS:96},
$\cL_6$ involves 111 (32) independent terms of even (odd)
intrinsic parity.}
would be needed at $\cO (p^6)$.
Higher--order terms in the chiral expansion
are much more sensitive
to the non-trivial aspects of the underlying QCD dynamics.
 
With $p \lap M_K \, (M_\pi)$,
we expect $\cO (p^4)$
corrections to the lowest--order amplitudes at the level
of $p^2/\Lambda_\chi^2 \lap 20\% \, (2\% )$.
We need to include those corrections if we aim to increase
the accuracy of the ChPT predictions beyond this level.
Although the number of free constants in $\cL_4$ looks
quite big, only a few of them
contribute to a given  observable.
In the absence of external fields, for instance,
the Lagrangian reduces to the first three terms; elastic
$\pi\pi$ and $\pi K$ scatterings are then sensitive to
$L_{1,2,3}$.
The two--derivative couplings $L_{4,5}$ generate mass corrections
to the meson decay constants (and
mass--dependent wave--function renormalizations).
Pseudoscalar masses are affected by the non-derivative
terms $L_{6,7,8}$;
$L_9$ is mainly responsible for the charged--meson 
electromagnetic radius
and $L_{10}$, finally, only contributes to amplitudes with at least
two external vector or axial--vector fields, like
the radiative semileptonic decay $\pi\to e\nu\gamma$.
 
Table~\ref{tab:Lcouplings} \cite{BEG:94}
summarizes the present status of the phenomenological determination
of the constants $L_i$.
The quoted numbers correspond to the
renormalized couplings, at a scale $\mu = M_\rho$.
The values of these couplings at any other 
renormalization scale can be
trivially obtained, through the logarithmic running implied by
\eqn{eq:renormalization}:
\be
L_i^r(\mu_2) \, = \, L_i^r(\mu_1) \, + \, {\Gamma_i\over (4\pi)^2}
\,\log{\left({\mu_1\over\mu_2}\right)} .
\label{eq:l_running}
\ee
%
 
\begin{table}[htb]
\caption{Phenomenological values of the
renormalized couplings $L_i^r(M_\rho)$.
The last column shows the source used to get this information.}
\label{tab:Lcouplings}
\begin{tabular}{@{}lll@{}}
\hline
$i$ & $L_i^r(M_\rho) \times 10^3$ & Source \\
\hline
1 & $\hphantom{-}0.4\pm0.3$ & $K_{e4}$, $\pi\pi\to\pi\pi$
\\
2 & $\hphantom{-}1.4\pm0.3$ & $K_{e4}$, $\pi\pi\to\pi\pi$
\\
3 & $-3.5\pm1.1$ & $K_{e4}$, $\pi\pi\to\pi\pi$
\\
4 & $-0.3\pm0.5$ &  Zweig rule
\\
5 & $\hphantom{-}1.4\pm0.5$ & $F_K : F_\pi$
\\
6 & $-0.2\pm0.3$ & Zweig rule
\\
7 & $-0.4\pm0.2$ & Gell-Mann--Okubo, $L_5$, $L_8$
\\
8 & $\hphantom{-}0.9\pm0.3$ & $M_{K^0} - M_{K^+}$, $L_5$,
$(m_s - \hat{m}) : (m_d-m_u)$
\\
9 & $\hphantom{-}6.9\pm0.7$ & $\langle r^2\rangle^\pi_V$
\\
10 & $-5.5\pm0.7$ & $\pi\to e\nu\gamma$
\\ \hline
\end{tabular}
\end{table}

Comparing the Lagrangians $\cL_2$ and $\cL_4$, 
one can make an estimate
of the expected size of the couplings $L_i$ in terms of the scale of
SCSB. Taking
$\Lambda_\chi \sim 4 \pi f_\pi \sim 1.2\, {\rm GeV}$,
one would get
\be
L_i \sim {f_\pi^2/4 \over \Lambda_\chi^2} \sim {1\over 4 (4 \pi)^2}
\sim 2\times 10^{-3} ,
\label{eq:l_size}
\ee
in reasonable agreement with the phenomenological values quoted in
table~\ref{tab:Lcouplings}.
This indicates a good convergence of the momentum expansion
below the  resonance region, i.e. $p < M_\rho$.

The chiral Lagrangian allows us to make a good book--keeping of
phenomenological information with a few couplings.
Once these couplings have been fixed,
we can predict many other quantities. In addition,
the information contained in table~\ref{tab:Lcouplings}
is very useful to easily test different QCD--inspired models.
Given any particular model aiming to correctly describe QCD at low
energies, we no longer need to make an extensive
phenomenological analysis to test its reliability; it suffices
to calculate the low--energy couplings predicted by the model, 
and compare them with the values in table~\ref{tab:Lcouplings}.
 
An exhaustive description of the chiral phenomenology
at $\cO (p^4)$  is beyond the scope of these lectures.
Instead, I will just present a few examples to
illustrate both the power and limitations of the ChPT techniques.
 
\medskip
\noindent {\it\thesubsection.1. Decay constants}
\smallskip

\noindent
In the isospin limit ($m_u = m_d = \hat m$),
the $\cO (p^4)$ calculation of the meson decay constants
gives \cite{GL:85}:
\beqn\label{eq:f_meson}
f_\pi & =  & f \left\{ 1 - 2\mu_\pi - \mu_K +
    {4 M_\pi^2\over f^2} L_5^r(\mu)
    + {8 M_K^2 + 4 M_\pi^2 \over f^2} L_4^r(\mu)
    \right\} \! , 
\no\\
f_K & =  & f \left\{ 1 - {3\over 4}\mu_\pi - {3\over 2}\mu_K
    - {3\over 4}\mu_{\eta_8}
    + {4 M_K^2\over f^2} L_5^r(\mu)
    + {8 M_K^2 + 4 M_\pi^2 \over f^2} L_4^r(\mu)
    \right\}\! , 
\no\\
f_{\eta_8} & =  & f \left\{ 1 - 3\mu_K +
    {4 M_{\eta_8}^2\over f^2} L_5^r(\mu)
    + {8 M_K^2 + 4 M_\pi^2 \over f^2} L_4^r(\mu)
    \right\}\! , 
\eeqn
where
\be
\mu_P \equiv {M_P^2\over 32 \pi^2 f^2} \,
\log{\left( {M_P^2\over\mu^2}\right)} .
\label{eq:mu_p}
\ee
The result depends on two $\cO (p^4)$ couplings, $L_4$ and $L_5$.
The $L_4$ term generates a universal shift of all meson decay
constants,
$\delta f^2 = 16 L_4 B_0 \langle\cM\rangle$,
which can be eliminated taking ratios.
From the experimental value \cite{LR:84}
\be
{f_K\over f_\pi} = 1.22\pm 0.01 \, ,
\label{eq:f_k_pi_ratio}
\ee
one can then fix $L_5(\mu)$; this gives the result quoted in
table~\ref{tab:Lcouplings}.
Moreover, one gets the absolute prediction 
\cite{GL:85}
\be
{f_{\eta_8}\over f_\pi} = 1.3 \pm 0.05 \, .
\label{eq:f_eta_pi_ratio}
\ee
Taking into account isospin violations, one can also predict
\cite{GL:85} a tiny
difference between $f_{K^\pm}$ and $f_{K^0}$, proportional to
$m_d-m_u$.

\medskip
\noindent {\it\thesubsection.2. Electromagnetic form factors}
\smallskip

\noindent
At $\cO (p^2)$ the electromagnetic coupling of the Goldstone bosons
is just the minimal one, obtained through the covariant derivative.
The next--order corrections generate a momentum--dependent
form factor:
\be
F^{\phi^\pm}_V(q^2) = 1 + {1\over 6} \,
\langle r^2 \rangle^{\phi^\pm}_V \, q^2 + \ldots \quad ;
\quad
F^{\phi^0}_V(q^2) =  {1\over 6} \,
\langle r^2 \rangle^{\phi^0}_V \, q^2 + \ldots
\label{eq:ff}
\ee
The meson electromagnetic radius
$\langle r^2 \rangle^\phi_V$
gets local contributions from the $L_9$ term,
plus logarithmic loop corrections \cite{GL:85}:
\beqn\label{eq:radius}
\langle r^2 \rangle^{\pi^\pm}_V & = & {12 L^r_9(\mu)\over f^2}
    - {1\over 32 \pi^2 f^2} \left\{
   2 \log{\left({M_\pi^2\over\mu^2}\right)}
    + \log{\left({M_K^2\over\mu^2}\right)} + 3 \right\} ,
\no\\
\langle r^2 \rangle^{K^0}_V & =  & - {1\over 16 \pi^2 f^2}
\,\log{\left({M_K\over M_\pi}\right) } ,
\\
\langle r^2 \rangle^{K^\pm}_V & = & \langle r^2 \rangle^{\pi^\pm}_V
   + \langle r^2 \rangle^{K^0}_V . 
\no
\eeqn
Since neutral bosons do not couple to the photon at tree level,
$\langle r^2 \rangle^{K^0}_V$
only gets a loop contribution, which is moreover finite
(there cannot be any divergence because there
exists no counter-term to renormalize it).
The predicted value,
$\langle r^2 \rangle^{K^0}_V = -0.04\pm 0.03 \, {\rm fm}^2$, is in
perfect agreement with the experimental determination \cite{MO:78}
$\langle r^2 \rangle^{K^0}_V = -0.054\pm 0.026 \, {\rm fm}^2$.
 
The measured electromagnetic pion radius,
$\langle r^2 \rangle^{\pi^\pm}_V = 0.439\pm 0.008 \, {\rm fm}^2$
\cite{AM:86},
is used as input to estimate the coupling $L_9$.
This observable provides  a good example of the importance of
higher--order local terms in the chiral expansion \cite{LE:89}.
If one tries to ignore the $L_9$ contribution, using instead some
{\it physical} cut-off $p_{\rm max}$ to regularize the  loops,
one needs $p_{\rm max}\sim 60 \, {\rm GeV}$,
in order to reproduce the experimental value; this is clearly
nonsense.
The pion charge radius is dominated by the $L^r_9(\mu)$
contribution, for any reasonable value of $\mu$.
 
The measured $K^+$ charge radius \cite{DA:82},
$\langle r^2 \rangle^{K^\pm}_V = 0.28\pm 0.07 \, {\rm fm}^2$,
has a larger experimental uncertainty.
Within present errors, it is in agreement with the parameter--free
relation in eq.~\eqn{eq:radius}.

\medskip
\noindent {\it\thesubsection.3. $K_{l3}$ decays}
\smallskip

\noindent
The semileptonic decays $K^+\to\pi^0 l^+ \nu_l$ and
$K^0\to\pi^- l^+ \nu_l$ are governed by the corresponding
hadronic matrix
elements of the vector current,
\be
\langle \pi| \bar s\gamma^\mu u |K\rangle = C_{K\pi} \,\left[
\left( P_K + P_\pi\right)^\mu \, f_+^{K\pi}(t) \, + \,
\left( P_K - P_\pi\right)^\mu \, f_-^{K\pi}(t) \right] ,
\label{eq:vector_matrix}
\ee
where $t\equiv (P_K-P_\pi)^2$,
$C_{K^+\pi^0} = -1/\sqrt{2}$ and $C_{K^0\pi^-} = -1$.
At lowest order, the two form factors reduce to trivial constants:
$f_+^{K\pi}(t) = 1$ and $f_-^{K\pi}(t) = 0$.
There is however a sizeable correction to $f_+^{K^+\pi^0}(t)$,
due to $\pi^0\eta$ mixing, which is
proportional to $(m_d-m_u)$,
\be
f_+^{K^+\pi^0}(0) \, = \,
1 + {3\over 4} \, {m_d-m_u\over m_s - \hat m}
\, = \, 1.017 \, .
\label{eq:fp_kp_p0}
\ee
This number should be compared with the experimental ratio
\be
{f_+^{K^+\pi^0}(0)\over f_+^{K^0\pi^-}(0)} \, = \,
1.028 \pm 0.010 \, .
\label{eq:expratio}
\ee
The $\cO (p^4)$ corrections to $f_+^{K\pi}(0)$ can be expressed in
a parameter--free manner in terms of the physical meson masses
\cite{GL:85}.
Including those contributions,
one gets the more precise values
\be
 f_+^{K^0\pi^-}(0) = 0.977 \, , \qquad \qquad
{f_+^{K^+\pi^0}(0)\over f_+^{K^0\pi^-}(0)} = 1.022 \, ,
\label{eq:fp_predictions}
\ee
which are in perfect agreement with the experimental result
\eqn{eq:expratio}.
The accurate ChPT calculation of these quantities allows us to
extract \cite{LR:84} the most precise determination
of the Cabibbo--Kobayashi--Maskawa matrix element $V_{us}$:
\be
|V_{us}| = 0.2196 \pm 0.0023 \, .
\label{eq:v_us}
\ee

At $\cO (p^4)$, the form factors get momentum--dependent contributions.
Since $L_9$ is
the only unknown chiral coupling occurring in $f_+^{K\pi}(t)$ at this
order, the slope
$\lambda_+$
of this form factor can be fully predicted:
\be
\lambda_+ \equiv {1\over 6 }\,\langle r^2\rangle^{K\pi}_V\, M_\pi^2
= 0.031\pm 0.003 \, .
\label{eq:slope}
\ee
This number is in excellent agreement with the experimental
determinations \cite{PDG:96}, 
$\lambda_+ = 0.0300\pm 0.0016$ ($K^0_{e3}$) and
$\lambda_+ = 0.0286\pm 0.0022$ ($K^\pm_{e3}$).

Instead of $f_-^{K\pi}(t)$, it is usual to parametrize the
experimental results in terms of the so-called
scalar form factor
\be
f_0^{K\pi}(t) = f_+^{K\pi}(t) +{t\over M_K^2 - M_\pi^2} f_-^{K\pi}(t) 
\, .
\label{eq:scalar_ff}
\ee
The slope of this form factor is determined by the constant $L_5$,
which in turn is fixed by $f_K/f_\pi$.
One gets the result \cite{GL:85}:
\be
\lambda_0 \equiv {1\over 6 }\,\langle r^2\rangle^{K\pi}_S\, M_\pi^2
= 0.017\pm 0.004 \, .
\label{eq:slope2}
\ee
The experimental situation concerning the
value of this slope is far from clear. 
The Particle Data Group \cite{PDG:96} quotes a world average
$\lambda_0 = 0.025\pm 0.006$.
 
\medskip
\noindent {\it\thesubsection.4. Meson and quark masses}
\smallskip

\noindent
The mass relations \eqn{eq:masses} get modified at $\cO (p^4)$.
The additional contributions depend on the low--energy constants
$L_4$, $L_5$, $L_6$, $L_7$ and $L_8$.
It is possible, however, to obtain one relation between the
quark and meson masses, which does not contain any of the $\cO (p^4)$
couplings.
The dimensionless ratios
\be
Q_1 \equiv {M_K^2 \over M_\pi^2} \, , \qquad\qquad
Q_2 \equiv {(M_{K^0}^2 - M_{K^+}^2)_{\mathrm{QCD}} 
    \over M_K^2 - M_{\pi}^2}\,  ,
\label{eq:q1q2_def}
\ee
get  the same $\cO (p^4)$ correction \cite{GL:85}:
\be
Q_1 = {m_s + \hat m \over 2 \hat m} \, \{ 1 + \Delta_M\} ,
\qquad\qquad
Q_2 = {m_d - m_u \over m_s - \hat m} \, \{ 1 + \Delta_M\} ,
\label{eq:q1q2}
\ee
where
\be
\Delta_M = - \mu_\pi + \mu_{\eta_8} + {8\over f^2}\,
(M_K^2 - M_\pi^2)\, \left[ 2 L_8^r(\mu) - L_5^r(\mu)\right] .
\label{eq:delta_M}
\ee
Therefore, at this order, the ratio $Q_1/Q_2$ is just given
by the corresponding ratio of quark masses,
\be
Q^2 \equiv {Q_1\over Q_2} =
{m_s^2 - \hat m^2 \over m_d^2 - m_u^2} \, .
\label{eq:Q2}
\ee
To a good approximation, eq.~\eqn{eq:Q2}
can be written as an ellipse, which constrains the quark--mass ratios:
\be
\left({m_u\over m_d}\right)^2 + {1\over Q^2}\,
\left({m_s\over m_d}\right)^2 = 1 \, .
\label{eq:ellipse}
\ee

Obviously, the quark--mass ratios \eqn{eq:Weinbergratios},
obtained at $\cO (p^2)$, satisfy this elliptic constraint.
At $\cO (p^4)$, however, it is not possible to make a separate
determination of $m_u/m_d$ and $m_s/m_d$ without having additional
information on some of the $L_i$ couplings.
 
In order to determine the individual quark--mass ratios
from eqs.~\eqn{eq:q1q2}, we would need to fix
the constant $L_8$. However, there is no way to find an observable
that isolates this coupling.
The reason is an accidental symmetry of the Lagrangian 
$\cL_2 + \cL_4$, which remains invariant under
the following simultaneous change \cite{KM:86}
of the quark--mass matrix and some of the chiral couplings:
\beqn\label{eq:kmsymmetry}
\cM' & = & \alpha \,\cM + \beta\, (\cM^\dagger)^{-1} \, \det\cM\, ,
\qquad B_0' \, = \, B_0 / \alpha\, ,
\no\\
L'_6 & = & L_6 - \zeta \, , \qquad
L'_7 \, = \, L_7 - \zeta \, , \qquad
L'_8 \, = \, L_8 + 2 \zeta \, , 
\eeqn
where $\alpha$ and $\beta$ are arbitrary constants, and
$\zeta = \beta f^2 / (32\alpha B_0)$.
The only information on the quark--mass matrix $\cM$ that we used
to construct the effective Lagrangian was that it transforms as
$\cM\to g_R \cM g_L^\dagger$.
The matrix $\cM'$ transforms in the same manner;
therefore, symmetry alone does not allow us to distinguish between
$\cM$ and $\cM'$.
Since only the product $B_0 \cM$ appears in the Lagrangian,
$\alpha$ merely changes the value of the constant $B_0$.
The term proportional to $\beta$ is a correction of $\cO (\cM^2)$;
when inserted in $\cL_2$, it generates a contribution to
$\cL_4$, which is reabsorbed by the redefinition
of the $\cO (p^4)$ couplings.
All chiral predictions will be invariant under the transformation
\eqn{eq:kmsymmetry}; therefore it is not possible to
separately determine the values of the quark masses and the
constants $B_0$, $L_6$, $L_7$ and $L_8$.
We can only fix those combinations of chiral couplings and masses
that remain invariant under \eqn{eq:kmsymmetry}.
 
We can resolve the ambiguity by obtaining
one additional information from outside the pseudoscalar--meson
chiral Lagrangian framework.
For instance, by analyzing the
isospin breaking in the baryon mass spectrum and the $\rho$--$\omega$
mixing \cite{GL:82}, it is possible to fix the ratio
\be
R\,\equiv\, {m_s - \hat m\over m_d - m_u}\, =\, 43.7\pm 2.7 \, .
\label{eq:r}
\ee
Inserting this number in \eqn{eq:Q2}, the two separate
quark--mass ratios can be obtained.
Moreover, one can then determine  $L_8$ from \eqn{eq:q1q2}.

The meson masses in \eqn{eq:q1q2_def} refer to pure QCD;
using the Dashen theorem \cite{DA:69}
$(\Delta M^2_K - \Delta M^2_\pi)_{\mathrm{em}}\equiv
(M^2_{K^+} - M^2_{K^0} - M^2_{\pi^+} + M^2_{\pi^0})_{\mathrm{em}} = 0$
to correct for the electromagnetic contributions,
the observed values of the meson masses give $Q = 24$.
Taking the conservative range
$(\Delta M^2_K - \Delta M^2_\pi)_{\mathrm{em}} =
(0.75\pm 0.75)\times 10^{-3} \, {\rm GeV}^2$
as an estimate of the violation of Dashen theorem at $\cO (e^2\cM )$,
one gets the corrected value $Q=22.7\pm 1.4$. This implies \cite{PI:95}:
\be
{m_s \over \hat m} = 22.6 \pm 3.3 \, ,
\qquad\qquad
{m_d - m_u \over 2 \hat m} \, = \, 0.25\pm 0.04 \, .
\label{eq:ms_m_ratio_2}
\ee

\subsection{The Role of Resonances in ChPT}
\label{subsec:resonances}
 
It seems rather natural to expect that the lowest--mass resonances,
such as $\rho$ mesons, should have an important impact on the physics
of the pseudoscalar bosons. 
Below the $\rho$ mass scale, the singularity associated with the
pole of the resonance propagator is replaced by the corresponding
momentum expansion; therefore, the exchange of virtual $\rho$ mesons generates
derivative Goldstone couplings proportional to powers of
$1/M_\rho^2$.

A systematic analysis of the role of resonances in the ChPT Lagrangian
was performed in ref.~\cite{EGPR:89}.
One writes first a general chiral--invariant Lagrangian
$\cL(U,V,A,S,P)$, describing the couplings of meson resonances
of the type $V(1^{--})$, $A(1^{++})$, $S(0^{++})$ and $P(0^{-+})$ to
the Goldstone bosons, at lowest--order in derivatives. The coupling
constants of this Lagrangian are phenomenologically extracted from
physics at the resonance mass scale. One has then an effective chiral
theory defined in the intermediate energy region. The
generating functional \eqn{eq:generatingfunctional} is given in this
theory by the path-integral formula
$$
\exp{\{i Z\}} \, = \,
\int \, {\cal D}U(\phi)\, \cD V \,\cD A \,\cD S \,\cD P
\, \exp{\left\{ i \int d^4x \,\cL(U,V,A,S,P) \right\}} .
$$
The integration of the heavy fields  leads to a low--energy
theory with only Goldstone bosons.
At lowest order, this integration can be explicitly performed by
expanding around the classical solution for the resonance fields. 
Expanding the resulting non-local action
in powers of momenta, one gets then the local ChPT Lagrangian.

The formal procedure to introduce higher--mass states in the
chiral Lagrangian was first discussed by Coleman \etal\ \cite{CO:69,CA:69}.
The wanted ingredient for a non-linear representation of the chiral
group is the compensating $SU(3)_V$ transformation $h(\phi,g)$ which
appears under the action of $G$ on the
coset representative $u(\phi)$
[see eqs.~\eqn{eq:h_def} to \eqn{eq:u_parametrization}]:
\be
u(\phi)\,\toG\, g_R\,u(\phi)\, h^\dagger(\phi,g)\, = \,
h(\phi,g)\,u(\phi)\, g_L^\dagger \, . 
\label{eq:h_def_2}
\ee

In practice, we shall only be                  
interested in resonances transforming as octets or singlets under               
$SU(3)_V$. Denoting  the resonance multiplets generically by                    
$R= \vect{\lambda}\vect{R}/\sqrt{2}$  (octet) and $R_1$ (singlet), the            
non-linear realization of $G$ is given by                                      
\be
R\,\toG\, h(\phi,g) \; R \;                        
h(\phi,g)^\dagger                                                                 
\, ,                                                                            
\qquad\qquad R_1\,\toG\, R_1 \, .                                               
\label{eq:R_transformation}
\ee
Since the action of $G$ on the octet field $R$
is local, we are led to define a covariant derivative                                                                      
\be                                                                    
\nabla_\mu R \,=\, \partial_\mu R + [\Gamma_\mu,R] \, ,
\label{eq:d_covariant}                         
\ee
with                                                                            
\be                                                        
\Gamma_\mu \,=\,                                                                
\frac{1}{2} \left\{ u^\dagger (\partial_\mu - ir_\mu) u +                   
  u(\partial_\mu - i\ell_\mu) u^\dagger \right\}                               
\label{eq:connection}                                                                             
\ee
ensuring the proper transformation                                              
\be                                                                            
\nabla_\mu R \,\toG\, h(\phi,g)\,                                                 
\nabla_\mu R \,\, h(\phi,g)^\dagger\, .
\label{eq:dc_transf}                                             
\ee
Without external fields, $\Gamma_\mu$ is the usual natural connection           
on  coset space.                                                                

To determine the resonance--exchange contributions to the effective             
chiral  Lagrangian, we need the lowest--order couplings
to the pseudoscalar Goldstones
which are linear in the resonance fields. 
It is useful to define objects transforming as $SU(3)_V$ octets:
\beqn\label{eq:octet_objects}
u_\mu &\equiv & i u^\dagger D_\mu U u^\dagger\, =\, u_\mu^\dagger\, ,                                                               
\no\\                                                      
\chi_\pm &\equiv & u^\dagger \chi u^\dagger \pm u \chi^\dagger u\, ,            
\\
f^{\mu\nu}_\pm & = & u F_L^{\mu\nu} u^\dagger \pm                    
u^\dagger   F_R^{\mu\nu} u \, .        \no                                  
\eeqn
Invoking $P$ and $C$ invariance, 
the relevant lowest--order Lagrangian can be written as
\cite{EGPR:89}
\be
\cL_{\mathrm{R}} \, =\,                  
\sum_{R=V,A,S,P} \left\{\cL_{\mathrm{Kin}}(R) + \cL_2(R)\right\}            
\, ,                                                                            
\label{eq:res_Lagrangian}                                                                            
\ee
with kinetic terms\footnote{
The vector and axial--vector mesons are described
in terms of antisymmetric tensor fields $V_{\mu\nu}$
and $A_{\mu\nu}$    \protect\cite{GL:84,EGPR:89}                             
instead of the more familiar vector fields.} 
%
\beqn\label{eq:kin_s} 
\cL_{\mathrm{Kin}}(R = V,A) & = & 
    - {1\over 2}\, \langle \nabla^\lambda R_{\lambda\mu}                      
\nabla_\nu R^{\nu\mu} -{M^2_R\over 2} \, R_{\mu\nu} R^{\mu\nu}\rangle
\no\\ &&\mbox{}
-{1\over 2}\, \partial^\lambda R_{1,\lambda\mu}
\partial_\nu R_1^{\nu\mu} +
{M^2_{R_1}\over 4} \, R_{1,\mu\nu} R_1^{\mu\nu} ,
\\
\cL_{\mathrm{Kin}}(R = S,P) & = & {1\over 2} \,
\langle \nabla^\mu R 
\nabla_\mu R - M^2_R R^2\rangle
+ {1\over 2}  \partial^\mu R_1 \partial_\mu R_1 -
 {M^2_{R_1}\over 2} R_1^2\, , 
\no                                                                        
\eeqn
where $M_R$, $M_{R_1}$ are the corresponding masses in the chiral               
limit.  The interactions $\cL_2(R)$ read                
\beqn\label{eq:R_int}                                                                             
\cL_2[V(1^{--})] & = &  {F_V\over 2\sqrt{2}} \,
     \langle V_{\mu\nu} f_+^{\mu\nu}\rangle +                                   
    {iG_V\over \sqrt{2}} \, \langle V_{\mu\nu} u^\mu u^\nu\rangle ,                                                                            
\no\\  
\cL_2[A(1^{++})] & = & {F_A\over 2\sqrt{2}} \,                                                                    
    \langle A_{\mu\nu} f_-^{\mu\nu} \rangle  ,                                
\\   
\cL_2[S(0^{++})]  & = &  c_d \, \langle S u_\mu            
u^\mu\rangle + c_m \, \langle S \chi_+ \rangle +                                
   \tilde c_d \, S_1 \, \langle u_\mu u^\mu \rangle +                              
    \tilde c_m \, S_1 \, \langle \chi_+\rangle  ,                                
\no\\    
\cL_2[P(0^{-+})] & = &  id_m \, \langle P \chi_-                                                                          
\rangle +  i \tilde d_m \, P_1 \langle \chi_-\rangle  .                          
\no   
\eeqn
All coupling constants are real.
The octet fields are written in the usual matrix notation                       
$$                                                                          
V_{\mu\nu} = {\vect{\lambda}\over\sqrt{2}}\vect{V}_{\mu\nu}  =            
\pmatrix{                                                       
{1\over\sqrt{2}}\rho^0_{\mu\nu} + {1\over \sqrt{6}}\omega_{8,\mu\nu} 
& \rho^+_{\mu\nu} & K^{*+}_{\mu\nu}
\cr                                                                 
\rho^-_{\mu\nu} & - {1\over\sqrt{2}}\rho^0_{\mu\nu} + 
{1\over \sqrt{6}}\omega_{8,\mu\nu} & K^{\, *0}_{\mu\nu}
\cr
K^{*-}_{\mu\nu} & \bar{K}^{*0}_{\mu\nu} 
& -{2\over \sqrt{6}}\omega_{8,\mu\nu}                                                             
}  ,                                                
$$
and similarly for the other octets. We observe that for $V$ and $A$             
only  octets can couple whereas both octets and singlets appear for             
$S$ and $P$  (always to lowest order $p^2$).

Vector--meson exchange generates contributions to
$L_1$, $L_2$, $L_3$, $L_9$ and $L_{10}$ \cite{EGPR:89,DRV:89}, while
$A$ exchange only contributes to $L_{10}$ \cite{EGPR:89}:
\beqn
L_1^V & = & {G_V^2\over 8 M_V^2}\, , \qquad L_2^V = 2 L_1^V ,
\qquad L_3^V = -6 L_1^V , 
\no\\  L_9^V & = & {F_V G_V\over 2 M_V^2}\, , \qquad
L_{10}^{V+A} = - {F_V^2\over 4 M_V^2} + {F_A^2\over 4 M_A^2} \, . 
\label{eq:vmd_results}
\eeqn

To fix the vector--meson parameters, one takes $M_V=M_\rho$,
$|F_V|= 154$ MeV (from $\rho^0\to e^+e^-$) and
$|G_V| = 53$ MeV (from the electromagnetic pion radius, i.e. from $L_9$)
\cite{EGPR:89}.
The axial parameters can be determined using the old Weinberg
sum rules \cite{WE:67b,deRafael}:
$F_A^2 = F_V^2 - f_\pi^2 = (123 \, {\rm MeV})^2$ and $M_A^2 = M_V^2
F_V^2/ F_A^2 = (968 \, {\rm MeV})^2$.
The resulting values of the $L_i$ couplings \cite{EGPR:89} 
are summarized in
table~\ref{tab:vmd}, which compares the different resonance--exchange
contributions with the phenomenologically determined values of
$L_i^r(M_\rho)$. 
The results shown in the table
clearly establish a chiral version of vector (and axial--vector) meson
dominance: whenever they can contribute at all, $V$ and $A$ exchange
seem to completely dominate the relevant coupling constants.

\begin{table}[htb]
\caption{$V$, $A$, $S$, $S_1$ and $\eta_1$ contributions to the
 coupling constants $L_i^r$ in units of $10^{-3}$.
 The last column shows the results obtained using the relations
 \protect\eqn{eq:vmdpred}.}
\label{tab:vmd}
\begin{tabular}{@{}llllllcll@{}}
\hline
 i & $\quad L_i^r(M_\rho)$ & $\hphantom{0.0}V$ & $A\,$ & $\quad S$ &
      $S_1$ & $\eta_1$ & Total & Total$^{c)}$
\\ \hline
 1 & $\hphantom{-}0.4\pm0.3$ &
      $\hphantom{-1}0.6$ & $0\hphantom{.0}$ & $-0.2$ &
      $0.2^{b)}$ & $0$ & $\hphantom{-}0.6$ & $\hphantom{-}0.9$
\\
 2 & $\hphantom{-}1.4\pm0.3$ &
      $\hphantom{-1}1.2$ & $0\hphantom{.0}$ & $\hphantom{-}0$ &
      $0\hphantom{.2^{b)}}$ & $0$ & $\hphantom{-}1.2$
      & $\hphantom{-}1.8$
\\
 3 & $-3.5\pm1.1$ &
      $\;\, -3.6$ & $0\hphantom{.0}$ & $\hphantom{-}0.6$ &
      $0\hphantom{.2^{b)}}$ & $0$ & $-3.0$ & $-4.9$
\\
 4 & $-0.3\pm0.5$ &
      $\hphantom{-1}0\hphantom{.0}$ & $0\hphantom{.0}$ & $-0.5$ &
      $0.5^{b)}$ & $0$ & $\hphantom{-}0.0$ & $\hphantom{-}0.0$
\\
 5 & $\hphantom{-}1.4\pm0.5$ &
      $\hphantom{-1}0\hphantom{.0}$ & $0\hphantom{.0}$ &
      $\hphantom{-}1.4^{a)}$ &
      $0\hphantom{.2^{b)}}$ & $0$ & $\hphantom{-}1.4$
      & $\hphantom{-}1.4$
\\
 6 & $-0.2\pm0.3$ &
      $\hphantom{-1}0\hphantom{.0}$ & $0\hphantom{.0}$ & $-0.3$ &
      $0.3^{b)}$ & $0$ & $\hphantom{-}0.0$ & $\hphantom{-}0.0$
\\
 7 & $-0.4\pm0.2$ &
      $\hphantom{-1}0\hphantom{.0}$ & $0\hphantom{.0}$ & $\hphantom{-}0$ &
      $0\hphantom{.2^{b)}}$ & $-0.3$ & $-0.3$ & $-0.3$
\\
 8 & $\hphantom{-}0.9\pm0.3$ &
      $\hphantom{-1}0\hphantom{.0}$ & $0\hphantom{.0}$ &
      $\hphantom{-}0.9^{a)}$ &
      $0\hphantom{.2^{b)}}$ & $0$ & $\hphantom{-}0.9$
      & $\hphantom{-}0.9$
\\
 9 & $\hphantom{-}6.9\pm0.7$ &
      $\hphantom{-1}6.9^{a)}$ & $0\hphantom{.0}$ & $\hphantom{-}0$ &
      $0\hphantom{.2^{b)}}$ & $0$ & $\hphantom{-}6.9$
      & $\hphantom{-}7.3$
\\
 10 & $-5.5\pm0.7$ &
      $-10.0$ & $4.0$ & $\hphantom{-}0$ &
      $0\hphantom{.2^{b)}}$ & $0$ & $-6.0$ & $-5.5$
\\ \hline
\end{tabular}
\centerline{\footnotesize
     $^{a)}$ Input. $\qquad$
     $^{b)}$ Large--$N_C$ estimate. $\qquad$
     $^{c)}$ With \eqn{eq:vmdpred}.}
\end{table}
 
There are different phenomenologically successful models in the
literature for $V$ and $A$ resonances (tensor--field description
\cite{GL:84,EGPR:89}, massive Yang--Mills
\cite{ME:88}, hidden gauge formulation \cite{BKY:88},
etc.). It can be shown \cite{EGLPR:89}   that all models are
equivalent  (i.e. they give the same contributions to the $L_i$), 
provided
they incorporate the appropriate QCD constraints at high energies.
Moreover, with additional QCD--inspired assumptions of high--energy
behaviour, such as an unsubtracted dispersion relation for the pion
electromagnetic form factor, all $V$ and $A$ couplings can be
expressed in terms of
$f_\pi$ and $M_V$ only \cite{EGLPR:89}:
\be
F_V = \sqrt{2} f_\pi\, , \quad G_V= f_\pi/\sqrt{2}\, , \quad
F_A=f_\pi\, , \quad M_A=\sqrt{2} M_V\, .
\label{eq:f_relations}
\ee
In that case, one has
\be
L_1^V = L_2^V/2 = - L_3^V/6 = L_9^V/8 = -L_{10}^{V+A}/6 =
f_\pi^2/(16 M_V^2) \, .
\label{eq:vmdpred}
\ee
The last column in table~\ref{tab:vmd} shows the predicted
numerical values of the $L_i$ couplings, using the relations
\eqn{eq:vmdpred}.
 
The exchange of scalar resonances
generates the contributions \cite{EGPR:89}:
\beqn\label{eq:s_exchange}
L_1^{S+S_1} & = & -{c_d^2\over 6 M_S^2} +
     {\tilde{c}_d^2\over 2 M_{S_1}^2} \, , 
\qquad
L_3^S = {c_d^2\over 2 M_S^2} \, , 
\no\\ 
L_4^{S+S_1} & = & -{c_d c_m\over 3 M_S^2} +
     {\tilde{c}_d \tilde{c}_m\over M_{S_1}^2} \, , 
\qquad\;
L_5^S = {c_d c_m\over M_S^2} \, , 
\\ 
L_6^{S+S_1} & = & -{c_m^2\over 6 M_S^2} +
     {\tilde{c}_m^2\over 2 M_{S_1}^2} \, , 
\qquad
L_8^S = {c_m^2\over 2 M_S^2} \, . 
\no
\eeqn
Since the experimental information is quite scarce in the scalar
sector, one needs to assume that the couplings $L_5$ and $L_8$ are
due exclusively to scalar--octet exchange, to determine the
couplings $c_d$ and $c_m$.
The $S_1$--exchange contributions can be expressed in terms of
the octet parameters using large--$N_C$ arguments. For $N_C=\infty$,
$M_{S_1}=M_S$, $|\tilde{c_d}|=|c_d|/\sqrt{3}$ and
$|\tilde{c_m}|=|c_m|/\sqrt{3}$ \cite{EGPR:89}; therefore,
octet-- and singlet--scalar exchange cancel in $L_1$, $L_4$ and $L_6$.
Taking $M_S=M_{a_0}=983$ MeV, one gets then 
the numbers in table~\ref{tab:vmd}. Although these results
 cannot be considered as a
proof for scalar dominance, they provide at least a convincing
demonstration of its consistency.

Neglecting the higher--mass
$0^{-+}$ resonances, the only  remaining meson--exchange is the one
associated with the $\eta_1$, which generates a sizeable contribution
to $L_7$ \cite{GL:85,EGPR:89}:
\be
L_7^{\eta_1} = - {\tilde{d}_m^2\over 2 M_{\eta_1}^2} \, .
\label{eq:etas_contrib}
\ee
The magnitude of this contribution can be calculated from the 
quark--mass expansion of $M_\eta^2$ and $M_{\eta'}^2$, which fixes 
the $\eta_1$
parameters in the large $N_c$ limit \cite{EGPR:89}:
$M_{\eta_1}=804$ MeV, $|\tilde{d}_m|=20$ MeV. 
The final result for $L_7$ is in close agreement
with its phenomenological value.
 
The combined resonance contributions appear to saturate the $L_i^r$
almost entirely \cite{EGPR:89}. Within the uncertainties of the
approach, there is no need for invoking any additional contributions.
Although the comparison has been made for $\mu=M_\rho$, a similar
conclusion would apply for any value of $\mu$ in the low--lying
resonance region between 0.5 and 1 GeV.

The observed resonance saturation can be understood with large--$N_C$
considerations. In the limit of a large number of colours, the QCD
amplitudes reduce to tree--level hadron exchanges \cite{Manohar};
loop effects being suppressed by powers of $1/N_C$.
Although in principle an infinite tower of resonance exchanges should
contribute to the low--energy chiral couplings, the dominant contributions
come from the lowest--mass states due to the $1/M_R^2$ suppression factor.
Nevertheless, $1/N_C$ corrections could be sizeable, 
specially in cases such as the scalar sector
where final--state interactions (loop effects) are known to be
important \cite{OO:97}.

\subsection{Short--Distance Estimates of ChPT Parameters}
\label{subsec:couplings}

All chiral couplings are in principle calculable from QCD.
Unfortunately, we are not able
at present to make such a first--principle computation.
Although the integral over the quark fields in
\eqn{eq:generatingfunctional}
can be done explicitly, we do not know how to
perform analytically the remaining integration over the
gluon fields.

Lattice calculations \cite{Gupta,Luscher,Martinelli}
offer a promising numerical tool to
investigate the matching between the underlying QCD theory and the
effective chiral Lagrangian; however, the present techniques
are not good enough to face this difficult problem in a
reliable way.
On the other side,
a perturbative evaluation of the gluonic contribution would
obviously fail in reproducing the correct dynamics of SCSB.
A possible way out is to parametrize phenomenologically the
SCSB and make a weak gluon--field expansion around the
resulting physical vacuum.
 
The simplest parametrization
is obtained by adding to the QCD Lagrangian the 
chiral invariant term \cite{ERT:90}
\be
\Delta\cL_{\mathrm{QCD}} = - M_Q \left( \bar q_R U q_L +
    \bar q_L U^\dagger q_R \right) ,
\label{eq:ERTmodel}
\ee
which serves to introduce the $U$ field, and a mass parameter
$M_Q$, which regulates the infra-red behaviour of the low--energy
effective action. In the presence of this term
the operator $\bar q q$ acquires a vacuum expectation value;
therefore, \eqn{eq:ERTmodel} is an effective way to
generate the order parameter due to SCSB.
Making a chiral rotation of the quark fields,
$Q_L \equiv u(\phi)\, q_L$, $Q_R \equiv u(\phi)^\dagger q_R$, 
with $U=u^2$,
the interaction \eqn{eq:ERTmodel} reduces to a
mass term for the {\it dressed} quarks $Q$; the parameter
$M_Q$ can then be interpreted as a
{\it constituent quark mass}.
 
The derivation of the low--energy effective chiral Lagrangian
within this framework has been extensively discussed in ref.~\cite{ERT:90}.
In the chiral and large--$N_C$ limits,
and including the leading gluonic contributions, one gets:
\beqn\label{eq:ERTresult}
\lefteqn{8 L_1 = 4 L_2 = L_9 = {N_C\over 48\pi^2}
  \left[ 1 + \cO\left(1/M_Q^6\right)\right] , } && \hfill
\\
\lefteqn{L_3 = L_{10} = -{N_C\over 96\pi^2}
   \left[ 1 + {\pi^2\over 5 N_C}
   {\langle{\alpha_s\over\pi}GG\rangle\over M_Q^4} +
\cO\left(1/M_Q^6\right)\right] . }
\no
\eeqn
Due to dimensional reasons, the leading contributions
to the $\cO (p^4)$ couplings only depend on
$N_C$ and geometrical factors.
It is remarkable  that $L_1$, $L_2$ and $L_9$
do not get any gluonic correction at this order; this
result is independent of the way SCSB has been
parametrized
($M_Q$ can be taken to be infinite).
Table~\ref{tab:ERT90} compares the predictions obtained with
only the leading term in \eqn{eq:ERTresult}
(i.e. neglecting the gluonic correction) with the
phenomenological determination of the
$L_i$ couplings.
The numerical agreement is quite impressive;
both the order of magnitude and the sign are correctly
reproduced (notice that this is just a free--quark result!).
Moreover, the gluonic corrections shift the values of
$L_3$ and $L_{10}$ in the right direction, making them
more negative.
 
\begin{table}[htb]
\caption{Leading--order ($\alpha_s=0$) predictions for the
$L_i$'s, within the QCD--inspired model \protect\eqn{eq:ERTmodel}.
The phenomenological values are shown in the second row 
for comparison.
All numbers are given in units of $10^{-3}$.}
\label{tab:ERT90}
\begin{tabular}{@{}lccccc@{}}
\hline
& $L_1$ & $L_2$ & $L_3$ & $L_9$ & $L_{10}$
\\ \hline
$L_i^{\mathrm{th}}(\alpha_s=0)$ & 0.79 & 1.58 & $-3.17$ & 6.33 & $-3.17$
\\
$L_i^r(M_\rho)$ & $0.4\pm0.3$ & $1.4\pm0.3$ & $-3.5\pm1.1$ &
$6.9\pm0.7$ & $-5.5\pm0.7$
\\ \hline
\end{tabular}
\end{table}
 
The results \eqn{eq:ERTresult}
obey almost all relations in \eqn{eq:vmdpred}.
Comparing the predictions for $L_{1,2,9}$ in 
eq.~\eqn{eq:vmdpred} with the QCD--inspired
ones in \eqn{eq:ERTresult},
one gets a quite good estimate of the $\rho$ mass:
\be
M_V = 2\sqrt{2}\pi f = 821 \, {\rm MeV} .
\label{eq:rho_mass}
\ee

Is it quite easy to prove that the interaction \eqn{eq:ERTmodel}
is equivalent to the mean--field approximation of the
Nambu--Jona-Lasinio model \cite{NJL:61},
where SCSB is triggered by four--quark operators.
It has been conjectured \cite{BBR:93}
that integrating out the quark and gluon fields of QCD,
down to some intermediate scale
$\Lambda_\chi$, gives rise to an extended Nambu--Jona-Lasinio
Lagrangian.
By introducing collective fields (to be identified later with
the Goldstone fields and
$S$, $V$, $A$ resonances) the model can be transformed into a
Lagrangian bilinear in the quark fields, which can therefore
be integrated out.
One then gets an effective Lagrangian,
describing the couplings of the pseudoscalar bosons to
vector, axial--vector and scalar resonances.
Extending the analysis beyond the mean--field approximation,
ref.~\cite{BBR:93} obtains predictions for
20 measurable quantities, including the $L_i$'s, in terms of only
4 parameters. The quality of the fits is quite impressive.
Since the model contains all resonances that are known to saturate
the $L_i$ couplings, it is not surprising that one gets an
improvement of the
mean--field--approximation results, specially for
the constants $L_5$ and $L_8$, which are sensitive to
scalar exchange.
What is more important, this analysis clarifies a
potential problem of double counting:
in certain limits the model approaches either the pure
quark--loop predictions \eqn{eq:ERTresult} or
the resonance--exchange results \eqn{eq:vmdpred},
but in general it interpolates between these two cases.

\subsection{$U(3)_L\otimes U(3)_R$ ChPT}
\label{subsec:U(3)}

In the large--$N_C$ limit the $U(1)_A$ anomaly 
\cite{AD:69,BJ:69,AB:69} is absent.
The massless QCD Lagrangian \eqn{eq:LQCD} has then a larger
$U(3)_L\otimes U(3)_R$ chiral symmetry, and there are
nine Goldstone bosons associated with the SCSB to the diagonal
subgroup $U(3)_V$. These Goldstone excitations can be
conveniently collected in the $3\times 3$ unitary matrix
\bel{eq:U_tilde}
\widetilde U(\phi) \,\equiv\,
\exp{\left\{ i{\sqrt{2}\over f}\,\widetilde\Phi \right\}}
\, , \qquad\quad
\widetilde\Phi\,\equiv\, {\eta_1\over\sqrt{3}}\, I_3 \, + \,
{\vect{\lambda}\over\sqrt{2}}\,\vect{\phi} \, .
\ee
Under the chiral group, $\widetilde U(\phi)$ transforms as
$\widetilde U\to g_R \widetilde U g_L^\dagger\; $ ($g_{R,L}\in U(3)_{R,L}$).
To lowest order in the chiral expansion, the interactions of the nine
Goldstone bosons are described by the Lagrangian \eqn{eq:lowestorder}
with $\widetilde U(\phi)$ instead of $U(\phi)$.
Notice that the $\eta_1$ kinetic term in
$\langle D_\mu\widetilde U D_\mu\widetilde U^\dagger\rangle$ decouples from
the $\phi$'s and the $\eta_1$ particle becomes stable in the
chiral limit.

To lowest non-trivial order in $1/N_C$, the chiral symmetry breaking
effect induced by the $U(1)_A$ anomaly can be taken into
account in the effective low--energy theory, through the term
\cite{DVV:80,WI:80,RST:80}
\bel{eq:anom_term}
\cL_{U(1)_A} \, = \, - {f^2 \over 4} {a \over N_C} \, 
\left\{ {i \over 2 } \left[\log{(\det{\widetilde U})} - \log{(\det{
\widetilde U^\dagger})}\right] \right\}^2  ,
\ee
which breaks $U(3)_L \otimes U(3)_R$ but preserves
$SU(3)_L \otimes SU(3)_R \otimes U(1)_V$. The
parameter $\, a \,$ has dimensions of mass squared and, with the
factor
$1/N_C$ pulled out, is booked to be of $\cO (1)$ in the large--$N_C$
counting rules. Its value is not fixed by symmetry requirements alone;
it depends crucially on the dynamics of instantons. In the presence
of the term \eqn{eq:anom_term}, 
the $\eta_1$ field becomes massive even in the
chiral limit:
\bel{eq:M_eta1}
M_{\eta_1}^2 = 3 {a\over N_C} + \cO (\cM) \, .
\ee

Owing to the large mass of the $\eta'$, the effect of the $U(1)_A$ anomaly
cannot be treated as a small perturbation. Rather, one should keep
the term \eqn{eq:anom_term} together with the lowest--order Lagrangian
\eqn{eq:lowestorder}. It is possible to build a consistent combined expansion
in powers of momenta, quark masses and $1/N_C$, by counting the
relative magnitude of these parameters as \cite{LE:96}:
\bel{eq:U(3)_counting}
\cM \sim 1/N_C \sim p^2 \sim \cO (\delta) \, .
\ee

A $U(3)_L \otimes U(3)_R$ description \cite{HLPT:97}
of the pseudoscalar particles, including the singlet $\eta_1$ field,
allows one to understand many properties of the $\eta$ meson
in a quite simple way.

A good example is provided by the electromagnetic decays 
$P\to\gamma\gamma$,
which are generated at $\cO (p^4)$ by the Wess--Zumino--Witten
\cite{WZ:71,WI:83} anomaly term in \eqn{eq:WZW}:
\bel{eq:PtoGG}
A(P\to\gamma\gamma) = - {N_C\over 3}\, {\alpha\over\pi f}\,
c_P \,\varepsilon^{\mu\nu\rho\sigma}\,
\epsilon_{1\mu} \epsilon_{2\nu} q_{1\rho} q_{2\sigma} \, ,
\ee
with
$c_{\pi^0} = 1$ and $c_{\eta_8} = 1/\sqrt{3}$.
The predicted decay rate of the neutral pion,
\bel{eq:pigg}
\Gamma(\pi^0\to\gamma\gamma) = \left( {N_C\over 3} \right)^2
{\alpha^2 M_{\pi^0}^3\over 64 \pi^3 f^2} =
7.73\;\mbox{\rm eV},
\ee
is in good agreement with the measured value,
$\Gamma(\pi^0\to\gamma\gamma) = (7.7 \pm 0.6)$~eV, providing
a nice confirmation of the non-abelian QCD anomaly.
However, the usual $SU(3)_L \otimes SU(3)_R$ description, where 
$\eta\approx\eta_8$, underestimates the
$\Gamma(\eta\to\gamma\gamma)$ decay rate by about a factor of three.

In the $U(3)_L \otimes U(3)_R$ framework, the $\eta_8$ mixes with the
$\eta_1$ (both fields share the same isospin and charge):
\bel{eq:eta_mix}
\left( \ba \eta \\ \eta' \ea \right) =
\left( \begin{array}{cc} \cos{\theta_P} & -\sin{\theta_P} \\
   \sin{\theta_P} & \cos{\theta_P} \ea  \right)
\left( \ba \eta_8 \\ \eta_1 \ea \right) .
\ee
The diagonalization of the isoscalar mass matrix implies
\cite{LE:96,HLPT:97} a  rather sizeable mixing
$\theta_P\approx -20^\circ$.
Taking the nonet version of the Wess--Zumino--Witten term \eqn{eq:WZW},
one gets 
$c_{\eta_1} = 2\sqrt{2}/\sqrt{3}$, which implies a $\eta\gamma\gamma$ coupling
$c_{\eta} = (\cos{\theta_P} - 2\sqrt{2}\sin{\theta_P})\, c_{\eta_8}
\approx 1.9\, c_{\eta_8}$; this
provides the needed enhancement to understand the experimental value of
$\Gamma(\eta\to\gamma\gamma)$.
The $\eta'\to\gamma\gamma$ decay rate is also well reproduced
by the predicted amplitude
$c_{\eta'} = (2\sqrt{2}\cos{\theta_P} + \sin{\theta_P})\, c_{\eta_8}$.
The accuracy of the predictions can be further improved with some
amount of symmetry breaking through
$f_\eta\not= f_{\eta'}\not= f_\pi$ 
from higher--order effects \cite{BI:93,HLPT:97}.

In the standard $SU(3)_L \otimes SU(3)_R$ ChPT, the $\eta'$ is integrated 
out and its effects are hidden in higher--order local couplings.
The fact that the singlet pseudoscalar does affect the $\eta$ dynamics
in a significative way is then reflected in the presence of important
higher--order corrections, which are more efficiently taken into
account within the $U(3)_L \otimes U(3)_R$ EFT \cite{PI:90}.

Deeply related to the $U(1)_A$ anomaly is the possible presence
of an additional term in the QCD Lagrangian,
\bel{eq:CPstrong}
\cL_\theta\, = \,\theta_0 \, {g^2 \over 64 \pi^2} \,
\varepsilon_{\mu \nu \rho \sigma}\,
G^{\mu\nu}_a(x) G^{a,\rho\sigma}(x) \, ,
\ee
with $\theta_0$, the so-called vacuum angle, a hitherto unknown
parameter.
This term violates {\sl P, T} and {\sl CP} and may lead
to observable effects in flavour conserving transitions.
A detailed discussion of this subject within ChPT can be found in
ref.~\cite{PR:91b}.

%
%

\section{Non-Leptonic Kaon Decays}
\label{sec:kaon}

Since the kaon mass is a very low energy scale, the theoretical
analysis of non-leptonic kaon decays is highly non-trivial. While the 
underlying flavour--changing weak transitions among the constituent 
quarks are associated with the $W$ mass scale, the 
corresponding hadronic amplitudes are governed by the long--distance
behaviour of the strong interactions, 
i.e. the confinement regime of QCD.

The standard short--distance approach to weak transitions 
(see section~\ref{subsec:wilson}.2) makes
use of the asymptotic freedom property of QCD
to successively integrate out the fields with heavy masses down to 
scales $\mu < m_c$.
Using the operator product expansion (OPE) and
renormalization--group techniques, one gets an effective $\Delta S=1$
Hamiltonian \cite{buras},
\bel{eq:sd_hamiltonian}
\cH_{\mathrm{eff}}^{\Delta S=1} \, = \, {G_F\over\sqrt{2}}
V_{ud}^{\phantom{*}} V_{us}^*\,
\sum_i c_i(\mu)\, Q_i \, + \, \mbox{\rm h.c.},
\ee
which is a sum of local four--fermion operators $Q_i$,
constructed with the light degrees of freedom
($u,d,s; e,\mu,\nu_l$), modulated by
Wilson coefficients $c_i(\mu)$
which are functions of the heavy ($W,t,b,c,\tau$) masses.
The overall renormalization scale $\mu$
separates the short-- ($M>\mu$) and long-- ($m<\mu$) distance
contributions,  which are contained in $c_i(\mu)$
and $Q_i$, respectively.
The physical amplitudes are of course independent of $\mu$;
thus, the explicit scale (and scheme) dependence of the Wilson
coefficients, should cancel exactly with the corresponding dependence
of the $Q_i$ matrix elements between on--shell states.

Our knowledge of the $\Delta S=1$ effective Hamiltonian has improved
considerably in recent years, thanks to the completion of the
next-to-leading logarithmic order calculation of the Wilson
coefficients \cite{buras}.
All gluonic corrections of $\cO(\alpha_s^n t^n)$ and 
$\cO(\alpha_s^{n+1} t^n)$ are already known,
where $t\equiv\log{(M/m)}$ refers to the logarithm of any ratio of 
heavy--mass scales ($M,m\geq\mu$). Moreover, the full $m_t/M_W$ dependence
(at lowest order in $\alpha_s$) has been taken into account.

Unfortunately, in order to predict the
physical amplitudes one is still confronted with the calculation of
the hadronic matrix elements of the quark operators.
This is a very difficult problem, which so far remains unsolved. 
The present technology to calculate low--energy matrix elements is not
yet developed to the degree of sophistication of perturbative QCD.
We have only been able to obtain rough estimates using
different approximations (vacuum saturation, $N_C\to\infty$ limit, 
QCD low--energy effective action, \ldots)
or applying QCD techniques (lattice, QCD sum rules) which suffer from
their own technical limitations.

   Below the resonance region ($\mu < M_\rho$) the strong interaction
dynamics can be better understood with global symmetry considerations. 
The effective ChPT formulation of the Standard Model is an ideal
framework to describe kaon decays \cite{Orsay,dR:95}. 
This is because in K decays the
only physical states which appear are pseudoscalar mesons, photons and
leptons, and because the characteristic momenta involved are small
compared to the natural scale of chiral symmetry breaking 
($\Lambda_\chi\sim 1$~GeV).

\begin{figure}[tbh]      
\setlength{\unitlength}{0.62mm} \centering
\begin{picture}(165,120)
\put(0,0){\makebox(165,120){}}
\thicklines
\put(10,105){\makebox(40,15){\large Energy Scale}}
\put(58,105){\makebox(36,15){\large Fields}}
\put(110,105){\makebox(40,15){\large Effective Theory}}
\put(8,108){\line(1,0){149}} {\large
\put(10,75){\makebox(40,27){$M_W$}}
\put(58,75){\framebox(36,27){$\ba W, Z, \gamma, g \\
     \tau, \mu, e, \nu_i \\ t, b, c, s, d, u \ea $}}
\put(110,75){\makebox(40,27){Standard Model}}

\put(10,40){\makebox(40,18){$\lsim m_c$}}
\put(58,40){\framebox(36,18){$\ba  \gamma, g  \; ;\; \mu ,  e, \nu_i  
             \\ s, d, u \ea $}} 
\put(110,40){\makebox(40,18){$\cL_{\mathrm{QCD}}^{(n_f=3)}$, \  
             $\cH_{\mathrm{eff}}^{\Delta S=1,2}$}}

\put(10,5){\makebox(40,18){$M_K$}}
\put(58,5){\framebox(36,18){$\ba\gamma \; ;\; \mu , e, \nu_i  \\ 
            \pi, K,\eta  \ea $}} 
\put(110,5){\makebox(40,18){ChPT}}
\linethickness{0.3mm}
\put(76,37){\vector(0,-1){11}}
\put(76,72){\vector(0,-1){11}}
\put(80,64.5){OPE} 
\put(85,27.5){\huge ?} }
\end{picture}
\caption{Evolution from $M_W$ to the kaon mass scale.
  \label{fig:eff_th}}
\end{figure}

Figure~\ref{fig:eff_th} shows a schematic view of the procedure used
to evolve down from $M_W$ to the kaon mass scale.
At the different energy regimes one uses different effective theories,
involving only those fields which are relevant at that scale.
The corresponding effective parameters (Wilson coefficients, chiral
couplings) encode the information on the heavy degrees of freedom 
which have been integrated out.
These effective theories are convenient realizations of the fundamental
Standard Model at a given energy scale 
(all of them give rise to the same generating functional and therefore to
identical predictions for physical quantities). From a technical point of
view, we know how to compute the effective Hamiltonian at the charm mass
scale. Much more difficult seems the attempt to derive the chiral
Lagrangian from first principles. The symmetry considerations only fix
the allowed chiral structures, at a given order in momenta, but leave
their corresponding coefficients completely undetermined.
The calculation of the chiral couplings from
the effective short--distance Hamiltonian, remains the main open problem
in kaon physics.

\subsection{Weak Chiral Lagrangian}

The effect of strangeness--changing non-leptonic
weak interactions with $\Delta S=1$ is incorporated 
in the low--energy chiral theory as a perturbation to
the strong effective Lagrangian $\cL_{\mathrm{eff}}(U)$. 
At lowest order in the number of derivatives,
the most general effective bosonic Lagrangian, with the same
$SU(3)_L\otimes SU(3)_R$ transformation properties as the short--distance
Hamiltonian \eqn{eq:sd_hamiltonian}, contains two terms:
\beqn\label{eq:lg8_g27}
\cL_2^{\Delta S=1} &=& -{G_F \over \sqrt{2}}  V_{ud}^{\phantom{*}} V_{us}^*
\Bigg\{ g_8  \,\langle\lambda L_{\mu} L^{\mu}\rangle 
 +
g_{27} \left( L_{\mu 23} L^\mu_{11} + {2\over 3} L_{\mu 21} L^\mu_{13}
\right) \Bigg\}
\no\\ &&\mbox{} +
\mbox{\rm h.c.} \, , 
\eeqn
where the matrix $L_{\mu}=i f^2 U^\dagger D_\mu U$  represents the octet of 
$V-A$
currents, and $\lambda\equiv (\lambda^6 - i \lambda^7)/2$ projects onto the
$\bar s\to \bar d$ transition [$\lambda_{ij} = \delta_{i3}\delta_{j2}$].
The chiral couplings $g_8$ and $g_{27}$ measure the strength of the two
parts of the effective Hamiltonian \eqn{eq:sd_hamiltonian} transforming as 
$(8_L,1_R)$ and $(27_L,1_R)$, respectively, under chiral rotations.
Their values can be extracted from $K \rightarrow 2 \pi$ decays
\cite{PGR:86}:
\bel{eq:g8_g27}
\vert g_8 \vert \simeq 5.1\, , \qquad\qquad 
\vert g_{27}/ g_8 \vert \simeq 1/18 \, .
\ee
The huge difference between these two couplings
shows the well--known
enhancement of the octet $\vert\Delta I\vert = 1/2$ transitions.
 
Using the Lagrangians \eqn{eq:lowestorder}  and \eqn{eq:lg8_g27}, 
the rates for decays 
like $K \rightarrow 3 \pi$ or $ K \rightarrow \pi \pi \gamma $ can be 
predicted
at $\cO(p^2)$ through a trivial tree--level calculation. However, the data
are already accurate enough for the next--order corrections to be sizeable.
Moreover, due to a mismatch between the minimum number of powers of momenta
required by gauge invariance and the powers of momenta that the 
lowest--order
effective Lagrangian can provide \cite{EPR:87a,EPR:87b,EPR:88}, 
the amplitude for any non-leptonic radiative
$K$ decay with at most one pion in the final state ($K \rightarrow \gamma
\gamma  , K \rightarrow \gamma l^+ l^- , K \rightarrow \pi \gamma \gamma ,
K \rightarrow \pi l^+ l^-$, \ldots)
vanishes to $\cO(p^2)$.
These decays are then sensitive to the non-trivial quantum field theory
aspects of ChPT.

Unfortunately, at $\cO (p^4)$ there is a very large number of 
possible terms,
satisfying the appropriate $(8_L,1_R)$ and $(27_L,1_R)$
transformation properties.
Using the $\cO (p^2)$ equations of motion obeyed by $U$ to reduce the
number of terms, 35 independent structures
(plus 2 contact terms involving external fields only)
remain in the octet
sector alone \cite{KMW:90,EC:90,EF:91,EKW:93}.
Restricting the attention to those terms that contribute to
non-leptonic amplitudes where the only external gauge fields
are photons, still leaves 22 relevant octet terms 
\cite{EKW:93}.
Clearly, the predictive power of a completely general chiral
analysis, using  only symmetry constraints, is rather limited.
Nevertheless, as we are going to see,
it is still possible to make predictions.

Due to the complicated interplay of electroweak and strong
interactions, the low--energy constants of the weak non-leptonic
chiral Lagrangian encode a much richer information than
in the pure strong sector.
These chiral couplings contain both long-- and short--distance
contributions, and some of them (like $g_8$) have in addition
a CP--violating imaginary part.
Genuine short--distance physics, such as the electroweak {\it penguin}
operators \cite{buras}, have their corresponding effective realization
in the chiral Lagrangian.
Moreover, there are four $\cO (p^4)$ terms containing an
$\varepsilon_{\mu\nu\alpha\beta}$ tensor,
which get a direct (probably dominant) contribution from
the chiral anomaly \cite{ENP:92,BEP:92}.
 
In recent years, there have been many attempts to estimate
these low--energy couplings using different approximations, such as
factorization \cite{PR:91a},
weak--deformation model \cite{EPR:90},
effective--action approach  \cite{PR:91a,BP:93},
or resonance exchange \cite{EKW:93,IP:92,dAP:97}.
Although more work in this direction is certainly needed,
a qualitative picture of the size of the different couplings
is already emerging.

\subsection{$K\to 2\pi, 3\pi$ Decays}
\label{subsec:kpp}
 
Imposing isospin and Bose symmetries, and keeping terms up to
$\cO (p^4)$, a general parametrization \cite{DD:79}
of the $K\to 3\pi$ amplitudes involves ten measurable parameters:
$\alpha_i$, $\beta_i$, $\zeta_i$, $\xi_i$, $\gamma_3$ and
$\xi'_3$, where $i=1,3$ refers to the $\Delta I =\frac{1}{2}, \frac{3}{2}$
pieces.
At $\cO (p^2)$, the quadratic slope parameters
$\zeta_i$, $\xi_i$ and $\xi'_3$ vanish; therefore the
lowest--order Lagrangian \eqn{eq:lg8_g27} predicts five
$K\to 3\pi$ parameters in terms of the two couplings
$g_8$ and $g_{27}$, extracted from $K\to 2 \pi$.
These predictions give the right qualitative pattern,
but there are sizeable differences with the
measured amplitudes.
Moreover, non-zero values for some of the slope parameters
have been clearly established experimentally.
 
The agreement is substantially improved at $\cO (p^4)$
\cite{KMW:91}.
In spite of the large number of unknown couplings in the general
effective $\Delta S=1$ Lagrangian,
only 7 combinations of these weak chiral constants
are relevant for describing the $K\to 2\pi$ and $K\to 3 \pi$
amplitudes \cite{KDHMW:92}.
Therefore, one has 7 parameters for 12 observables, which
results in 5 relations.
The extent to which these relations are satisfied provides
a non-trivial test of chiral symmetry at the four-derivative level.
The results of such a test \cite{KDHMW:92} are shown in
table~\ref{tab:KDHMW92}, where the 5 conditions have been
formulated as predictions for the 5 slope parameters.
The comparison is very successful for the two
$\Delta I = \frac{1}{2}$ parameters, but
the data are not good enough to say anything conclusive about
the other three $\Delta I = \frac{3}{2}$ predictions.

\begin{table}
\caption{Predicted and measured values of the quadratic
slope parameters in the $K\to 3\pi$ amplitudes 
\protect\cite{KDHMW:92}.
All values are given in units of $10^{-8}$.}
\label{tab:KDHMW92}
\begin{tabular}{@{}ccc@{}}
\hline
Parameter & Experimental value & Prediction
\\ \hline
$\zeta_1$ & $-0.47\pm0.15$ & $-0.47\pm0.18$ \\
$\xi_1$ & $-1.51\pm0.30$ & $-1.58\pm0.19$ \\
$\zeta_3$ & $-0.21\pm0.08$ & $-0.011\pm0.006$ \\
$\xi_3$ & $-0.12\pm0.17$ & $\hphantom{-}0.092\pm0.030$ \\
$\xi'_3$ & $-0.21\pm0.51$ & $-0.033\pm0.077$ \\ \hline
\end{tabular}
\end{table}

The $\cO (p^4)$ analysis of these decays has also clarified the
role of long--distance effects ($\pi\pi$ rescattering)
in the dynamical enhancement of amplitudes with $\Delta I = \frac{1}{2}$.
The $\cO (p^4)$ corrections give indeed a sizeable
constructive contribution, which results \cite{KMW:91}
in a fitted value for $|g_8|$ that is about $30\%$ smaller
than the lowest--order determination \eqn{eq:g8_g27}.
While this certainly goes in the right direction,
it also shows that the bulk of the enhancement mechanism
comes from a different source.

\subsection{Radiative $K$ Decays}
\label{subsec:radiative}
 
Owing to the constraints of electromagnetic gauge invariance, 
radiative
$K$ decays with at most one pion in the final state do not occur
at $\cO (p^2)$.
Moreover, only a few terms of the octet $\cO (p^4)$ Lagrangian
are relevant for this kind of processes
\cite{EPR:87a,EPR:87b,EPR:88}:
\beqn\label{eq:lweak}
\cL_4^{\Delta S=1,{\mathrm{em}}} & \doteq & 
-{G_F \over \sqrt{2}}  V_{ud}^{\phantom{*}} V_{us}^* \, g_8 
\,\bigg\{
-{i e \over f^2} F^{\mu\nu} \,\left[
   w_1 \,\langle Q \lambda L_\mu L_\nu\rangle  
 + w_2 \,\langle Q L_\mu \lambda L_\nu\rangle \right]
\no\\ &&\qquad\qquad\qquad\;\;\,
\mbox{} + e^2 f^2 w_4\, F^{\mu\nu} F_{\mu\nu}
 \,\langle\lambda QU^\dagger QU\rangle \bigg\}
 + \mbox{\rm h.c.}  
\eeqn
The small number of unknown chiral couplings allows us to
derive useful relations among different processes and
to obtain definite predictions. The absence of a tree--level
$\cO (p^2)$ contribution makes the final results very sensitive to the
loop structure of the amplitudes.

\medskip
\noindent {\it\thesubsection.1. $K_S\to\gamma\gamma$}
\smallskip

\begin{figure}[tbh]
\centerline{\epsfig{file=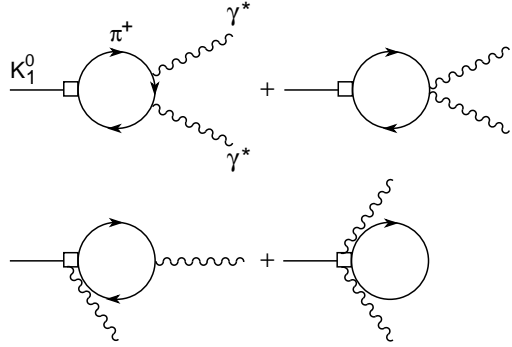,width=7.0cm}}
\caption{Feynman diagrams for $K_1^0\to\gamma^*\gamma^*$.}
\label{fig:ksgg}
\end{figure}

\noindent
The symmetry constraints do not allow any direct tree--level
$K_1^0\gamma\gamma$ coupling at $\cO(p^4)$
($K^0_{1,2}$ refer to the CP--even and CP--odd eigenstates, 
respectively).
This decay proceeds then
through a loop of charged pions as shown in
fig.~\ref{fig:ksgg} (there are
similar diagrams with charged kaons in the loop, but
their sum is proportional to
$M^2_{K^0} - M^2_{K^+}$ and therefore can be neglected).
Since there are no possible counter-terms to renormalize
divergences, the one--loop amplitude is necessarily finite.
Although each of the four diagrams in fig.~\ref{fig:ksgg}
is quadratically divergent, these divergences cancel in the sum.
The resulting prediction \cite{dAE:86,GO:87},
$\mbox{\rm Br}(K_S\to\gamma\gamma) = 2.0 \times 10^{-6}$,
is in very good agreement
with the experimental measurement \cite{BA:95,BU:87}:
\bel{eq:ksgg}
\mbox{\rm Br}(K_S\to\gamma\gamma) \, = \,
(2.4 \pm 0.9) \times 10^{-6} \, . 
\ee

\medskip
\noindent {\it\thesubsection.2. $K_L\to\gamma\gamma$}
\smallskip

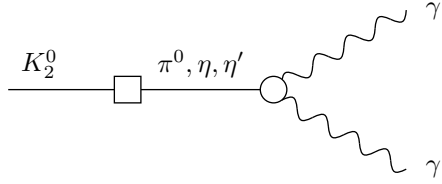
\begin{figure}[htb]
\centering
\begin{picture}(170,80)
\Line(0,40)(45,40)
\Line(45,40)(100,40)
\Photon(150,70)(103,42)34
\Photon(150,10)(103,36)34
\BBoxc(45,40)(10,10)
\GCirc(100,40)51
\Text(5,50)[l]{$K_2^0$}
\Text(73,50)[c]{$\pi^0,\eta,\eta'$}
\Text(158,70)[l]{$\gamma$}
\Text(158,10)[l]{$\gamma$}
\end{picture}
\caption{Feynman diagram for $K_2^0\to\gamma^*\gamma^*$.}
\label{fig:klgg}
\end{figure}

\noindent
At $\cO (p^4)$, the $K_2^0\to\gamma^*\gamma^*$ decay amplitude,
\bel{eq:Kgg}
 A(K_L\to\gamma^*\gamma^* ) = c(q_1^2,q_2^2)\; \varepsilon^{\mu\nu\rho\sigma}\,
 \epsilon_{1\mu} \epsilon_{2\nu} q_{1\rho} q_{2\sigma} \, ,
\ee
proceeds through
a tree--level $K_2^0\to\pi^0,\eta,\eta'$ transition, followed by
$\pi^0,\eta,\eta'\to\gamma\gamma$ vertices.
The lowest--order chiral prediction, can only generate a constant
form factor $c(q_1^2,q_2^2)$; it thus corresponds to the decay into
on-shell photons ($q_1^2=q_2^2=0$) \cite{ENP:92}:
\bel{eq:c_Kgg}
 c(0,0) =  {G_F\over\sqrt{2}} V^{\phantom{*}}_{ud} V_{us}^*
   \, {2 g_8 \alpha f  \over\pi (1 - r_\pi^2)}\,
 c_{\mathrm{red}} \, ,
\ee
\beqn\label{eq:cred}
c_{\mathrm{red}} &=& 1 - {(1 - r_\pi^2)\over 3(r_\eta^2-1)} 
  \, (c_\theta -2\sqrt{2} s_\theta)\, (c_\theta +2\sqrt{2}\rho_n s_\theta)
\no\\ && \mbox{}
 + {(1 - r_\pi^2)\over 3(r_{\eta'}^2-1)} \,
  (2\sqrt{2} c_\theta + s_\theta) \, (2\sqrt{2} \rho_n c_\theta - s_\theta)
  \, ,
\eeqn
where
$r_P^2\equiv M_P^2/M_{K_L}^2$, $c_\theta \equiv\cos{\theta_P}$ and
$s_\theta \equiv\sin{\theta_P}$.
We have factored out the contribution of the pion
pole, which normalizes the dimensionless reduced amplitude
$c_{\mathrm{red}}$. The second and third terms in  $c_{\mathrm{red}}$
correspond to the $\eta$ and $\eta'$
contributions respectively. Nonet symmetry (which is exact in the
large--$N_C$ limit) has been assumed in the electromagnetic
$2\gamma$ vertices; this is known to provide a quite good description
of the anomalous $P\to 2\gamma$ decays ($P=\pi^0, \eta, \eta'$).
Possible deviations of nonet symmetry in the non-leptonic weak vertex are
parametrized through $\rho_n\not= 1$.

 In the standard $SU(3)_L\otimes SU(3)_R$ ChPT, the $\eta'$ contribution
is absent  and $\theta_P=0$; therefore,
$c_{\mathrm{red}}\propto (3 M_\eta^2 + M_\pi^2 - 4 M_K^2)$, which vanishes
owing to the Gell-Mann--Okubo mass relation.
The physical $K_L\to\gamma\gamma$ amplitude is then a higher--order 
---$O(p^6)$--- effect in the chiral counting, which makes difficult to perform
a reliable calculation.

 The situation is quite different if one uses instead the
$U(3)_L\otimes U(3)_R$ EFT, 
including the singlet $\eta_1$ field.
Taking $s_\theta = -1/3$ ($\theta \approx -19.5^\circ$), the $\eta$--pole
contribution in eq.~(\ref{eq:cred}) is proportional to $(1-\rho_n)$ and
vanishes in the nonet--symmetry limit; the large and positive $\eta'$
contribution results then in $c_{\mathrm{red}}=1.80$ for $\rho_n=1$.
With $0\leq\rho_n\leq 1$, the $\eta$ and $\eta'$ contributions interfere
destructively and $c_{\mathrm{red}}$ is dominated by the pion pole. One
would get $c_{\mathrm{red}}\simeq 1$ for $\rho_n\simeq 3/4$.

The measured $K_L\to\gamma\gamma$ decay rate \cite{PDG:96} corresponds to
$|c(0,0)| = (3.51\pm 0.05)\times 10^{-9}\;\mbox{\rm GeV}^{-1}$.
Taking into account the 30\% reduction of the $|g_8|$ value at $\cO (p^4)$
(this sizeable shift results mainly from the constructive $\pi\pi$ rescattering
contribution, which is obviously absent in $K_L\to\gamma\gamma$),
this implies 
$c_{\mathrm{red}}^{\mathrm{exp}} =  (1.19\pm 0.16)$.

Leaving aside numerical details, we can safely conclude that the
physical $K_L\to\gamma\gamma$ amplitude, with on-shell photons, 
is indeed dominated by the pion pole ($c_{\mathrm{red}}\sim 1$). 
Although the exact numerical prediction is sensitive to several small
corrections ($\rho_n\not= 1$, $f_\pi\not= f_\eta\not= f_{\eta'}$, 
$s_\theta\not=-1/3$) and therefore is quite uncertain,
the needed cancellation between
the $\eta$ and $\eta'$ contributions arises in a natural way and can be
fitted easily with a reasonable choice of symmetry--breaking parameters.

\medskip
\noindent {\it\thesubsection.3. $K_{S,L}\to\mu^+\mu^-$}
\smallskip

\noindent
A straightforward chiral analysis \cite{EP:91}
shows that, at lowest order in momenta, the only allowed
tree--level $K^0\mu^+\mu^-$ coupling corresponds to the
CP--odd state $K_2^0$.
Therefore, the $K_1^0\to\mu^+\mu^-$ transition can only be
generated by a finite non-local two--loop contribution.
The explicit calculation \cite{EP:91} gives:
\bel{eq:ksmm_ratios}
{\Gamma(K_S\to\mu^+\mu^-)\over\Gamma(K_S\to\gamma\gamma)}
= 2\times 10^{-6}, \qquad
{\Gamma(K_S\to e^+ e^-)\over\Gamma(K_S\to\gamma\gamma)}
= 8\times 10^{-9},
\ee
well below the present (90\% CL) experimental upper limits  
\cite{GJ:73,BL:94}:    
Br$(K_S\to\mu^+\mu^-) < 3.2\times 10^{-7}$,
Br$(K_S\to e^+e^-) < 2.8\times 10^{-6}$.
Although, in view of
the smallness of the predicted ratios,
this calculation seems quite academic, it has important
implications for CP--violation studies.

\begin{figure}[tbh]
\centering
\begin{picture}(160,80)
\Line(0,40)(50,40)
\Photon(102,70)(55,45)34
\Photon(102,10)(55,35)34
\BBoxc(50,40)(10,10)
\ArrowLine(140,10)(102,10)
\ArrowLine(102,10)(102,70)
\ArrowLine(102,70)(140,70)
\Text(5,50)[l]{$K^0_{1,2}$}
\Text(150,70)[l]{$\mu^-$}
\Text(150,10)[l]{$\mu^+$}
\Text(75,70)[c]{$\gamma$}
\Text(75,10)[c]{$\gamma$}
\end{picture}
\caption{Electromagnetic loop contribution to 
$K^0_{1,2}\to \mu^+ \mu^-$.
The $K_1^0 \gamma^* \gamma^*$ vertex is generated through 
the one--loop diagrams shown in fig.~\protect\ref{fig:ksgg},
while the $K_2^0 \gamma^* \gamma^*$ transition proceeds through the
tree--level amplitude in fig.~\protect\ref{fig:klgg}.}
\label{fig:ksmm}
\end{figure}
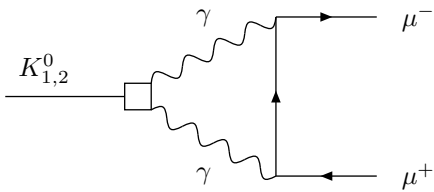

The longitudinal muon polarization $\cP_L$
in the decay $K_L\to\mu^+\mu^-$ is an interesting measure of 
CP violation.
As for every CP--violating observable in the neutral kaon system,
there are in general two different kinds of contributions to $\cP_L$:
indirect CP violation through the small 
$K_1^0$ admixture of the $K_L$
($\varepsilon$ effect), and direct CP violation in the 
$K_2^0\to\mu^+\mu^-$ decay amplitude.
 
   In the Standard Model, the direct CP--violating amplitude is
induced by Higgs exchange with an effective one--loop flavour--changing
$\bar s d H$ coupling \cite{BL:86}.
The present lower bound
on the Higgs mass    
implies a conservative upper limit
$|\cP_{L,\mathrm{Direct}}| < 10^{-4}$.
Much larger values, $\cP_L \sim O(10^{-2})$, appear quite naturally
in various extensions of the Standard Model \cite{GN:90,MO:93}.
It is worth emphasizing that $\cP_L$ is especially
sensitive to the presence of light scalars with CP--violating
Yukawa couplings. Thus, $\cP_L$ seems to be a good signature to look
for new physics beyond the Standard Model; for this to be the case,
however, it is very important to have a good quantitative
understanding of the Standard Model prediction to allow us to infer,
from a measurement of $\cP_L$, the existence of a new CP--violation
mechanism.
 
  The chiral calculation of the $K_1^0\to\mu^+\mu^-$ amplitude
allows us to make a reliable estimate
of the contribution to $\cP_L$ due to $K^0$--$\bar K^0$ mixing
\cite{EP:91}:
\bel{eq:p_l}
1.9 \, < \, |\cP_{L,\varepsilon}| \times 10^3 \Biggl( 
{2 \times 10^{-6} \over
\mbox{\rm Br}(K_S\to\gamma\gamma)} \Biggr)^{1/2} \, < 2.5 \, .
\ee
Taking into account
the present experimental errors in $\mbox{\rm Br}(K_S\to\gamma\gamma)$  and
the inherent theoretical uncertainties due to uncalculated
higher--order corrections,
one can conclude that experimental indications for
$|\cP_L|>5\times 10^{-3}$ would constitute clear evidence
for additional
mechanisms of CP violation beyond the Standard Model.

\begin{figure}[htb]
\centering
\begin{picture}(160,80)   
\Line(0,40)(45,40)
\Line(45,40)(100,40)
\ArrowLine(140,70)(100,40)
\ArrowLine(100,40)(140,10)
\BBoxc(45,40)(10,10)
\GCirc(100,40)51
\Text(100,40)[c]{\large $\times$}
\Text(5,50)[l]{$K_2^0$}
\Text(73,50)[c]{$\pi^0,\eta,\eta'$}
\Text(150,72)[l]{$\mu^+$}
\Text(150,10)[l]{$\mu^-$}
\end{picture}
\caption{Local counter-term contribution
to $K_L\rightarrow\mu^+\mu^-$.}
\label{fig:Kmm_local}
\end{figure}
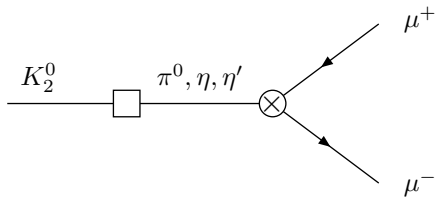

The calculation of the CP--conserving $K_2\to\mu^+\mu^-$ amplitude
is more difficult.
There are well--known short--distance contributions \cite{buras}
(electroweak penguins and box diagrams), which are sensitive to
the presence of a virtual top quark and could be used to improve our
knowledge on the quark--mixing factor $V_{td}$.
Unfortunately, this process is dominated by the electromagnetic long--distance
contribution in fig.~\ref{fig:ksmm}.
Moreover, the measured rate \cite{AK:95,HE:95}
\begin{equation}
\label{eq:exp}
  {\rm Br}(K_L\to\mu^+\mu^- ) = (7.2\pm 0.5) \times 10^{-9}
\end{equation}
appears to be completely saturated by the absorptive contribution
from the on-shell  $2\gamma$ intermediate state,
\bel{eq:abs}
  \mbox{\rm Br}(K_2\to\mu^+\mu^-)_{\mathrm{Abs}} = 
  (7.07\pm 0.18)\times 10^{-9} \, ,
\ee
which leaves very small room to accommodate the dispersive contribution:
$\mbox{\rm Br}(K_2\to\mu^+\mu^-)_{\mathrm{Dis}} = (0.1\pm 0.5)\times 10^{-9}$.

The ChPT calculation of this long--distance amplitude is not easy, because
the $K_2\to\gamma^*\gamma^*$ vertex is quite uncertain and, moreover,
there is now an unknown local $K_2\mu^+\mu^-$ counter-term,
which renormalizes the divergent photon loop.
Nevertheless, it is still possible to compute the ratio
$\mbox{\rm Br}(K_2\to\mu^+\mu^-)/\mbox{\rm Br}(K_2\to\gamma\gamma)$
in the large--$N_C$ limit \cite{DP:98}.
At leading order in $1/N_C$,
the $K_2\to\gamma^*\gamma^*$ transition occurs through the
$\pi^0,\eta,\eta'$ poles, as shown in fig.~\ref{fig:klgg}. Therefore, the
problematic electromagnetic loop is actually the same
governing the decays $\pi^0,\eta,\eta'\to l^+ l^-$,
and the unknown local contribution in fig.~\ref{fig:Kmm_local}
can be fixed from the measured rates for these transitions.

The chiral analysis of $K_2\to\mu^+\mu^-$ \cite{DP:98} shows that
the experimentally observed small dispersive amplitude
fits perfectly well within the large--$N_C$ description of
this process. Moreover, it allows to extract a constraint on the
short--distance contribution, which can be translated into direct
information on the top mass and the quark--mixing factors \cite{DP:98}:
\bel{eq:Chi_SD}
\delta\chi_{\mathrm{SD}} \approx 1.7\, \left(\rho_0 -\bar\rho\right)\;
\left({M_t(M_t^2)\over 170\;\mbox{\rm GeV}}\right)^{1.56}\;
  \left({\left| V_{cb}\right|\over 0.040}\right)^2
= 2.2^{+1.1}_{-1.3} \, ,
\ee
where $\rho_0\approx 1.2$ and $\bar\rho\equiv \rho\, (1 - \lambda^2/2)$,
with $\rho$ and $\eta$ the usual quark--mixing parameters in the
Wolfenstein parametrization.
This constraint is in good agreement with the present
information from other weak transitions \cite{buras}, 
$|\bar\rho|\leq 0.3$, which implies
$\delta\chi_{\mathrm{SD}}\approx 1.8\pm 0.6$.

\medskip
\noindent {\it\thesubsection.4. $K\to\pi\gamma\gamma$}
\smallskip

\noindent
The most general form of the $K\to\pi\gamma\gamma$ amplitude
depends on four independent invariant
amplitudes $A(y,z)$, $B(y,z)$, $C(y,z)$ and $D(y,z)$,
where $y\equiv|p_K\cdot(q_1-q_2)|/M_K^2$ and $z=(q_1+q_2)^2/M_K^2$
 \cite{EPR:88}:
\beqn\label{eq:a_b_def}
\lefteqn{{\cal A}[K(p_K)\to\pi(p_\pi)\gamma(q_1)\gamma(q_2)]\, =\, 
    \epsilon_\mu(q_1) \,\epsilon_\nu(q_2) \, \Biggl\{ 
    {A\over M^2_K}\,
    \Bigl(  q_2^\mu q_1^\nu - q_1\cdot q_2 \, g^{\mu\nu}\Bigr) 
\Biggr. }\no \\
&& \Biggl. \mbox{}
+ {2 B\over M^4_K}\,
\Bigl(p_K\cdot q_1 \, q_2^\mu p_K^\nu
+ p_K\cdot q_2\, q_1^\nu p_K^\mu
- q_1\cdot q_2 \,  p_K^\mu p_K^\nu  
\Bigr.\Biggr. \no\\ && \Biggl.\Bigl. \mbox{}\qquad\quad\mbox{}
- p_K\cdot q_1\, p_K\cdot q_2 \, g^{\mu\nu}\Bigr)
\, + \, {C\over M^2_K}\,\varepsilon^{\mu\nu\rho\sigma} q_{1\rho}q_{2\sigma}
\Biggr.\\ && \Biggl. \mbox{}
+ {D\over M^4_K}\,\left[
\varepsilon^{\mu\nu\rho\sigma}\left(
p_K\cdot q_2\, q_{1\rho} + p_K\cdot q_1\, q_{2\rho}\right) p_{K\sigma}
\right.\Biggr.\no\\ && \Biggl.\left.\mbox{}\qquad\quad\mbox{}
+ \left( p_K^\mu \varepsilon^{\nu\alpha\beta\gamma} +
  p_K^\nu \varepsilon^{\mu\alpha\beta\gamma}\right)
p_{K\alpha}q_{1\beta}q_{2\gamma}\right]
 \Biggr\} . \no
\eeqn
In the limit where CP is conserved, the amplitudes A  and B contribute 
to $K_2\to\pi^0\gamma\gamma$ whereas $K_1\to\pi^0\gamma\gamma$
involves the other two amplitudes C and D. All four amplitudes
contribute to $K^+\to\pi^+\gamma\gamma$.
Only $A(y,z)$ and $C(y,z)$ are non-vanishing to 
lowest non-trivial order, $\cO (p^4)$, in ChPT.

\begin{figure}[thb]
\centerline{
\begin{minipage}{.46\linewidth}
\centerline{\epsfig{file=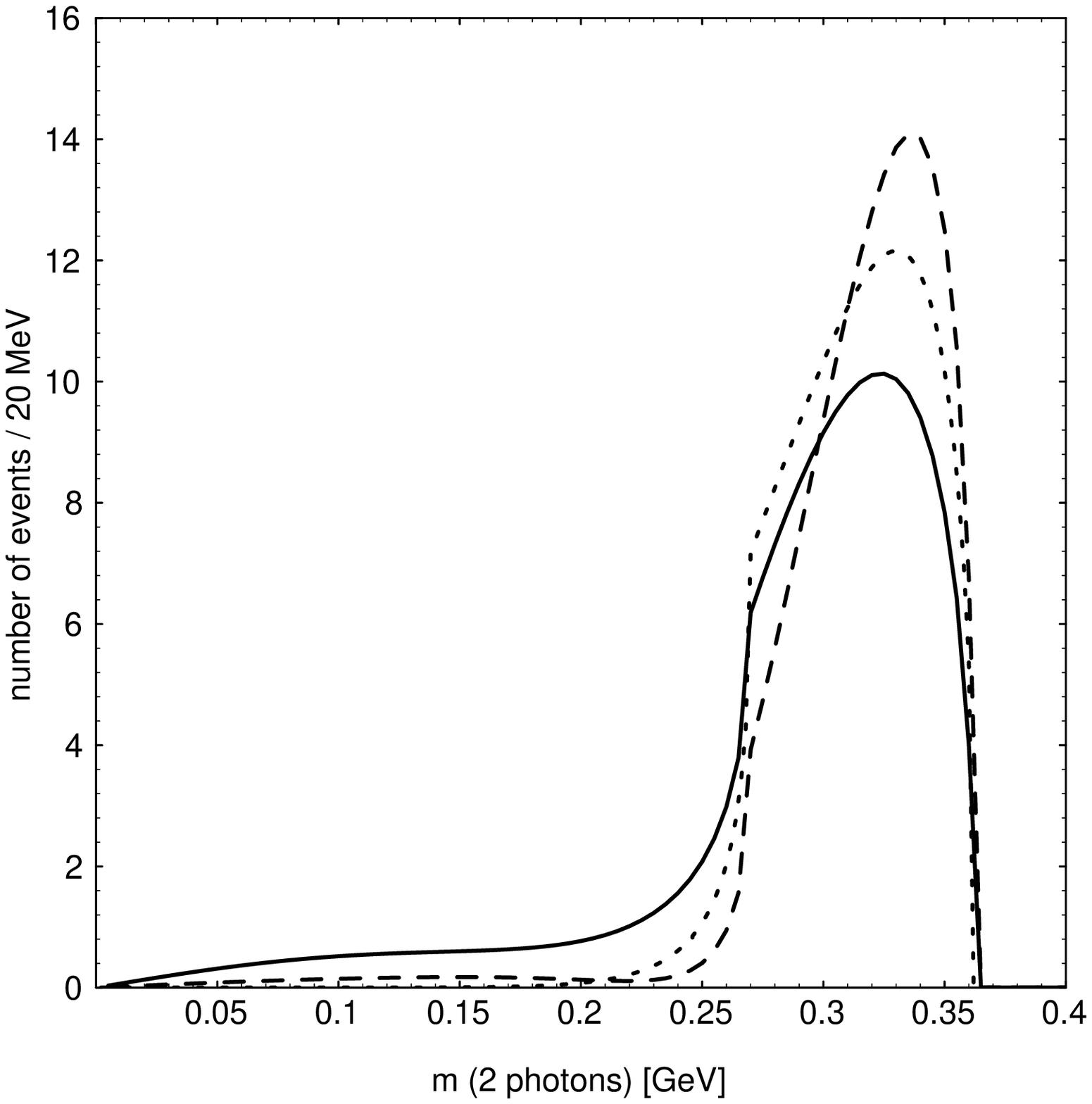,width=.9\linewidth,height=6.8cm}}
\caption{$2\gamma$--invariant--mass distribution for
$K_L\to\pi^0\gamma\gamma$:
$\cO (p^4)$ (dotted curve),
$\cO (p^6)$ with $a_V=0$ (dashed curve),
$\cO (p^6)$ with $a_V=-0.9$ (full curve).
The spectrum is normalized to the 50 unambiguous
events of NA31 \protect\cite{BA:92} (without acceptance corrections).}
\label{fig:spectrum}
\end{minipage}
\hspace{0.2cm}
\begin{minipage}{.505\linewidth}
\hbox{}\vspace*{0.27cm}
\centerline{\epsfig{file=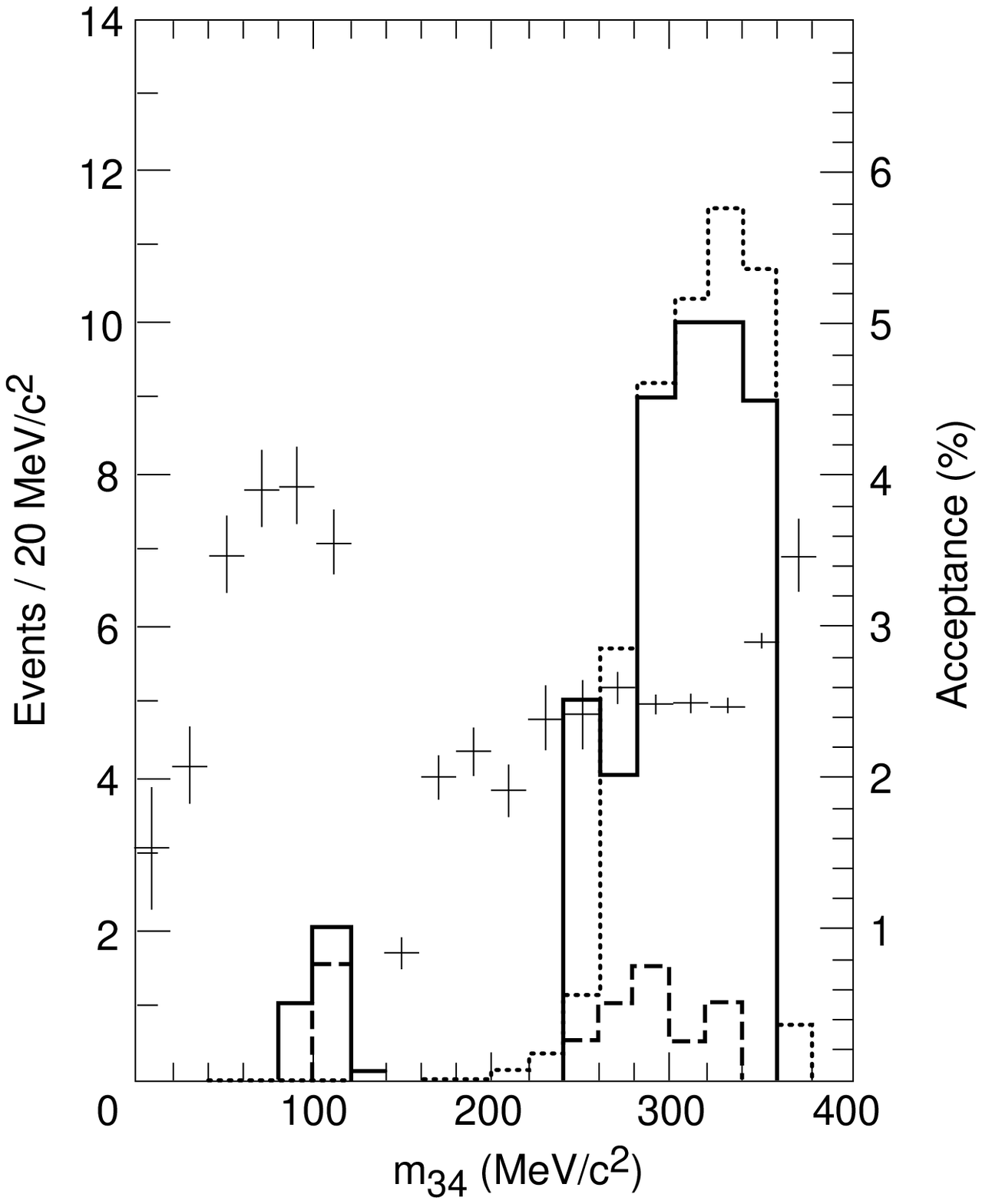,width=\linewidth,height=6cm}}
\caption{Measured
\protect\cite{BA:92}
$2\gamma$--invariant--mass distribution for
$K_L\to\pi^0\gamma\gamma$ (solid line).
The dashed line shows the estimated background.
The experimental acceptance is given by the crosses.
The dotted line simulates the $\cO (p^4)$ ChPT prediction.}
\label{fig:spectrum_NA31}
\end{minipage}
}
\vfill
\end{figure}
 
Again, the symmetry constraints do not allow any
tree--level contribution to $K_2\to\pi^0\gamma\gamma$
from $\cO (p^4)$ terms in the Lagrangian.
The $A(y,z)$ amplitude is therefore determined by a
finite loop calculation \cite{EPR:87b}.
The relevant Feynman diagrams are analogous to the ones in
fig.~\ref{fig:ksgg}, but with an additional $\pi^0$ line
emerging from the weak vertex;
charged kaon loops also give a small contribution in this case.
Due to the large absorptive $\pi^+\pi^-$ contribution,
the spectrum in the invariant mass of the two photons
is predicted \cite{EPR:87b,CdA:88}
to have a very characteristic behaviour
(dotted line in fig.~\ref{fig:spectrum}),
peaked at high values of $m_{\gamma\gamma}$.
The agreement with the measured two--photon distribution \cite{BA:92},
shown in fig.~\ref{fig:spectrum_NA31},
is remarkably good.
However, the $\cO (p^4)$ prediction for the rate \cite{EPR:87b,CdA:88},
$\mbox{\rm Br}(K_L \rightarrow \pi^0 \gamma \gamma) = 0.67\times 10^{-6}$,
is smaller than the experimental value \cite{BA:92,PA:91}:
\bel{eq:br_klpgg}
\mbox{\rm Br}(K_L \rightarrow \pi^0 \gamma \gamma )
\, = \, (1.70 \pm 0.28) \times 10^{-6} \, . 
\ee

Since the effect of the amplitude $B(y,z)$ first appears at
$\cO (p^6)$, one should worry about the size of the next--order
corrections. A na\"{\i}ve vector--meson--dominance
(VMD) estimate \cite{SE:88,MI:89,FR:89,HS:93}
through the decay chain
$K_L\to\pi^0,\eta,\eta'\to V \gamma\to\pi^0\gamma\gamma$
results in a sizeable contribution to $B(y,z)$.
However, this type of calculation predicts a photon
spectrum peaked at low values of $m_{\gamma\gamma}$,
in strong disagreement with experiment.
As first emphasized in ref.~\cite{EPR:90},
there are also so--called direct weak contributions
associated with $V$ exchange, which cannot be written as a strong
VMD amplitude with an external weak transition.
Model--dependent estimates of this direct contribution \cite{EPR:90,dAP:97}
suggest a strong cancellation with the
na\"{\i}ve vector--meson--exchange effect;
but the final result is unfortunately quite uncertain.
 
A detailed calculation of the most important $\cO (p^6)$ 
corrections has been performed in ref.~\cite{CEP:93}.
In addition to the VMD contribution, the unitarity corrections
associated with the two--pion intermediate state
(i.e. $K_L\to\pi^0\pi^+\pi^-\to\pi^0\gamma\gamma$) have been
included \cite{CEP:93,CdAM:93}.
Figure~\ref{fig:spectrum} shows the resulting photon spectrum
for $a_V=0$ (dashed curve) and $a_V=-0.9$ (full curve),
where $a_V$ parametrizes the size of the VMD amplitude.
The corresponding branching ratio is:
\bel{eq:br_pred_p6}
\mbox{\rm Br}(K_L\to\pi^0\gamma\gamma) \, = \, \cases{
0.67\times 10^{-6} , & \quad $O(p^4)$, \cr
0.83\times 10^{-6} , & \quad $O(p^6), \, a_V=0\, $, \cr
1.60\times 10^{-6} , & \quad $O(p^6), \, a_V=-0.9\, $. }
\ee
The unitarity corrections by themselves raise the rate only
moderately. Moreover, they produce an even more pronounced
peaking of the spectrum at large $m_{\gamma\gamma}$, which
tends to ruin the success of the $\cO (p^4)$ prediction.
The addition of the $V$ exchange contribution restores again
the agreement.
Both the experimental rate and the spectrum
can be simultaneously reproduced with  $a_V = -0.9$.
A more complete unitarization of the $\pi$--$\pi$ intermediate
states \cite{KH:94}, 
including the experimental $\gamma\gamma\to\pi^0\pi^0$
amplitude, increases the $K_L\to\pi^0\gamma\gamma$ decay width 
some 10\%, leading to a slightly smaller value of $|a_V|$.

For the charged decay $K^+\to\pi^+\gamma\gamma$, the sum of all
1--loop diagrams gives also a finite $O(p^4)$ amplitude
$A(y,z)$. However, chiral symmetry allows in addition for a direct
tree--level contribution proportional to the 
renormalization--scale--invariant constant \cite{EPR:88}
\bel{eq:c_deff}
\hat c = 32 \pi^2 \left[ 4 \left( L_9 + L_{10}\right)
-{1\over 3} \left( w_1 + 2 w_2 + 2 w_4 \right)\right] .
\ee
There is also a contribution to $C(y,z)$, generated by
the chiral anomaly \cite{EPR:88}.
Since $\hat c$ is unknown, ChPT alone cannot predict 
$\Gamma(K^+\to\pi^+\gamma\gamma)$; nevertheless, it gives, up to
a twofold ambiguity, a precise correlation between the rate and the
spectrum. Moreover, one can derive the lower bound \cite{EPR:88}
Br$(K^+\to\pi^+\gamma\gamma)\geq 4\times 10^{-7}$.

From na\"{\i}ve power--counting arguments one expects $\hat c\sim O(1)$,
although $\hat c = 0$ has been obtained in some models \cite{EPR:90}.
The shape of the $z$ distribution is very sensitive to $\hat c$
and, for reasonable values of this parameter, is predicted \cite{EPR:88}
again
to peak at large $z$ due to the rising absorptive part of the
$\pi\pi$ intermediate state.
An analysis of the main $\cO (p^6)$ corrections \cite{dAP:96},
analogous to the one previously performed for the $K_L$ decay mode
\cite{CEP:93,CdAM:93},
suggests that the unitarity corrections generate a sizeable
($\sim 30$--40\%) increase of the decay width.

The recent results of the BNL-E787 experiment
\cite{787:97a} show indeed a clear enhancement of events
at large $z$, in nice agreement with the theoretical expectations.
The value of $\hat c$ obtained from the data is
$\hat c = 1.8\pm 0.6$.
Assuming the predicted chiral spectrum,
this implies Br$(K^+\to\pi^+\gamma\gamma) = (1.1\pm 0.3)\times 10^{-6}$.

\medskip
\noindent {\it\thesubsection.5. $K\to\pi l^+ l^-$}
\smallskip

\noindent
The  $\cO (p^4)$  calculation of $K^+\to\pi^+ l^+ l^-$
and $K_S\to\pi^0 l^+ l^-$ involves a divergent loop,
which is renormalized by the $\cO (p^4)$ Lagrangian.
The decay amplitudes can then be written \cite{EPR:87a}
as the sum of a calculable loop
contribution plus an unknown combination of chiral couplings,
\beqn\label{eq:wp_w0}
w_+ & =  & 
   -{1\over 3} (4\pi)^2 \left[w_1^r + 2 w_2^r - 12 L_9^r\right]
  -{1\over 3} \log{\left(M_K M_\pi/\mu^2\right)} ,
\no\\
w_S  & =  & -{1\over 3} (4\pi)^2 \left[w_1^r - w_2^r\right]
  -{1\over 3} \log{\left(M_K^2/\mu^2\right)} , 
\eeqn
where $w_+$, $w_S$
refer to the decay of the $K^+$ and $K_S$ respectively.
These constants are expected to be of $\cO (1)$ by
na\"{\i}ve power--counting arguments.
The logarithms have been included to compensate the
renormalization--scale dependence of the chiral couplings,
so that $w_+$, $w_S$ are observable quantities.
If the final amplitudes are required to transform as
octets, then $w_2 = 4 L_9$, implying
$w_S = w_+ + {1\over 3}\log{\left(M_\pi/M_K\right)}$.
It should be emphasized
that this relation goes beyond the usual requirement
of chiral invariance.
 
The measured $K^+\to\pi^+ e^+ e^-$ decay rate determines \cite{EPR:87a}
two possible solutions for $w_+$.
The twofold ambiguity can be solved, looking to
the shape of the invariant--mass distribution of the final lepton
pair, which is regulated by the same parameter $w_+$.
A fit to the BNL--E777 data \cite{AL:92} gives
\bel{eq:omega}
w_+ = 0.89{\,}^{+0.24}_{-0.14}\, ,
\ee
in agreement with model--dependent
theoretical estimates \cite{EPR:90,BP:93}.
Once $w_+$ has been fixed, one can
predict \cite{EPR:87a}
the rates and Dalitz--plot distributions
of the related modes
$K^+\to\pi^+ \mu^+ \mu^-$,
$K_S\to\pi^0 e^+ e^-$ and $K_S\to\pi^0 \mu^+ \mu^-$.
The recent BNL-787 measurement \cite{787:97b}
Br$(K^+\to\pi^+ \mu^+ \mu^-) = (5.0\pm 0.4\pm 0.9)\times 10^{-8}$,
is in excellent agreement with the theoretical prediction
Br$(K^+\to\pi^+ \mu^+ \mu^-) = (6.2{\,}^{+0.8}_{-0.6})\times 10^{-8}$.

\medskip
\noindent {\it\thesubsection.6. $K_L\to\pi^0 e^+ e^-$}
\smallskip

\noindent
 The rare decay $K_L \rightarrow \pi^0 e^+ e^-$
is an interesting process
in looking for new CP--violating signatures.
If CP were an exact symmetry,
only the CP--even state $K_1^0$ could decay 
via one--photon emission, while
the decay of the CP--odd state $K_2^0$ would proceed through a 
two--photon intermediate state and, therefore,
its decay amplitude would be suppressed
by an additional power of $\alpha$.
When CP violation is taken into account,
however, an $\cO (\alpha)$ $K_L \rightarrow \pi^0 e^+ e^-$ decay
amplitude is induced, both through the small
$K_1^0$ component of the $K_L$
($\varepsilon$ effect) and through direct CP violation in the
$K_2^0 \rightarrow \pi^0 e^+ e^-$ transition.
The electromagnetic suppression of the CP--conserving amplitude then
makes it plausible that this decay is
dominated by the CP--violating contributions.
 
  The short--distance analysis of the product of 
weak and electromagnetic
currents allows a reliable calculation of the direct CP--violating
$K_2^0 \rightarrow \pi^0 e^+ e^-$ amplitude.
The corresponding branching ratio has been estimated
\cite{buras} to be:
\bel{eq:direct}
\mbox{\rm Br}(K_L \rightarrow \pi^0 e^+ e^-)\Big|_{\mathrm{Direct}}
= (4.5\pm 2.6) \times 10^{-12} .
\ee

The indirect CP--violating amplitude induced by the 
$K_1^0$ component of
the $K_L$ is given by the $K_S \rightarrow \pi^0 e^+ e^-$ amplitude
times the CP--mixing parameter $\varepsilon$.
Using the octet relation between $w_+$ and $w_S$,
the determination of the parameter $\omega_+$ in 
\eqn{eq:omega}
implies
\bel{eq:indirect}
\mbox{\rm Br}(K_L \rightarrow \pi^0 e^+ e^-)\Big|_{\mathrm{Indirect}}
 \le  1.5 \times 10^{-12}.
\ee
Comparing this  value with \eqn{eq:direct},
we see that the direct
CP--violating contribution is expected to be larger than the
indirect one. This is very different from the situation in
$K \rightarrow \pi \pi$, where the contribution due to mixing
completely dominates.
 
Using the computed  $K_L\to\pi^0\gamma\gamma$ amplitude,
one can estimate the CP--conserving two--photon exchange contribution
to $K_L\to\pi^0e^+e^-$,
by taking the absorptive part due to the two--photon 
discontinuity as an
educated guess of the actual size of the complete amplitude.
At $\cO (p^4)$, the $K_L\to\pi^0e^+e^-$ decay
amplitude is
strongly suppressed (it is proportional to $m_e$), owing to the
helicity structure of the $A(y,z)$ term \cite{EPR:88,DHV:87}.
This helicity suppression is, however, no longer true at 
the next order in the chiral expansion. 
The $\cO(p^6)$ estimate \cite{CEP:93} of the amplitude
$B(y,z)$ gives rise to
\bel{eq:klpee_p6}
\mbox{\rm Br}(K_L \rightarrow \pi^0 \gamma^* \gamma^* \rightarrow 
\pi^0 e^+ e^-)
\sim\, \cases{
0.3 \times 10^{-12}, & \quad $a_V=0\, $, \cr
1.8  \times 10^{-12}, & \quad $a_V=-0.9\, $. }
\ee

Thus, the decay width seems to be dominated by the CP--violating
amplitude, but the CP--conserving contribution could also be
important. Notice that if both amplitudes were comparable
there would be a sizeable CP--violating energy asymmetry between the 
$e^-$ and the $e^+$ distributions \cite{SE:88,HS:93,DG:95}.

   The present experimental upper bound \cite{HA:93},
\bel{eq:klpee_exp}
\mbox{\rm Br}(K_L \rightarrow \pi^0 e^+ e^-)\Big|_{\mathrm{Exp}}
 < 4.3 \times 10^{-9} \qquad (90\% \mbox{\rm CL}) ,
\ee
is still far away from the expected Standard Model signal,
but the prospects
for getting the needed sensitivity of around $10^{-12}$ in
the next few years are rather encouraging.
To be able to interpret a future experimental measurement of
the decay rate as a (direct) CP--violating signature,
it is first necessary, however,
to pin down more precisely the actual
size of the three different components of the decay amplitude \cite{Orsay}.

%
%

\section{Heavy Quark Effective Theory}
\label{sec:hqet}

The chiral symmetries of massless QCD are not relevant for
heavy quarks.
There is, however, another approximate limit of QCD
which turns out to be rather useful: the infinite--mass limit.

The dynamical simplifications which occur in the heavy--mass
limit can be easily understood by looking back to the more
familiar atomic physics.
The quantum mechanical properties of an electron in the
Coulomb potential of an atomic nucleus
are regulated by the reduced mass
$m_e M/(m_e + M)\approx m_e \ll M$, where $M$ is the heavy
nuclear mass.
Therefore, different isotopes ($M\not=M'$) of the same
atom ($Z=Z'$) have the same chemical properties to a very
good approximation (isotopic symmetry).
Moreover, atoms with nuclear spin $S$ are $(2S+1)$ degenerate
in the limit $M\to\infty$ (spin symmetry).

The QCD analog is slightly more complicated, but the general idea
is the same.
The quarks confined inside hadrons exchange momentum of a magnitude
of about $\Lambda\sim M_p/3 \approx 300$ MeV. The scale $\Lambda$
characterize the typical amount  by which quarks are off-shell;
it also determines the  hadronic size
$R_{\mathrm{had}}\sim 1/\Lambda$.
If we consider a {\it heavy--light} hadron composed of one heavy quark $Q$
and any
number of light constituents, the light quark(s) is (are) very far
off-shell by an amount of order $\Lambda$.
However, if $M_Q\gg\Lambda$, the heavy quark is almost on-shell
and its Compton wavelength $\lambda_Q\sim 1/M_Q$
is much smaller than the hadronic size $R_{\mathrm{had}}$.

Although the quark interactions change the momentum
of $Q$ by $\delta P_Q\sim\Lambda$, its velocity only changes
by a negligible amount,
$\delta v_Q\sim\Lambda/M_Q \ll 1$.
Thus, $Q$ moves approximately with constant velocity.
 In the hadron rest frame, the heavy quark is almost at rest
and acts as a static source of gluons. It is surrounded by a 
complicated, strongly interacting cloud of light quarks,
antiquarks and gluons, sometimes referred to as the
{\it brown muck}.
To resolve the quantum numbers of the heavy quark would
require a hard probe with $Q^2\gsim M_Q^2$; however,
the soft gluons coupled to the {\it brown muck} can only resolve
larger distances of order $R_{\mathrm{had}}$.
The light hadronic constituents are blind to the
flavour and spin orientation of the heavy quark;
they only feel its colour field which extends over large distances
because of confinement.
Thus, in the infinite--$M_Q$ limit,
the properties of heavy--light hadrons are independent
of the mass ({\it flavour} symmetry) and spin 
({\it spin} symmetry) of the heavy source of colour \cite{IW:89}.

In order to put these qualitative arguments 
within a more formal framework, let us write the heavy quark
momentum as
\bel{eq:P_Q}
P_Q^\mu \equiv M_Q v^\mu + k^\mu \, ,
\ee
where $v^\mu$ is the hadron four-velocity 
($v^2=1$) and $k^\mu$ the
{\it residual} momentum of order $\Lambda$.
In the limit $M_Q\to\infty$ with $v^\mu$ kept fixed
\cite{IW:89}, the QCD Feynman rules simplify
considerably \cite{GR:90}. The heavy quark propagator becomes
\bel{eq:Q_propagator}
{i\over \slashchar{P}_Q - M_Q} \, = \, {i\over v\cdot k}\, 
{1+ \slashchar{v}\over 2} + \cO(1/M_Q) \, .
\ee
The factors $P_\pm\equiv (1\pm \slashchar{v})/2$ are
energy projectors ($P_\pm^2=P_\pm$, $P_\pm P_\mp =0$).
Thus, the propagator is independent of $M_Q$ and only the 
positive--energy projection of the
heavy quark field propagates. Moreover, since
$P_+\gamma^\mu P_+ = P_+ v^\mu P_+$, the quark--gluon
vertex reduces to
\bel{eq:Q_vertex}
i g \left({\lambda^a\over 2}\right) \gamma^\mu
\quad\longrightarrow\quad
i g \left({\lambda^a\over 2}\right) v^\mu \, .
\ee
The resulting interaction is then independent of the heavy--quark
spin.

These Feynman rules can be easily incorporated into an effective
Lagrangian. Making  the field redefinition
\bel{eq:hv_def}
Q(x)\approx \e^{-iM_Q v\cdot x}\, h_v^{(Q)}(x) \, ,
\ee
where 
$h_v^{(Q)}=P_+ h_v^{(Q)} = \slashchar{v} h_v^{(Q)}$
(i.e., we are only considering the positive--energy projection
of the heavy--quark spinor),
the heavy--quark Lagrangian becomes \cite{EH:90,GE:90}
\bel{eq:L_Q}
\cL_{\mathrm{QCD}}^{(Q)} \, =\, \bar Q \, (i\,\slashchar{D} - M_Q)\, Q
\,\approx\, \bar h_v^{(Q)} \, i\, (v\cdot D)\, h_v^{(Q)} \, ,
\ee
showing explicitly that the interaction is independent of
the mass and spin of the heavy quark.
The corresponding equation of motion is:
\bel{eq:motion}
i \,\slashchar{D} Q = M_Q Q \quad\longrightarrow\quad
i \, (v\cdot D) h_v^{(Q)} =0 \, .
\ee

The redefinition \eqn{eq:hv_def} scales out the rapidly varying part
of the heavy--quark field.
The phase factor removes the {\it kinetic}
piece $M_Q v^\mu$ from the heavy--quark momentum, so that in 
momentum space a derivative acting on $h_v^{(Q)}$ just
produces the {\it residual} momentum $k^\mu$.
Notice that $h_v^{(Q)}$ is a two--component spinor,
which destroys a quark $Q$ but does not create the corresponding
antiquark; pair creation does not occur in the
Heavy Quark Effective Theory (HQET).

\subsection{Spectroscopic Implications}

Let us denote $s_l$ the total spin of the light degrees of
freedom in a hadron containing a single heavy quark $Q$.
In the $M_Q\to\infty$ limit, the dynamics is independent of
the heavy--quark spin. Therefore, there will be two
degenerate hadronic states with $J=s_l\pm\frac{1}{2}$.
For $Q\bar q$ mesons the ground state has negative parity
and $s_l=1/2$, giving a doublet  of degenerate spin--zero and
spin--one mesons.
The measured charm and bottom spectrum \cite{PDG:96} shows indeed that
this is true to a quite good approximation:
\bel{D_B_spectrum}\begin{array}{lll}
M_{D^*} - M_D = (142.12\pm0.07)\,\mbox{\rm MeV} , &\;\; &
\dfrac{(M_{D^*} - M_D)}{M_D} \approx 8\%  ,
\\
M_{B^*} - M_B = (45.7\pm 0.4)\,\mbox{\rm MeV} , &&
\dfrac{(M_{B^*} - M_B)}{M_B} \approx 0.9\%  .
\ea\ee
The infinite--mass limit works much better for the
bottom, although the result is also good in the charm case.
We expect these mass splittings to get corrections of the
form
$M_{P^*}-M_P\approx a/M_Q$; this gives
the refined prediction
$M_{B^*}^2 - M_B^2 \approx M_{D^*}^2 - M_D^2$,
which is in very good agreement with the data \cite{PDG:96}:
\bel{eq:D_B_masses}
M_{D^*}^2 - M_D^2 \approx 0.55 \:\mbox{\rm GeV}^2 , \qquad\quad
M_{B^*}^2 - M_B^2 \approx 0.48 \:\mbox{\rm GeV}^2 .
\ee

The first excitation with $s_l=\frac{3}{2}$, would correspond
to a degenerate $(1^+,2^+)$ doublet, which has been already
identified \cite{PDG:96} in the charm sector:
\bel{D_1,2_spectrum}
M_{D^*_2} - M_{D_1} \approx 37\,\mbox{\rm MeV} , \qquad
(M_{D^*_2} - M_{D_1})/ M_{D_1} \approx 1.5\%  .
\ee
For the beauty spectrum, one then expects
\bel{eq:B*_split}
M_{B^*_2}^2 - M_{B_1}^2 \approx M^2_{D^*_2} - M^2_{D_1}
\approx 0.18\:\mbox{\rm GeV}^2 .
\ee
%


%
%

\subsection{Effective Lagrangian}

The infinite--mass limit provides a very useful starting point to
analyze the physics of heavy quarks. Moreover, it is possible to 
estimate $1/M_Q$ corrections in a systematic way, by using the
appropriate EFT methods.

Using the energy projectors $P_\pm = (1\pm\slashchar{v})/2$ 
we can decompose the heavy quark field in two pieces,
\bel{eq:Q_split}
Q(x)\equiv (P_+ + P_-) \, Q(x) \equiv
\e^{-iM_Q v\cdot x}\, \left(h_v^{(Q)}(x) + H_v^{(Q)}(x)\right)\, ,
\ee
where we have extracted the leading
quark--mass dependence through the explicit phase factor.
Because of the energy projectors, the new fields satisfy
$\slashchar{v}\, h_v^{(Q)} = h_v^{(Q)}$ and
$\slashchar{v}\, H_v^{(Q)} = - H_v^{(Q)}$.
In the hadron rest frame, $v^\mu = (1,\vect{0}\,)$, 
$P_\pm = (1\pm\gamma_0)/2$;
thus, $h_v^{(Q)}(x)$ and $H_v^{(Q)}(x)$ correspond to the
upper and lower components of $Q(x)$, respectively.
The field $h_v^{(Q)}(x)$ annihilates a heavy quark with velocity $v^\mu$,
while $H_v^{(Q)}(x)$ creates a heavy antiquark with the same velocity.

In order to define the HQET, we should
{\it integrate out} $H_v^{(Q)}(x)$ because, at the energy scale we are 
interested in ($k\ll M_Q$), heavy antiquarks cannot be produced.
This is
slightly more tricky than the usual integration of a heavy field in
EFT, since only the lower component of $Q(x)$ is {\it integrated out}.
What we want to do is more similar to a non-relativistic approximation,
but keeping the full power of Lorentz covariance.
Notice that the field redefinition \eqn{eq:Q_split} is only adequate for
describing a heavy quark. If one wants to study the physics
of a heavy antiquark, one should use instead
\bel{eq:Q_antisplit}
Q(x)\equiv (P_- + P_+) \, Q(x) \equiv
\e^{iM_Q v\cdot x}\, \left(h_v^{-(Q)}(x) + H_v^{-(Q)}(x)\right)\, .
\ee
The antiquark formalism is identical to the quark one, 
with the replacements $v^\mu\to -v^\mu$
and $h_v^{(Q)}(x)\to h_v^{-(Q)}(x)$.

With the redefinition \eqn{eq:Q_split},
the heavy--quark Lagrangian becomes
\beqn\label{eq:L_hH}
\cL_{\mathrm{QCD}}^{(Q)} & =& 
\left(\bar h_v^{(Q)} + \bar H_v^{(Q)}\right)\,
\left[ i\,\slashchar{D} - 2 M_Q P_-\right]\,
\left( h_v^{(Q)} +  H_v^{(Q)}\right)
\no\\ & = &
\bar h_v^{(Q)} \, i\, (v\cdot D)\, h_v^{(Q)}
- \bar H_v^{(Q)} \,\left( i\, v\cdot D + 2 M_Q \right)\, H_v^{(Q)}
\no\\ &&\mbox{}
+ \bar h_v^{(Q)} \, i\,\slashchar{D}_\perp\,  H_v^{(Q)}
+ \bar H_v^{(Q)} \, i\,\slashchar{D}_\perp\,  h_v^{(Q)}
\, ,
\eeqn
where
$D^\mu_\perp\equiv D^\mu - v^\mu (v\cdot D)$ 
is the component of the Dirac operator orthogonal to the velocity,
i.e. $v\cdot D_\perp = 0$,
and we have used the relations
$P_\pm\gamma^\mu P_\pm = \pm P_\pm v^\mu P_\pm$
and
$P_\mp \slashchar{D} P_\pm = P_\mp \slashchar{D}_\perp P_\pm$.
In the hadron rest frame,
$D^\mu_\perp = (0,\vect{D})$ contains just the space components of the 
covariant derivative.

The field $h_v^{(Q)}$ describes a massless degree of freedom, while
$H_v^{(Q)}$  corresponds to fluctuations with twice the heavy quark mass.
The third and fourth terms in \eqn{eq:L_hH}, which mix the two fields,
describe quark--antiquark creation and annihilation. A virtual heavy
quark propagating forward in time can turn into a virtual antiquark
propagating backward in time and then turn back into a quark.
Since there is no energy to produce on-shell quark--antiquark pairs,
the virtual fluctuation into the intermediate 
$h_v^{(Q)}h_v^{(Q)}\bar H_v^{(Q)}$ state can only
propagate over a very short distance
$\Delta x \sim 1/M_Q$.

At the classical level, we can eliminate the field $H_v^{(Q)}$ using the
QCD equation of motion
$(i\,\slashchar{D} - M_Q)\, Q = 0$, which in terms of the
$h_v^{(Q)}$ and $H_v^{(Q)}$ fields takes the form
\bel{eq:EqMot}
i\,\slashchar{D}h_v^{(Q)} + 
\left(i\,\slashchar{D} - 2 M_Q\right)\, H_v^{(Q)} = 0 \, .
\ee
Multiplying it by $P_\pm$, this equation gets projected into two
different pieces:
\be\label{eq:EqMotPro}
i \, v\cdot D h_v^{(Q)} = - i \,\slashchar{D}_\perp H_v^{(Q)} ;
\qquad\;
\left( i \, v\cdot D + 2 M_Q \right) H_v^{(Q)} = i\,\slashchar{D}_\perp 
h_v^{(Q)} .
\ee
The second shows explicitly that $H_v^{(Q)}\sim \cO (1/M_Q)$:
\bel{eq:HQ}
H_v^{(Q)} = {1\over\left( i\, v\cdot D + 2 M_Q -i\epsilon\right)}
\, i \,\slashchar{D}_\perp h_v^{(Q)} .
\ee
Inserting \eqn{eq:HQ} back into \eqn{eq:L_hH}, one gets
the Lagrangian:
\bel{eq:LHQET}
\cL_{\mathrm{eff}} =
\bar h_v^{(Q)} \, i\, (v\cdot D)\, h_v^{(Q)} +
\bar h_v^{(Q)} \, i\slashchar{D}_\perp
{1\over\left( i v\cdot D + 2 M_Q -i\epsilon\right)}\,
i \slashchar{D}_\perp h_v^{(Q)} .
\ee
The second term corresponds to the virtual quark--antiquark fluctuations
of $\cO (1/M_Q)$.

This Lagrangian can be obtained in a more elegant way,
manipulating the QCD generating functional. The functional integration
over the $H_v^{(Q)}$ field is Gaussian
and can be explicitly performed. One gets the classical action, given
by the Lagrangian  \eqn{eq:LHQET}, times the determinant of the Dirac 
operator,
\bel{eq:determ}
\det \left( i v\cdot D + 2 M_Q -i\epsilon\right)^{1/2}
= \exp\left\{{1\over 2} {\rm tr} \left[ 
\log \left( i v\cdot D + 2 M_Q -i\epsilon\right)\right]\right\} ,
\ee
which is a quantum effect.
However, by choosing the axial gauge $v\cdot G=0$, one can easily
see that \eqn{eq:determ} is just an irrelevant constant \cite{MRR:92,ST:92}
(this result is of course gauge independent).

\subsection{$1/M_Q$ Expansion}

Because of the phase factor in \eqn{eq:Q_split}, the x--dependence of 
the effective field $h_v^{(Q)}$ is rather weak. Derivatives acting on
$h_v^{(Q)}$ produce powers of the small momentum $k^\mu$. Therefore,
the non-local HQET Lagrangian \eqn{eq:LHQET}
can be expanded in powers of $D/M_Q$.

Using the identity
\beqn\label{eq:iden}
P_+ i \slashchar{D}_\perp i \slashchar{D}_\perp P_+
&= &
 P_+ \left\{ \left( i \slashchar{D}_\perp \right)^2 +{1\over 2}
 \left[ i \slashchar{D} , i \slashchar{D} \right]\right\} P_+
\no\\ & = &
 P_+ \left\{ \left( i \slashchar{D}_\perp \right)^2 +{g\over 2}
 \sigma_{\alpha\beta} G^{\alpha\beta} \right\} P_+,
\eeqn
with
$G^{\alpha\beta}\equiv {\lambda^a\over 2} G^{\alpha\beta}_a$
the gluon field strength tensor,
one finds \cite{EH:90b,FGL:91}
\beqn\label{eq:expL}
\cL_{\mathrm{HQET}} &=&
\bar h_v^{(Q)} \, i\, (v\cdot D)\, h_v^{(Q)} +
{1\over 2 M_Q} \bar h_v^{(Q)} \, \left( i\,\slashchar{D}_\perp\right)^2
h_v^{(Q)} 
\no\\ && \mbox{}
+ {g\over 4 M_Q} \bar h_v^{(Q)} \, \sigma_{\alpha\beta}
G^{\alpha\beta} h_v^{(Q)}
+ \cO(1/M_Q^2) .
\eeqn

The physical meaning of the two $\cO(1/M_Q)$ operators
is rather transparent in the rest frame
[$\slashchar{D}_\perp = (0,\vect{D})$; 
$P_+ \sigma_{0i} P_+ = 0$]:
\beqn\label{eq:rfid}
O_{\mathrm{kin}}&\equiv & {1\over 2 M_Q} \bar h_v^{(Q)} \, 
\left( i\,\slashchar{D}_\perp\right)^2 h_v^{(Q)}
\quad\longrightarrow\quad
-{1\over 2 M_Q} \bar h_v^{(Q)} \, \left( i\,\vect{D}\right)^2 h_v^{(Q)} ,
\no\\
O_{\mathrm{mag}}&\equiv & {g\over 4 M_Q} \bar h_v^{(Q)} \, 
\sigma_{\alpha\beta} G^{\alpha\beta} h_v^{(Q)}
\quad\longrightarrow\quad
-{g\over M_Q} \bar h_v^{(Q)} \, \vect{S}\cdot\vect{B}_{\mathrm{c}}\,
h_v^{(Q)} .
\eeqn
The first operator is just the gauge--covariant extension of the
kinetic energy associated with the off-shell residual momentum of
the heavy quark.
The second operator is the non-abelian analog of the QED Pauli
term, which describes the interaction
of the heavy--quark spin with the gluon field.
Here,
$B^i_{\mathrm{c}} \equiv -{1\over 2}\epsilon^{ijk} G^{jk}$
are the components of the colour magnetic field and
\bel{eq:spin}
\vect{S}\equiv {1\over 2}\gamma_5\gamma^0\vect{\gamma} =
{1\over 2}\left( \bat \vect{\sigma} & 0 \\ 0 & \vect{\sigma}\ea\right)
\ee
is the usual spin operator, which satisfies
\bel{eq:spinprop}
\left[S^i,S^j\right] = i \epsilon^{ijk} S^k \, , \qquad\qquad
\left[\slashchar{v}, S^i\right] = 0 \, .
\ee
Thus, the heavy--quark spin symmetry is broken at $\cO(1/M_Q)$
by this chromomagnetic hyperfine interaction.

Using the expression \eqn{eq:HQ} for $H_v^{(Q)}$, obtained from
the equation of motion, one can also derive a $1/M_Q$ expansion
for the full heavy--quark field $Q(x)$:
\beqn\label{eq:Q_exp}
Q(x) & = &  \e^{-iM_Q v\cdot x}\, \left[
1 + {1\over \left(i\, v\cdot D + 2 M_Q - i\epsilon\right)} 
i\,\slashchar{D}_\perp\right] h_v^{(Q)}(x) 
\no\\ & = & \e^{-iM_Q v\cdot x}\, \left(
1 + {i\,\slashchar{D}_\perp\over 2 M_Q} + \cdots \right) h_v^{(Q)}(x) .
\eeqn
This relation tells us how to construct (at tree level) the HQET operators.
For instance, the vector current
$V^\mu = \bar q\gamma^\mu Q$, composed of a heavy quark and a light
antiquark, is represented in the HQET by the expansion
\beqn\label{eq:V_exp}
V^\mu(x) &=& \e^{-iM_Q v\cdot x}\, \bar q(x)\gamma^\mu\left(
1 + {i\,\slashchar{D}_\perp\over 2 M_Q} + \cdots \right) h_v^{(Q)}(x) 
\no\\ &\equiv &
 \e^{-iM_Q v\cdot x} \, V^\mu(x)_{\mathrm{HQET}}.
\eeqn

\subsection{Renormalization and Matching}

\begin{figure}[tbh]      
\setlength{\unitlength}{0.6mm} \centering
\begin{picture}(160,160)
\thicklines

\put(65,150){\makebox(30,10){\large Large $\mu$}}
\put(20,132){\makebox(40,10){\Large QCD}}
\put(20,118){\makebox(40,10){\large $(M_c = 0 \; ; \; M_b = 0)$}}
\put(95,125){\makebox(60,10){\large Renormalization Group}}

\put(65,100){\makebox(30,10){\large $\mu = M_b$}}
\multiput(10,105)(5,0){11}{\line(1,0){2.5}}
\multiput(97.5,105)(5,0){6}{\line(1,0){2.5}}
\put(125,100){\makebox(40,10){\large Matching}}
\put(20,82){\makebox(40,10){\Large HQET${}_b$}}
\put(20,68){\makebox(40,10){\large $(M_c =0 \; ; \; M_b = \infty )$}}
\put(95,75){\makebox(60,10){\large Renormalization Group}}

\put(65,50){\makebox(30,10){\large $\mu = M_c$}}
\multiput(10,55)(5,0){11}{\line(1,0){2.5}}
\multiput(97.5,55)(5,0){6}{\line(1,0){2.5}}
\put(125,50){\makebox(40,10){\large Matching}}
\put(20,32){\makebox(40,10){\Large HQET${}_{b,c}$}}
\put(20,18){\makebox(40,10){\large $(M_c =\infty \; ; \; M_b = \infty )$}}
\put(95,25){\makebox(60,10){\large Renormalization Group}}
\put(65,0){\makebox(30,10){\large Low $\mu$}}

\linethickness{0.3mm}
\put(80,148){\vector(0,-1){36}}
\put(80,98){\vector(0,-1){36}}
\put(80,48){\vector(0,-1){36}}

\end{picture}
\caption{Evolution from high to low scales in heavy--quark physics.
\label{fig:HQETevolution}}
\end{figure}
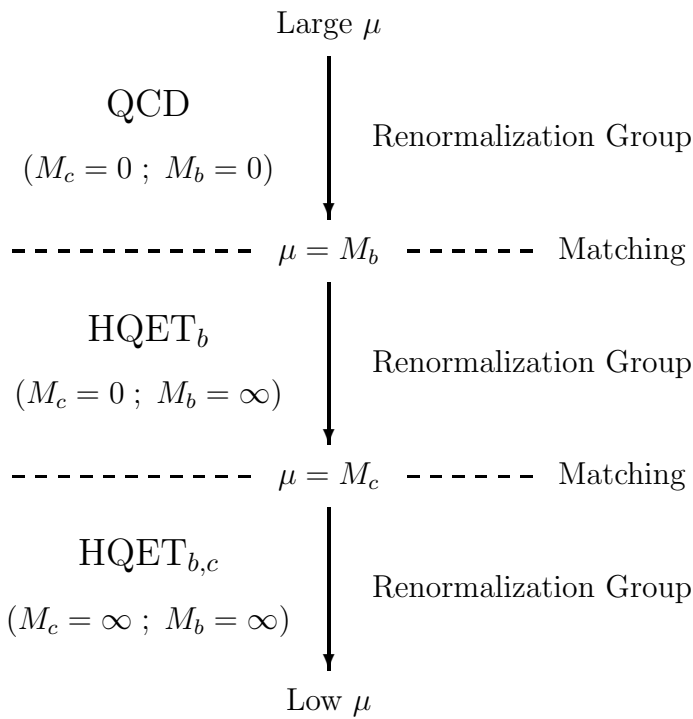

The general procedure to evolve down in energy is shown in
fig.~\ref{fig:HQETevolution}. One starts with the full QCD
theory at a high scale, where the $b$ quark can be considered
light (massless in first approximation). Using the renormalization group,
one goes down up to $\mu = M_b$, where the {\it small}
component of the $b$--quark field is {\it integrated out},
and the matching between QCD and the resulting HQET takes place.
Below $M_b$, one makes use of the HQET for the $b$ quark, until the
scale $M_c$ is reached. One can then perform a further integration
of the {\it small} components also for the charm quark, and change to
a different HQET where both the $b$ and the $c$ are considered heavy.

The numerical accuracy of the HQET predictions will be of course different
in the two HQETs, owing to the different masses of the bottom
and charm quarks. While the $1/M_b$ expansion is expected to work very 
well, corrections of $\cO(1/M_c)$ could be large in many cases.

A detailed study of renormalization and matching in HQET is beyond the 
scope of these lectures (this subject is covered by M.B. Wise
\cite{Wise}).
In the following, we are just going to illustrate how things work in
practice, through the calculation of a HQET current.

\medskip
\noindent {\it\thesubsection.1. Wave--function and vertex renormalization}
\smallskip

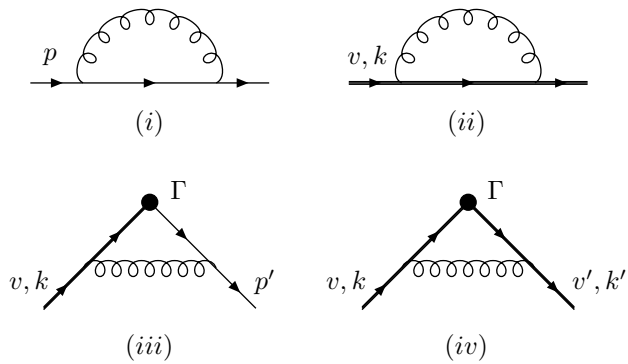
\begin{figure}[tbh]
\centering
\begin{picture}(210,125)
\ArrowLine(0,100)(20,100)\ArrowLine(20,100)(70,100)
\ArrowLine(70,100)(90,100)
\GlueArc(45,100)(25,0,180)37
\Text(5,110)[l]{$p$}\Text(45,85)[c]{$(i)$}

\ArrowLine(120,100)(140,100)\ArrowLine(140,100)(190,100)
\ArrowLine(190,100)(210,100)
\Line(120,99.5)(210,99.5)\Line(120,100.5)(210,100.5)
\GlueArc(165,100)(25,0,180)37
\Text(120,110)[l]{$v,k$}\Text(165,85)[c]{$(ii)$}

\ArrowLine(5,15)(20,30)\ArrowLine(20,30)(45,55)
\Line(5,14.5)(45,54.5)\Line(5,15.5)(45,55.5)
\ArrowLine(45,55)(70,30)\ArrowLine(70,30)(85,15)
\Gluon(20,30)(70,30)37
\Text(-8,25)[l]{$v,k$}\Text(85,25)[l]{$p'$}
\Text(45,0)[c]{$(iii)$} \Text(53,60)[l]{$\Gamma$}
\GCirc(45,55)30

\ArrowLine(125,15)(140,30)\ArrowLine(140,30)(165,55)
\Line(125,14.5)(165,54.5)\Line(125,15.5)(165,55.5)
\ArrowLine(165,55)(190,30)\ArrowLine(190,30)(205,15)
\Line(165,54.5)(205,14.5)\Line(165,55.5)(205,15.5)
\Gluon(140,30)(190,30)37
\Text(112,25)[l]{$v,k$}\Text(205,25)[l]{$v',k'$}
\Text(165,0)[c]{$(iv)$} \Text(173,60)[l]{$\Gamma$}
\GCirc(165,55)30

\end{picture}
\caption{Wave--function and vertex renormalization diagrams.}
\label{fig:divdiag}
\end{figure}

\noindent
The calculation of loop diagrams in HQET involves Feynman integrals which
look rather different than the ones appearing in the full fermion theory.
The heavy--quark propagators introduce velocity--dependent denominators,
which can be combined with the normal Feynman propagators, using the
identity \cite{GE:91}:
\bel{eq:Feyid}
{1\over (q^2)^n (q\cdot v)^m} \, = \,
{(n+m-1)!\over (n-1)! \, (m-1)! } \,\int_0^\infty\,d\lambda\;
{2^m \lambda^{m-1}\over (q^2 + 2\lambda\, q\cdot v)^{n+m}} .
\ee

It is a good exercise to perform the one--loop wave--function 
renormalization of the heavy quark. We are also going to need the
vertex renormalization of heavy--light ($\bar q \Gamma Q$) and
heavy--heavy ($\bar Q'\Gamma Q$) currents:
\bel{eq:Zrendef}
q_B = Z_q^{1/2}\, q_R \, ; \qquad\qquad
\Gamma_B = Z_\Gamma\, \Gamma_R \, .
\ee
The relevant Feynman diagrams are shown in fig.~\ref{fig:divdiag}.
The calculation is rather simple, because we only need to compute
the divergent pieces. One finds:

\begin{enumerate}

\item Quark self-energy:

\bel{eq:qsf}
-i\,\Sigma(p) \;\dot=\; -i\, \slashchar{p}\,
 {\alpha_s\,\mu^{2\epsilon}\over 3\pi\hat{\epsilon}}
 \qquad\longrightarrow\qquad
 Z_q \;\dot=\; 1 + {\alpha_s\,\mu^{2\epsilon}\over 3\pi\hat{\epsilon}} \, ;
\ee

\item Heavy--quark self-energy:

\bel{eq:Qsf}
-i\,\Sigma(v,k) \;\dot=\; i\, (v\cdot k)\,
 {2\alpha_s\,\mu^{2\epsilon}\over 3\pi\hat{\epsilon}}
 \qquad\longrightarrow\qquad
 Z_{h_v} \;\dot=\; 1 - {2\alpha_s\,\mu^{2\epsilon}\over 3\pi\hat{\epsilon}} \, ;
\ee

\item  Heavy--light vertex:

\bel{eq:qQvertex}
i\, V_\Gamma \;\dot=\; -i\,\Gamma\,
 {\alpha_s\,\mu^{2\epsilon}\over 3\pi\hat{\epsilon}}
 \qquad\longrightarrow\qquad
 Z_\Gamma \;\dot=\; 1 - {\alpha_s\,\mu^{2\epsilon}\over 3\pi\hat{\epsilon}} \, ;
\ee

\item  Heavy--heavy vertex:

\bel{eq:QQvertex}
i\, \tilde{V}_\Gamma \;\dot=\; i\,\Gamma\,
 {2\alpha_s\,\mu^{2\epsilon}\over 3\pi\hat{\epsilon}}
 \,\omega\, r(\omega)
 \quad\;\longrightarrow\quad\;
 \tilde{Z}_\Gamma \;\dot=\; 1 + 
 {2\alpha_s\,\mu^{2\epsilon}\over 3\pi\hat{\epsilon}}
 \,\omega\, r(\omega) \, ;
\ee
where
\bel{eq:r_factor}
r(\omega) \equiv {\log{\left(\omega + \sqrt{\omega^2 - 1}\right)}\over
  \sqrt{\omega^2 - 1}}
  \, ; \qquad\qquad \omega\equiv v\cdot v' \, .
\ee
\end{enumerate}

The detailed calculations can be found in ref.~\cite{GE:91}.

\medskip
\noindent {\it\thesubsection.2. Currents in HQET}
\smallskip

\noindent
Let us consider the current
\bel{eq:QCDcurr}
J_\Gamma \, =\, \bar c \,\Gamma\, b \, ,
\ee
where $\Gamma = \gamma^\mu$ (vector) or $\gamma^\mu\gamma_5$ 
(axial--vector).
When the small components of the $b$ quark are integrated out, this current
should be matched to its HQET realization, which at lowest order takes
the form\footnote{
At next-to-leading order there are additional effective operators involved
\protect\cite{NE:94}.
}
\bel{eq:cBcurrent}
J_\Gamma\quad\longrightarrow\quad
C(\mu)\; \bar c \,\Gamma\, h_v^{(b)} \, .
\ee

In full QCD the vector and axial--vector currents do not get renormalized
(the vertex  and wave--function renormalizations compensate each other);
but this is no--longer true in the HQET \cite{VS:87,PW:88}:
\bel{eq:ZetaHQET}
Z_{J_\Gamma} = Z_q^{1/2} \, Z_{h_v}^{1/2}\, Z_\Gamma \;\dot=\;
1 - {\alpha_s\, \mu^{2\epsilon}\over 2\pi \hat{\epsilon}} 
\qquad\longrightarrow\qquad
\gamma_{J_\Gamma} \;\dot=\; - {\alpha_s\over\pi} \, .
\ee
Therefore,
\bel{eq:C(mu)_HQET}
C(\mu) \approx 
 C(M_b) \;
\left( {\alpha_s(\mu^2)\over\alpha_s(M_b^2)}
\right)^{-1/\beta_1^{(n_f=4)}}   
\approx\,
\left( {\alpha_s(\mu^2)\over\alpha_s(M_b^2)}\right)^{6/25}\, .
\ee
Notice that the relevant QCD $\beta$ function is defined in the
theory with $n_f=4$ light quarks only.

If one considers also the charm quark as heavy, the current should be
matched again into a different HQET where the small components of both 
the $c$ and the $b$ have been integrated out:
\bel{eq:CBcurrent}
J_\Gamma\quad\longrightarrow\quad
\tilde{C}(\mu)\; \bar h_{v'}^{(c)} \,\Gamma\, h_v^{(b)} \, .
\ee
Since two different velocities are now involved, the relevant
renormalization factor $\tilde{Z}_{J_\Gamma}$ and the associated
anomalous dimension are functions of $v\cdot v'$:
\beqn\label{eq:ZetaHQET2}
\tilde{Z}_{J_\Gamma} &=& Z_{h_v}\, \tilde{Z}_\Gamma \;\dot=\;
1 + {2\alpha_s\, \mu^{2\epsilon}\over 3\pi \hat{\epsilon}} \,
\left[\omega\, r(\omega) - 1\right]
\, , \no\\
\tilde{\gamma}_{J_\Gamma}(\omega) &\,\dot=\, & 
{4\over 3} \, {\alpha_s\over\pi} \,
\left[\omega\, r(\omega) - 1\right] \, .
\eeqn
Thus,
\beqn\label{eq:C(mu)_HQET2}
\tilde{C}(\mu) & \,\approx\, & \tilde{C}(M_c) \;
 \left( {\alpha_s(\mu^2)\over\alpha_s(M_c^2)}
\right)^{\tilde{\gamma}^{(1)}_{J_\Gamma}(\omega)/\beta_1^{(n_f=3)}}
 \no\\ &\,\approx\, &
\left( {\alpha_s(M_c^2)\over\alpha_s(M_b^2)}\right)^{6/25} \,
\left( {\alpha_s(\mu^2)\over\alpha_s(M_c^2)}
\right)^{-{8\over 27} \left[\omega\, r(\omega) - 1\right] } .
\eeqn
The anomalous dimension vanishes for $v\cdot v'=1$, i.e.
$\tilde{\gamma}_{J_\Gamma}(1) = 0$. Therefore,
the heavy--heavy current does not get renormalized when the velocities
are equal.

\subsection{Hadronic Matrix Elements}

In order to compute physical quantities we need to evaluate hadronic
matrix elements of the HQET operators. This is again a difficult
non-perturbative problem. Nevertheless, we can derive  relations
among different matrix elements using the flavour and spin symmetries.

It is convenient to work with a mass--independent normalization 
for the meson states; i.e., to redefine the hadronic states as
\bel{eq:had_def}
| H(v)\rangle \equiv {1\over \sqrt{M_H}} \, |H(p)\rangle \, ,
\ee
with the normalization
$\langle H(v')| H(v)\rangle = 2 v^0 (2\pi)^3
\delta^{(3)}(\vec{p}-\vec{p}^{\,\prime})$.

The implications of the HQET symmetries can be derived in a rather
simple way by using a covariant tensor representation of the states
with definite transformation properties under the Lorentz group and
the heavy--quark spin--flavour symmetry \cite{GE:91,FGGW:90,Bj:91,FA:92}.

Let us consider the lowest $Q\bar q$ multiplet ($s_l=1/2$),
which contains a doublet of degenerate spin--zero and spin--one mesons
$H\equiv [ P(0^-), V(1^-)]$.
Knowing their transformation symmetry properties, we can build
appropriate wave functions to represent the states:
\beqn\label{eq:wave_f}
\lefteqn{
P(v)\propto\langle 0|h^{(Q)}_v \bar q|P(v)\rangle \,\sim\, 
- P_+\,\gamma_5 \, , 
} && \no\\ \lefteqn{
V(v,\epsilon)\propto\langle 0|h^{(Q)}_v \bar q|V(v,\epsilon )\rangle\,\sim\,
  P_+\,\slashchar{\,\epsilon\,}  , }
&&
\eeqn
where $\epsilon$ is the polarization of the vector meson
($\epsilon^*\cdot\epsilon = -1$, $v\cdot\epsilon =0$).
Since the two states are related by symmetry transformations, let us
introduce a combined wave function $\cM(v)$ that represents
both $P(v)$ and $V(v,\epsilon)$:
\beqn\label{eq:M_state}
\cM(v) &\equiv & P_+ \,\left[ - a \gamma_5 + 
  \sum_\epsilon b_\epsilon \slashchar{\,\epsilon\,}\,\right]
\, , \no\\
\overline{\cM}(v) &\equiv & \gamma^0 H(v)^\dagger \gamma^0 =
\left[ a^* \gamma_5 + 
 \sum_\epsilon b_\epsilon^* \slashchar{\,\epsilon\,}^*\right]\,P_+ \, .
\eeqn
Because of the positive--energy projector these states satisfy
$\slashchar{v} \,\cM(v) = \cM(v)$, 
$\overline{\cM}(v)\, \slashchar{v}  = \overline{\cM}(v)$,
$\cM(v) = P_+ \,\cM(v)\, P_-$ and 
$\overline{\cM}(v) = P_-\, \overline{\cM}(v)\, P_+$.
The coefficients $a$ and $b_\epsilon$ are labels which indicate a particular
meson state ($a=1$, $b_\epsilon=0$ for the pseudoscalar state;
$a=0$, $b_\epsilon=\delta_{\epsilon\epsilon_0}$ for the vector state with
polarization $\epsilon_0$). 

To compute the hadronic matrix element of a given operator $\cO$, one
replaces the hadronic states by the appropriate wave functions and builds
the most general object with the same symmetry structure as $\cO$.
For instance,  the norm of the meson states can be evaluated through
\beqn\label{eq:gen_norm}
\langle M(v)| M(v)\rangle &=& {\rm tr}\left[ \overline{\cM}(v) \,\cM(v)\,
\left(A + B \,\slashchar{v} + \cdots\right)\right] 
\no\\
 & = &  N \; {\rm tr}\left[ \overline{\cM}(v)\, \cM(v)\right]
= - 2 N \left(|a^2|+\sum_\epsilon |b_\epsilon|^2\right) \,  .
\eeqn
All possible Lorentz--invariant combinations ($1$, $\slashchar{v}$, 
$\slashchar{v}\slashchar{v}$, \ldots) should be included. Since
$\cM(v)\,\slashchar{v}= -\cM(v)$ and $\slashchar{v}\slashchar{v}=1$,
in this case all structures reduce to the identity operator.
Thus, there is only an arbitrary factor $N$ which fixes the global
normalization. This result shows that the relative normalization
of the pseudoscalar and vector states in eq.~\eqn{eq:M_state} is correct.

Let us now consider the matrix element of a quark current 
$\bar h^{(Q')}_{v'}\,\Gamma\, h^{(Q)}_{v\phantom{'}}$, which changes a heavy 
quark $Q$ into another heavy quark $Q'$. Lorentz covariance
forces the amplitude to be proportional to
$\overline{\cM}'(v')\,\Gamma\, \cM(v)$. 
This structure should be multiplied by
an arbitrary function of all Lorentz invariants $\Xi(v,v')$, which contains
the long--distance dynamics associated with the light degrees of freedom.
The flavour and spin heavy--quark symmetries require that
$\Xi$ should be independent of the spins and masses of the heavy quarks, as
well as of the Dirac structure of the current.
Hence, it can only be a function of the meson velocities (and of the
renormalization scale $\mu$); moreover, it should transform as a scalar 
with even parity. Thus,
\bel{eq:Xi_form}
\Xi(v,v') = \Xi_1 + \Xi_2 \,\slashchar{v} + \Xi_3 \,\slashchar{v}' 
+ \Xi_4 \,\slashchar{v} \,\slashchar{v}' \, .
\ee
Within the trace
$\, {\rm tr}\left[ \overline{\cM}'(v')\,\Gamma \cM(v) \,\Xi(v,v')\right]$,
the $\slashchar{v}$ operators can be eliminated, 
using the projection properties of the meson wave functions:
\bel{eq:Xi_form2}
\Xi(v,v') \longrightarrow \Xi_1 - \Xi_2 - \Xi_3 + \Xi_4 
\equiv - \xi(v\cdot v') \, .
\ee
Therefore,
\beqn\label{eq:gen_matrix}
\lefteqn{
\langle M'(v')|\bar h^{(Q')}_{v'}\,\Gamma\, h^{(Q)}_{v\phantom{'}}| M(v)\rangle = 
-\xi(v\cdot v')\;
{\rm tr}\left[ \overline{\cM}'(v')\,\Gamma\, \cM(v)\right] 
} && \no\\ && \qquad =
-\xi(v\cdot v')\; {\rm tr}\Bigg[ {1+\slashchar{v}'\over 2}\,\Gamma\,
{1+\slashchar{v}\over 2}
\Bigg(-a a'^* + \sum_{\epsilon\epsilon'} b^{\phantom{'}}_\epsilon
 b'^*_{\epsilon'}
\slashchar{\,\epsilon\,}\slashchar{\,\epsilon\,}'^*
\\ && \qquad\qquad\qquad\qquad\qquad\qquad
- a \sum_{\epsilon'} b'^*_{\epsilon'} \,\gamma_5 \slashchar{\,\epsilon\,}'^*
+  a'^* \sum_\epsilon b^{\phantom{'}}_\epsilon \slashchar{\,\epsilon\,}\gamma_5
\Bigg)\Bigg] \, . \no
\eeqn
This equation summarizes in a compact way the consequences of the
HQET symmetries. All current matrix elements are given in terms of the
same (unknown) function $\xi(v\cdot v')$,
which is usually called the Isgur--Wise function.
Taking the appropriate $a$ and $b_\epsilon$ labels, one easily
derives the explicit expressions for the matrix elements which are relevant
in semileptonic $B\to D$ decays \cite{IW:89}:
\beqn\label{eq:IWmat}
\lefteqn{
\langle P(v')|\bar h^{(Q')}_{v'}\gamma^\mu h^{(Q)}_{v\phantom{'}}| P(v)\rangle =
\xi(v\cdot v')\, \left( v+v'\right)^\mu \, ,
}&&\no\\  \lefteqn{
\langle V(v',\epsilon')|\bar h^{(Q')}_{v'}\gamma^\mu h^{(Q)}_{v\phantom{'}}| 
P(v)\rangle  =
 i\,\xi(v\cdot v')\, \varepsilon^{\mu\nu\alpha\beta}
\,\epsilon'^*_\nu v'_\alpha v_\beta \, ,
}\\  \lefteqn{
\langle V(v',\epsilon')|\bar h^{(Q')}_{v'}\gamma^\mu\gamma_5 
h^{(Q)}_{v\phantom{'}}|  P(v)\rangle 
=   \xi(v\cdot v')\, \left[\epsilon'^{*\mu}
\left(1 + v\cdot v'\right) - v'^\mu\, \left(v\cdot\epsilon'^*\right)
\right]  .
} &&\no
\eeqn

We have seen before in eq.~\eqn{eq:vector_matrix} that the hadronic
matrix element $\langle P'| \bar c\,\gamma^\mu b | P\rangle$
depends on two general form factors $f_+(q^2)$ and $f_-(q^2)$.
In the HQET formalism this would correspond to the existence of two
different Lorentz structures $(v+v')^\mu$ and $(v-v')^\mu$.
However, since
$\bar h^{(Q')}_{v'} \,\left(\slashchar{v}-\slashchar{v}'\right)\,
h^{(Q)}_v = 0$,
%
%
there is no term proportional to $(v - v')^\mu$.
The non-perturbative problem is then reduced to a single form factor,
which only depends on the relative velocity
$(v-v')^2 = 2 (1 - v\cdot v')$.
Moreover, spin symmetry relates this matrix element with the ones
governing the $P\to V'$ transition, which involve four 
(one vector and three axial--vector) independent additional form factors.
In the infinite--mass limit the six $P\to P'$ and $P\to V'$
form factors 
(and the $V\to P',V'$ ones)
are given in terms of the universal function $\xi(v\cdot v')$.

The flavour symmetry allows us to pin down also the
normalization of the Isgur--Wise function.
When $v'=v$, the vector current 
$J^\mu = \bar h^{(Q')}_{v} \gamma^\mu h^{(Q)}_{v}=
\bar h^{(Q')}_v v^\mu h^{(Q)}_{v}$
is conserved:
\bel{eq:conv_j_v}
\partial_\mu J^\mu \, = \,\bar h^{(Q')}_v\, (v\cdot D)\, h^{(Q)}_v +
\bar h^{(Q')}_v\, (v\cdot\stackrel{\leftarrow}{D})\, h^{(Q)}_v = 0 \, ,
\ee
since $(v\cdot D)\, h^{(Q)}_v=0$ by the equation of motion.
This current conservation explains why 
the corresponding anomalous dimension vanishes at equal velocities.
The associated conserved charge
\bel{eq:H_charge}
N_{Q'Q}\,\equiv\,\int d^3x\; J^0(x) \, =\,
\int d^3x\; \bar h^{(Q')\dagger}_v\, h^{(Q)}_v
\ee
is a generator of the flavour symmetry. Acting over a $Q\bar q$ meson, it
replaces a quark $Q$ by a quark $Q'$: \
$N_{Q'Q} | P(v) \rangle = | P'(v) \rangle$. 
Therefore, it satisfies
\bel{eq:N_norm}
\langle P'(v) | N_{Q'Q} | P(v) \rangle =
\langle P'(v) |  P'(v) \rangle =
2 v^0 (2\pi)^3 \delta^{(3)}(\vect{0}\,) \, .
\ee
Comparing this relation with the $P\to P'$ matrix element in eq.~\eqn{eq:IWmat}
(taking $\mu=0$ and integrating over $d^3x$),
one gets the important result:
\bel{eq:IW_norm}
\xi(1) = 1 \, .
\ee

Notice, that the light-- and heavy--quark symmetries
allow us to pin down the normalization of the corresponding
form factors at rather different kinematical points.
For massless (or equal--mass) quarks, the conservation of the vector current
fixes $f_+(q^2)$ at zero momentum transfer.
The heavy--quark limit, however, provides information on the point
of zero recoil for the final meson. Since 
\bel{eq:vv_kinem}
v\cdot v' = {M_P^2 + M_{P'}^2 - q^2 \over 2 M_P M_{P'}} \, ,
\ee
the equal--velocity
regime corresponds to the maximum momentum transfer to the final leptons
in the $P\to P' l\nu_l$ decay:
$q^2_{\mathrm{max}} = (M_P-M_{P'})^2$.

The physical picture behind \eqn{eq:IW_norm} is quite easy to understand.
The $P\to P'$ transition is induced by the action of an external vector
current coupled to the heavy quark. Before the action of the current,
the non-perturbative {\it brown muck} orbits around the heavy quark $Q$ which
acts as a (static in the rest frame)
colour source; the whole system moves with a velocity $v$.
The effect of the current is to replace instantaneously the quark $Q$ by
a quark $Q'$ moving with velocity $v'$. If $v=v'$ nothing happens;
the light quarks are unable to
realize that a heavy--quark transition has taken place,
because the interaction is flavour independent. However,
if $v\not=v'$ the {\it brown muck} suddenly feels itself interacting
with a moving coulour source. The soft--gluon exchanges needed to rearrange
the light degrees of freedom into a final meson moving with velocity
$v'$ generate a form factor suppression $\xi(v\cdot v')$,
which can only depend on the Lorentz boost $\omega = v\cdot v'$
connecting the rest frames of the initial and final mesons.
The flavour symmetry guarantees that this form factor is a universal
function independent of the heavy mass.

\subsection{$V_{cb}$ Determination}

The result \eqn{eq:IW_norm} is of fundamental importance as it allows us
to perform a clean determination of the quark--mixing factor 
$|V_{cb}|$ with the decays
$B\to D^* l \bar\nu_l$ and $B\to D l \bar\nu_l$.
The $B\to D^*$ transition is particularly useful \cite{NE:91}, because 
it has a large branching ratio 
and the corresponding hadronic matrix element does not receive any $1/M_Q$ 
correction  \cite{LU:90} at zero recoil;
corrections to the infinite--mass limit are then of order
$1/M_Q^2$.

The differential decay distribution is proportional to 
$|V_{cb}|^2\, |\cF(v_B\cdot v_{D^*})|^2$, where
the form factor $\cF(\omega)$ coincides with
$\xi(\omega)$, up to symmetry--breaking corrections of order
$\alpha_s(M_Q^2)$ and $\Lambda^2/M_Q^2$.
The calculated short--distance QCD corrections 
and the estimated
$1/M_Q^2$ contributions 
result in \cite{NE:98}
\bel{eq:F(1)}
\cF(1) = 0.91\pm 0.03\, . 
\ee
The measurement of the $D^*$ recoil spectrum has been performed by several
experiments.
Extrapolating the data to the zero--recoil point and using 
eq.~\eqn{eq:F(1)}, 
a quite accurate determination of $V_{cb}$ is obtained.
The present world average is \cite{capri}:
\bel{eq:V_cb_excl}
|V_{cb}|\, = \,  0.038\pm 0.003 \, .
\ee
%

%
%

\section{Electroweak Chiral Effective Theory}
\label{sec:EWChEFT}

In spite of the spectacular success of the Standard Model (SM), we
still do not really understand the dynamics
underlying the electroweak symmetry breaking
$SU(2)_L\otimes U(1)_Y\to U(1)_{\mathrm{QED}}$.
The Higgs mechanism provides a renormalizable way
to generate the $W$ and $Z$ masses and, therefore, their
longitudinal degrees of freedom.
However, an experimental verification of this mechanism
is still lacking.
 
The scalar sector of the SM Lagrangian
can be written in the form
\bel{eq:l_sm}
\cL(\Phi) = {1\over 2} \langle D^\mu\Sigma^\dagger D_\mu\Sigma\rangle
- {\lambda\over 16} \left(\langle\Sigma^\dagger\Sigma\rangle
- v^2\right)^2 ,
\ee
where
\be
\Sigma \equiv \left( \bat
\Phi^{0*} & \Phi^+ \\  -\Phi^- & \Phi^0
\ea\right)
\label{eq:sigma_matrix}
\ee
and $D_\mu\Sigma$ is the usual gauge--covariant derivative
\be
D_\mu\Sigma \equiv \partial_\mu\Sigma
- i g \,\widehat{W}_\mu \,\Sigma + i g' \,\Sigma \,\widehat{B}_\mu \, ,
\quad
\widehat{W}_\mu \equiv
{\vect{\tau}\over 2}\, \stackrel{\rightarrow}{W}_\mu \, ,
\quad
\widehat{B}_\mu\equiv {\tau_3\over 2}\, B_\mu \, .
\label{eq:d_sigma}
\ee
In the limit where the coupling $g'$ is neglected,
$\cL(\Phi)$ is invariant
under global $G\equiv SU(2)_L\otimes SU(2)_C$ transformations
($SU(2)_C$ is the so-called custodial symmetry group),
\be
\Sigma \, \toG \,
g_L \,\Sigma\, g_C^\dagger , \qquad\qquad
g_{L,C}  \in SU(2)_{L,C} \, .
\label{eq:sigma_transf}
\ee

Performing a polar decomposition,
\be
\Sigma(x)  =  {1\over\sqrt{2}}
\left[ v + H(x) \right] \, U(\phi(x)) \, , \qquad
U(\phi)  =  \exp{\left\{ i \vect{\tau} \,
\vect{\phi} / v \right\} } \, , 
\label{eq:polar}
\ee
in terms of the Higgs field $H$ and the Goldstones
$\vect{\phi}$,
and taking the limit $\lambda\gg 1$ (heavy Higgs),
we can rewrite \cite{AB:80}
$\cL(\Phi)$ in the standard chiral form:
\be
\cL(\Phi) = {v^2\over 4}\,
\langle D_\mu U^\dagger D^\mu U \rangle  + 
\cO\left( H/ v \right) ,
\label{eq:sm_goldstones}
\ee
with
$D_\mu U \equiv \partial_\mu U
- i g \,\widehat{W}_\mu\, U + i g'\, U \,\widehat{B}_\mu$.

In the unitary gauge $U=1$, this $\cO (p^2)$ Lagrangian
reduces to the usual bilinear gauge--mass term:
\bel{eq:gauge_mass}
\cL(\Phi)\quad \stackrel{U=1}{\longrightarrow} \quad
M_W^2 \, W_\mu^\dagger W^\mu + {M_Z^2\over 2}\,  Z_\mu Z^\mu
\, ,
\ee
where
$Z^\mu \equiv \cos{\theta_W} W_3^\mu - \sin{\theta_W} B^\mu$,
$M_W = M_Z \,\cos{\theta_W} = v g/2$ and
$\tan{\theta_W} = g'/ g$.

Equation \eqn{eq:sm_goldstones}
is the universal model--independent interaction of the
Goldstone bosons induced by the assumed pattern of SCSB,
\bel{eq:pat_sb}
SU(2)_L\otimes SU(2)_C \quad\longrightarrow\quad SU(2)_{L+C} .
\ee
The scattering of electroweak Goldstone bosons
(or equivalently longitudinal gauge bosons)
is then described by the same formulae as
the scattering of pions, changing $f$ by $v$
\cite{CO:74,LQT:77,CG:85}.
To the extent that the present data are still not very sensitive to
the virtual Higgs effects, we have only tested up to now
the symmetry properties of the scalar sector encoded in
eq.~\eqn{eq:sm_goldstones}.
 
In order to really prove the particular scalar dynamics
of the SM, we need to test the
model--dependent part involving the Higgs field $H$.
If the Higgs turns out to be too heavy to be directly
produced (or if it does not exist at all),
one could still investigate the higher--order effects
by applying the standard chiral expansion techniques.

\subsection{Effective Lagrangian}

In the electroweak SM, the SCSB is realized linearly, 
through a scalar field which acquires a non-zero vacuum expectation value.
The spectrum of physical particles contains then
not only the massive vector bosons but also a neutral scalar Higgs field
which must be relatively light.

In a more general scenario, the electroweak SCSB can be parametrized 
in terms of an effective Lagrangian which contains the SM
gauge symmetry realized non-linearly \cite{AB:80,AP:80,LO:80}.
Only the known light degrees of freedom (leptons, quarks and gauge bosons)
appear in this effective Lagrangian, which does not include any
Higgs field.
Owing to its similarity with ChPT, this electroweak EFT is sometimes
called the chiral realization of the SM. 
With a particular choice of the parameters
of the Lagrangian, it includes the SM, as long
as the energies involved are small compared with the Higgs mass.
In addition it can also 
accommodate any model that reduces to the SM at low energies as
happens in many technicolour scenarios \cite{Chivukula}.
The price to be paid for this general parametrization is
the appearance of many couplings which
must be determined from experiment or computed in a more fundamental theory.

The lowest--order effective Lagrangian can be written in the following
way:
\bel{eq:L0_EW}
\cL_{\mathrm{EW}} = \cL_B + \cL_{\psi} + \cL_Y ,
\ee
where
\be
\cL_B = -\frac{1}{2}\, \langle
\widehat{W}_{\mu\nu}\widehat{W}^{\mu\nu} +
\widehat{B}_{\mu\nu}\widehat{B}^{\mu\nu}\rangle
+ \frac{v^2}{4}\, \langle D_{\mu} U^\dagger\, D^{\mu} U\rangle ,
\ee
with
\beqn\label{eq:FT_def}
\lefteqn{
\widehat{W}_{\mu\nu}\equiv 
{i\over g}\,\left[ \left(\partial_\mu - i g \widehat{W}_\mu\right) ,
\left(\partial_\nu - i g \widehat{W}_\nu\right) \right] =
{\vect{\tau}\over 2} \, \stackrel{\rightarrow}{W}_{\mu\nu}\, .}
&&\qquad\no\\
\lefteqn{
\widehat{B}_{\mu\nu}\equiv 
\partial_\mu\widehat{B}_\nu - \partial_\nu\widehat{B}_\mu =
{\tau_3\over 2}\, B_{\mu\nu} \, . }
\eeqn
$\cL_{\psi}$ is the usual fermionic kinetic Lagrangian and
\be
\cL_Y = -\bar{q}_L \, U \,\cM_q\, q_R -
  \bar{l}_L \, U \,\cM_l\, l_R + \mbox{\rm h.c.} \, ,
\ee
where $\cM_q$  ($\cM_l$) is a $2 \times 2$ block--diagonal matrix containing
the $3 \times 3$ mass matrices of the up and down quarks 
(neutrinos and charged leptons) and $q_{L,R}$ ($l_{L,R}$) are
doublets containing the up and down quarks (leptons) for the three families
in the weak basis.

The Lagrangian \eqn{eq:L0_EW} is invariant under local $SU(2)_L\otimes U(1)_Y$
gauge transformations:
\beqn\label{eq:gauge_transf}
\lefteqn{
\Psi_L \;\longrightarrow\; g_L\, \Psi_L , \quad
\Psi_R \;\longrightarrow\; g_R\, \Psi_R \,\quad (\Psi=q,l) , \quad
U \;\longrightarrow\; g_L\, U\, g_R^\dagger , } && \no\\
\lefteqn{
\widehat{W}_\mu \;\longrightarrow\; g_L\, \widehat{W}_\mu\, g_L^\dagger
  + {i\over g}\, g_L\,\partial_\mu g_L^\dagger , \qquad
\widehat{W}_{\mu\nu}\;\longrightarrow\; g_L\, \widehat{W}_{\mu\nu}\,
 g_L^\dagger , } && \\
\lefteqn{
\widehat{B}_\mu \;\longrightarrow\; \widehat{B}_\mu
  + {i\over g'}\, g_R\,\partial_\mu g_R^\dagger , \qquad\qquad
\widehat{B}_{\mu\nu}\;\longrightarrow\;  \widehat{B}_{\mu\nu} , 
} &&   \no
\eeqn
where
\bel{eq:LR_def}
g_L\equiv \exp{\left\{ i\vect{\alpha}\,{\vect{\tau}\over 2}\right\} } ,
\qquad\qquad
g_R\equiv \exp{\left\{ i \beta\,{\tau_3\over 2}\right\} } .
\ee

The lowest--order operators just fix the values of the $Z$ and $W$ masses at
tree level and do not carry any information on the underlying SCSB physics.
Therefore, in order to extract some 
information on new physics, we must study the effects coming 
from higher--order terms in the effective Lagrangian.
At the next order, that is containing at most four derivatives,
the  most general $CP$ and $SU(2)_L\otimes U(1)_Y$ invariant 
effective chiral Lagrangian 
with only gauge bosons and Goldstone fields,\footnote{
We only discuss a chiral EFT for the bosonic sector and 
assume the fermion couplings to be given by \protect\eqn{eq:L0_EW}.
Possible modifications of the fermionic couplings of the gauge bosons
have been investigated in refs.~\protect\cite{PZ:90,PPZ:91}.
}
%
\bel{eq:EWHOL}
\cL_{\mathrm{EW}}^{(4)} = \sum_{i=0}^{14} a_i\, O_i ,
\ee
contains 15 independent operators \cite{AP:80,LO:80}:
\beqn\label{eq:EWoper}
\lefteqn{O_0 = {v^2\over 4}\, \langle T V_\mu\rangle^2 ,}
& \qquad\qquad\qquad\qquad\qquad\qquad\qquad \qquad
& \no\\ \lefteqn{
O_1 = i{g g'\over 2}\, B_{\mu\nu}\, \langle T \widehat{W}^{\mu\nu}\rangle ,
} &&
O_2 = -i {g'\over 2}\, B_{\mu\nu}\, \langle T \left[V^\mu,V^\nu\right]\rangle ,
\no\\ \lefteqn{
O_3 = -g \,\langle\widehat{W}_{\mu\nu} \left[V^\mu,V^\nu\right] \rangle ,
} &&
O_4 = \langle V_\mu V_\nu \rangle\, \langle V^\mu V^\nu \rangle ,
\no\\ \lefteqn{
O_5 = \langle V_\mu V^\mu \rangle^2 ,
} &&
O_6 = \langle V_\mu V_\nu \rangle\, \langle T V^\mu \rangle\, 
  \langle T  V^\nu \rangle ,
\no\\ \lefteqn{
O_7 = \langle V_\mu V^\mu \rangle\, \langle T  V_\nu \rangle^2 ,
} &&
O_8 = {g^2\over 4}\, \langle T \widehat{W}_{\mu\nu}\rangle^2 ,
\\ \lefteqn{
O_9 = -{g\over 2}\, \langle T \widehat{W}_{\mu\nu}\rangle
  \, \langle T \left[V^\mu,V^\nu\right]\rangle ,
} &  &
O_{10} = \left\{ \langle T V_\mu\rangle\, \langle T V_\nu\rangle\right\}^2 ,
\no\\ \lefteqn{
O_{11} = \langle \left( D_\mu V^\mu\right)^2\rangle ,
} &&
O_{12} = \langle T D_\mu D_\nu V^\nu\rangle\, \langle T V^\mu\rangle ,
\no\\ \lefteqn{
O_{13} = {1\over 2}\,\langle T D_\mu V_\nu\rangle^2 ,
} &&
O_{14} = -i g \varepsilon^{\mu\nu\rho\sigma}\,
 \langle\widehat{W}_{\mu\nu} V_\rho\rangle\, \langle T V_\sigma\rangle .   
\no
\eeqn
We have introduced the combinations
\bel{eq:V_defi}
T\equiv U\,\tau^3\, U^\dagger ,
\quad
V_\mu\equiv D_\mu U\, U^\dagger ,
\quad
D_\mu V_\nu \equiv \partial_\mu V_\nu - i g \, \left[ \widehat{W}_\mu, 
V_\nu\right] ,
\ee
which transform as
\bel{eq:T_transf}
T\;\longrightarrow\; g_L\, T\, g_L^\dagger ,
\quad
V_\mu\;\longrightarrow\; g_L\, V_\mu\, g_L^\dagger ,
\quad
D_\mu V_\nu \;\longrightarrow\; g_L\, D_\mu V_\nu\, g_L^\dagger .
\ee
Notice that all the operators are invariant under parity, except $O_{14}$.

For massless fermions, the equations of motion for the gauge fields
imply
$\partial_\mu \langle T V^\mu\rangle = 0$
and
$D_\mu V^\mu=0$.
As a consequence,
$O_{11}= O_{12}= 0$ and
$O_{13} = -{g'^2\over 4} B_{\mu\nu} B^{\mu\nu} + O_1 - O_4 + O_5 - O_6
 + O_7 + O_8$.
Therefore, as long as one only considers light fermions ($m_\psi \ll v$),
the operators $O_{11}$, $O_{12}$ and $O_{13}$ can be eliminated from the
Lagrangian.

The physical meaning of the different operators is more transparent in the
unitary gauge, $U=1$, where all invariants reduce to polynomials of
the gauge fields.
The operators $O_0$,  $O_1$, $O_8$, $O_{11}$, $O_{12}$ and $O_{13}$
contain bilinear terms in the gauge fields;
therefore, the usual electroweak oblique corrections are sensitive
to \cite{FE:93} $a_0$, $a_1 + a_{13}$ and $a_8 + a_{13}$:
\beqn\label{eq:oblique}
\Delta r & \dot= & -2\, {\cos^2{\theta_W}\over\sin^2{\theta_W} } \, a_0 +
  \left( 1 -  {\cos^2{\theta_W}\over\sin^2{\theta_W} }\right) 
  g^2\, \left( a_8 + a_{13}\right)
  - 2 g^2 \left( a_1 + a_{13}\right) ,
\no\\
\Delta \rho & \dot= &  2 a_0 ,
\\
\Delta k & \dot= &  {
    2 \cos^2{\theta_W}\, a_0 + g^2\, \left( a_1 + a_{13}\right)
    \over \sin^2{\theta_W} - \cos^2{\theta_W}}\,
\, .\no
\eeqn
Here, $\Delta r$, $\Delta\rho$ and $\Delta k$ are the standard
parameters containing the
corrections induced by the gauge self-energies into the
$M_W$--$G_F$ relation, the neutral-- and charged--currents ratio,
and the leptonic vector coupling of the $Z$ boson, respectively
\cite{Treille}.

On the other hand, the operators 
$O_2$,  $O_3$, $O_9$, and $O_{14}$
parametrize  the trilinear non-abelian gauge couplings that are 
tested at LEP2.
Finally, $O_4$,  $O_5$, $O_6$, $O_7$ and $O_{10}$
contain only quartic terms in the gauge boson fields;
we could think to fix them, at least in principle,
by means of scattering experiments among gauge vector mesons at LHC
\cite{DH:89,DHPRU:95,DR:91,DV:91,BDV:93,BAG:94,BEGMNZ:98}.
All these operators contributing to three-- and four--point Green
functions modify the oblique corrections at the one--loop level
\cite{HT:90,GL:91,GE:91b,DEH:91,EH:92,dRGMH:92,HV:93,BEG:96},
which allows to put some (weak) upper limits on their couplings
($a_i\lsim 0.1$).

\begin{figure}[tbh]
\centerline{
\begin{minipage}{.45\linewidth}
  \epsfig{file=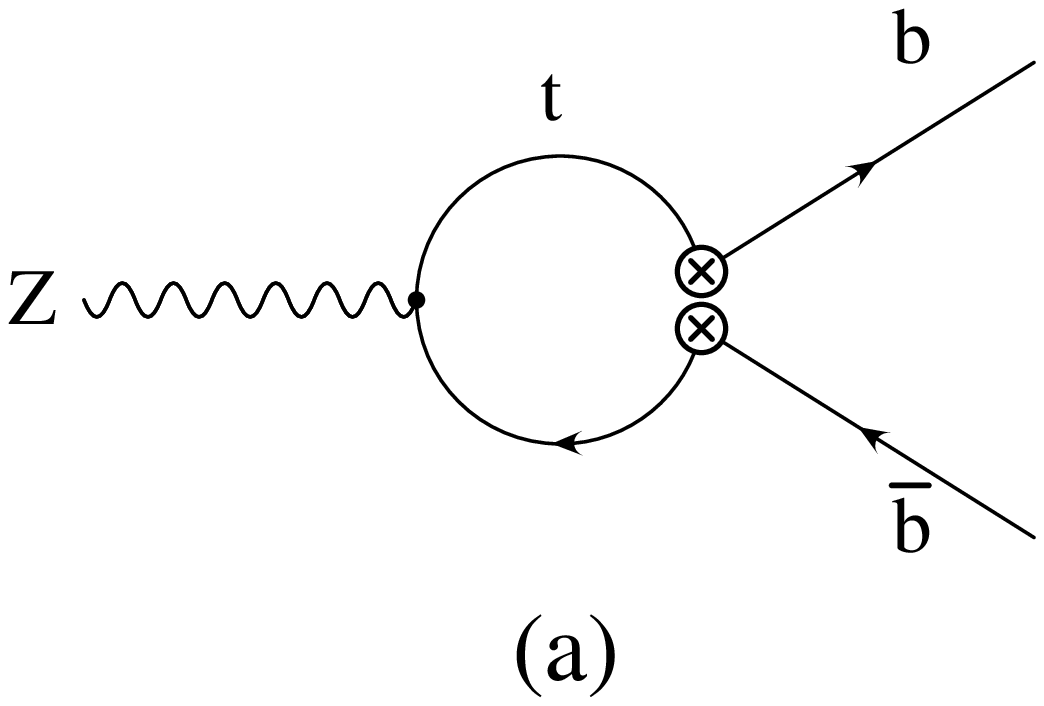,width=\linewidth,height=3.7cm}
\end{minipage}
\hspace{1cm}
\begin{minipage}{.42\linewidth}
  \epsfig{file=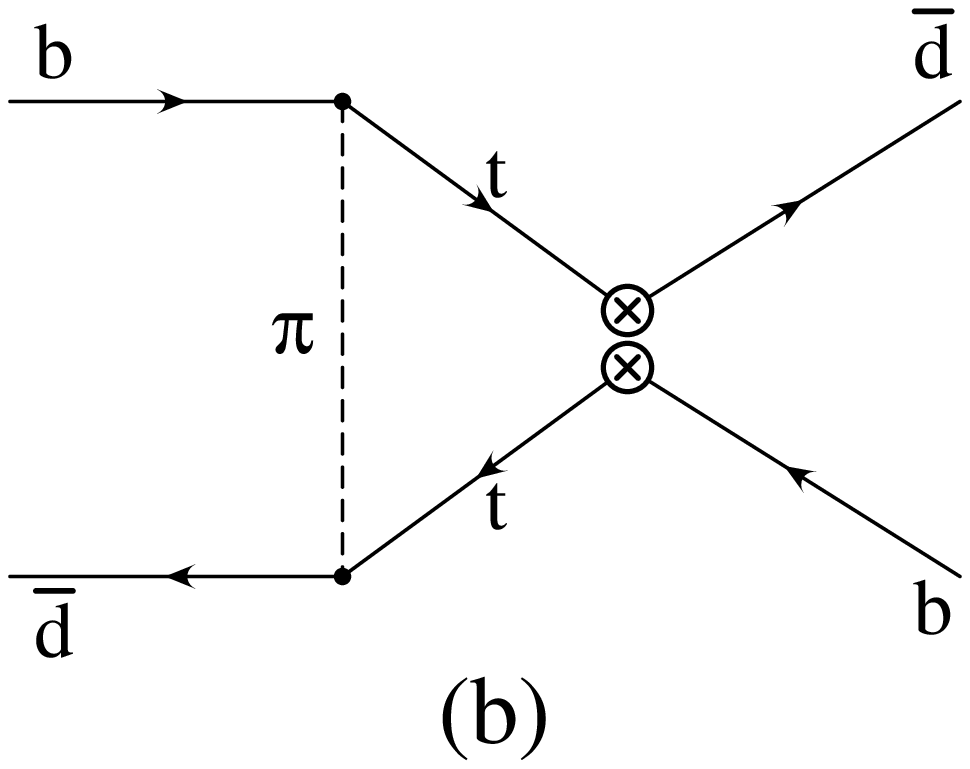,width=\linewidth,height=3.7cm}
\end{minipage} 
}
\caption{Contribution of the effective operator $O_{11}$
to $Z\rightarrow b\bar{b}$ (a) and
$B$--$\bar{B}$ mixing (b).}
\label{fig:O11}
\end{figure}

The couplings $a_{11}$ and $a_{12}$ remain untested because,
although quadratic in the Goldstone fields, 
they do not contribute to the one--loop oblique corrections.
They only involve the longitudinal components of the gauge bosons
and can be eliminated, using the classical equations of motion, if fermion
masses are neglected.
However, keeping the terms proportional to the top quark mass
and making use of the equations of motion,
the operator $O_{11}$ turns out \cite{BCPS:97} to be equivalent
to a four--fermion operator proportional to $M_t^2$:
\bel{eq:O114f}
O_{11} \;\dot=\; \frac{g^4}{8 M_W^4}\, M_t^2\,
\left((\bar{t}\gamma_5t)^2 -4
\sum_{i,j} (\bar{d}_{iL} t_R)(\bar{t}_R d_{jL})\, 
V^{\phantom{*}}_{tj} V^*_{ti}\right)~.
\ee
Therefore, $O_{11}$ affects the $Z\bar b b$ vertex, the $B^0$--$\bar B^0$
mixing, and the CP--violating parameter $\varepsilon_K$, generating
interesting correlations among the hard $M_t^4\,\log{M_t^2}$
corrections to these observables \cite{BCPS:97}; this allows us to
derive an $\cO(10\%)$ upper bound on $a_{11}$.
Similar corrections are induced on rare $B$ and $K$ decays \cite{BU:97}.

\subsection{Matching Conditions}

The SM gives definite predictions for the
chiral couplings of the $\cO (p^4)$ electroweak Lagrangian,
which could be tested in future experiments.
Table~\ref{eq:EWcouplings} shows \cite{FE:93}
the corresponding values of these 
couplings, for three different limits of the SM:
1) a very large Higgs mass; 2) a fourth generation
with a light lepton doublet ($M_l\leq M_Z$) and heavy
degenerate quarks ($M_{t'}= M_{b'}\gg M_Z$), and 3) a heavy top quark.
In the first case, the operators $O_{11}$, $O_{12}$ and $O_{13}$ have 
been eliminated with the equations of motion; the $O_{13}$ contribution
is then included in the couplings $a_1$, $a_4$, $a_5$, $a_6$, $a_7$ and
$a_8$.

\begin{table}[htb]
\caption{Electroweak chiral coefficients, in units of $1/(16\pi^2)$,
for different limits of the SM. }
\label{eq:EWcouplings}
\begin{tabular}{@{}cccc@{}}
\hline
& $M_H\to\infty$  & $M_{t',b'}\to\infty$  & $M_t\to\infty$
\\
& \protect\cite{LO:80,EM:95,HM:94} & \protect\cite{HF:84} & 
\protect\cite{FMM:92}
\\ \hline
$a_0$ & $ -{3\over 4} g'^2 \left[ \log{\left( {M_H/\mu}\right)} - 
 {5\over 12} \right]$ & \phantom{-}0 & $\frac{3}{2} \frac{M_t^2}{v^2}$
\\ \\
$a_1$ & $-{1\over 6} \log{(M_H/\mu)} + {5\over 72}$ & $-\frac{1}{2}$ &
   ${1\over 3} \log{\left( {M_t/\mu}\right)} - {1\over 4}$
\\  \\
$a_2$ & $ -{1\over 12}  \log{(M_H/\mu)} +  {17\over 144}$ 
& $-\frac{1}{2}$ &
   ${1\over 3}  \log{\left( {M_t/\mu}\right)} - {3\over 4}$
\\ \\
$a_3$ & ${1\over 12} \log{(M_H/\mu)} - {17\over 144}$ & 
$\phantom{-}\frac{1}{2}$ &
   $\phantom{-}\frac{3}{8}$
\\  \\
$a_4$ & ${1\over 6} \log{(M_H/\mu)} -  {17\over 72}$ & 
$\phantom{-}\frac{1}{4}$ &
   $\phantom{-}\log{\left( {M_t/\mu}\right)} - {5\over 6}$
\\ \\
$a_5$ & ${2\pi^2v^2\over M_H^2} + {1\over 12}  \log{\left( {M_H/\mu}\right)} - 
{79\over 72} + {9\pi\over 16\sqrt{3}}$ & $-\frac{1}{8}$ &
   $ - \log{\left( {M_t/\mu}\right)}  + {23\over 24}$
\\  \\
$a_6$ & 0 & \phantom{-}0 &    
$ - \log{\left( {M_t/\mu}\right)} + {23\over 24}$
\\  \\
$a_7$ & 0 & \phantom{-}0 &    
$ \phantom{-}\log{\left( {M_t/\mu}\right)} - {23\over 24}$
\\  \\
$a_8$ & 0 & \phantom{-}0 &    
$\phantom{-}\log{\left( {M_t/\mu}\right)}  - {7\over 12}$
\\  \\
$a_9$ & 0 & \phantom{-}0 &    
$\phantom{-}\log{\left( {M_t/\mu}\right)} - {23\over 24}$
\\  \\
$a_{10}$ & 0 & \phantom{-}0 &    $- \frac{1}{64}$
\\ \\ 
$a_{11}$ & --- & $- \frac{1}{2}$ &    $- \frac{1}{2}$
\\ \\ 
$a_{12}$ & --- & \phantom{-}0 &    $- \frac{1}{8}$
\\ \\ 
$a_{13}$ & --- & \phantom{-}0 &    $- \frac{1}{4}$
\\ \\ 
$a_{14}$ & 0 & \phantom{-}0 &    $\phantom{-}\frac{3}{8}$
\\ \hline
\end{tabular}
\end{table}

In the two considered heavy--quark cases, the light--fermion loops of the
resulting low--energy theory induce a gauge anomaly, because there
is an incomplete fermion generation which destroys the delicate
anomaly cancellation of the SM. Therefore, the effective theory
should also include a corresponding Wess--Zumino term \cite{WZ:71,WI:83},
whose gauge variation cancels exactly the anomaly produced by the
light fermions \cite{FE:93,HF:84,FMM:92,LSY:91}.

The couplings of the chiral effective Lagrangian
contain the interesting dynamical information on any underlying
electroweak theory, consistent with the gauge symmetries of the SM.
It remains to be seen whether the experimental determination
of the higher--order electroweak chiral couplings will confirm
the renormalizable SM Lagrangian, 
or will constitute an evidence of new physics.

\subsection{Non-Decoupling}

 The {\bf decoupling theorem} \cite{AC:75} states that 
{\it the low--energy 
effects of heavy particles are either suppressed by inverse powers of
the heavy masses, or they get absorbed into renormalizations of the
couplings and fields of the EFT obtained by removing the heavy particles.}

We have already seen how decoupling works in QED and QCD. However,
the effective couplings given in table~\ref{eq:EWcouplings}
show that heavy particles do not decouple in the electroweak
theory. The Higgs contributions increase logarithmically with the Higgs
mass, while a heavy top induces hard corrections which increase 
quadratically with $M_t$. The effects of a heavy fourth--generation
quark doublet do not increase with the quark masses, but leave a non-zero
constant correction at low energies.

The decoupling theorem has been proved \cite{AC:75} to be valid for
theories with an exact gauge symmetry. However, it is not necessarily
satisfied in theories with spontaneously broken gauge symmetries.
The non-decoupling effects originate in the different nature of the mass
terms. Whereas in theories with exact gauge symmetry, such as QED or QCD,
mass terms are gauge invariant, in the spontaneously broken case masses
are generated through the symmetry--breaking mechanism and, therefore,
are associated with interaction terms.

In order to have decoupling, the dimensionless couplings should not grow
with the heavy masses. Otherwise, the mass suppression induced by the
heavy--particle propagators can be compensated by the mass enhancement
provided by the interaction vertices, with an overall non-vanishing effect.
This is precisely what happens with the electroweak interaction.

In the SM, the boson and fermion masses are proportional to the scale
of SCSB:
\bel{eq:mass_gener}
M_W = M_Z \,\cos{\theta_W} = {g\over 2}\, v , \qquad
M_H = \sqrt{\lambda\over 2}\, v , \qquad
M_f = - {y_f\over\sqrt{2}}\, v .
\ee
The different mass scales are generated by the different dimensionless 
couplings 
appearing in these relations: the gauge coupling $g$ for the vector mesons,
the scalar potential coupling $\lambda$ for the Higgs and a different
Yukawa coupling $y_f$ for each fermion.

There are two different ways of taking the large--mass limit \cite{FE:93}.
The simplest alternative is to keep the couplings fixed and let the
scalar vacuum expectation value $v$ go to infinity. In this case, all
massive particles become heavy. Moreover, the electroweak interactions 
mediated by the heavy fields do indeed decouple, as we saw before
with the effective Fermi Hamiltonian:
\bel{eq:G_F_v_rel}
{G_F\over\sqrt{2}} = {g^2\over 8 M_W^2} = {1\over 2 v^2} 
\quad\longrightarrow \quad
0\, .
\ee

The large--mass limits considered in table~\ref{eq:EWcouplings} correspond
to a second and more interesting possibility, where only some masses are
taken to be heavy. In this case, the scalar vacuum expectation value
remains fixed and the large--mass limit actually means that some
couplings become large.
Decoupling is obviously no-longer true in such scenario. For instance,
in the limit $g\to\infty$ with $v$ fixed, the Fermi coupling remains
invariant in spite of the fact that $M_W\to\infty$.

The limit of a heavy Higgs is achieved with a large scalar
self-coupling $\lambda$. The Goldstone modes of the electroweak
SCSB, which correspond to the longitudinal polarization of the gauge
bosons, are then in a strong interaction regime.
The failure of the decoupling property
shows up in the effective electroweak chiral couplings $a_i$,
which are not suppressed by the Higgs mass.
Owing to the custodial $SU(2)_C$ symmetry of the scalar potential,
the dependence on $M_H$ is only logarithmic (screening theorem)
at one--loop \cite{VE:77}.
Power--like corrections are, however, possible at higher orders.

\begin{figure}[tbh]
\centerline{\epsfig{file=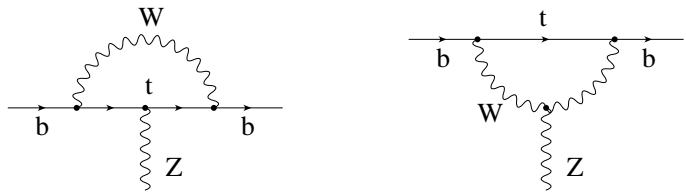,width=9cm}}
\caption{$M_t$--dependent corrections to the $Z\bar b b$ vertex.}
\label{fig:zbb}
\end{figure}

A heavy top quark implies a large Yukawa coupling $y_t$.
Therefore, the interactions of the top with the
Higgs and the Goldstones are strong in that case.
This generates a hard $M_t^2$ contribution to the $Z$ and $W$
self-energies \cite{VE:77}, which shows up in the chiral coefficient $a_0$.
Another interesting manifestation of non-decoupling \cite{BPS:88}
appears in the $Zb\bar b$ vertex, which gets one--loop $M_t^2$ corrections 
\cite{BPS:88,ABR:86,BH:88,LS:90} generated by 
the exchange of a virtual (longitudinal) $W$ boson between the two fermionic
legs. This hard contribution does not have any quark--mixing factor
suppression because $V_{tb}\approx 1$.
Another related effect is the $M_t^2$ factor in eq.~\eqn{eq:O114f},
generated by $O_{11}$ through the equations of motion \cite{BCPS:97}.

The non-decoupling of heavy particles implies that low--energy experiments
can be sensitive to large mass scales, which cannot be kinematically
accessed. Thus, the high--precision measurements performed at LEP and SLC
have been able to extract information on the top and the Higgs
\cite{Treille}. Notice that the screening of the one--loop $M_H$ 
dependences is the reason why the Higgs mass is so difficult to
pin down.
The top quark contributions play also a very important role in
flavour--changing transitions and CP--violation phenomena
\cite{buras}.

%
%

\section{Summary}
\label{sec:summary}

EFT is a very powerful tool to analyze physics at low energies,
without having to solve the details of dynamics at higher energy scales.
One does not need to know whether there are supersymmetric particles
in the 1~TeV region in order to understand the interactions of electrons
and photons at energies of the order of $m_e$.
Our problems formulating a consistent theory of quantum gravity at the
Planck scale do not prevent us from having a rather successful description
of physics at the electroweak scale.
Even if the fundamental QED is very well known, a non-relativistic
formulation of the electromagnetic interaction turns out to be more useful
in atomic physics and chemistry.

The main motivation behind the EFT framework is simplicity. 
Once the appropriate variables describing the relevant physics at the
scale considered have been identified, a useful approximate description
can be formulated.
Dimensional analysis allows us to estimate the size of possible
corrections, and to organize them in such a way that only a minimum number
need to be calculated, to reach a given accuracy.
Problems involving widely separated scales can be investigated with the
help of the renormalization group.

Symmetries are always a very important handle to develop a predictive
EFT. They restrict the form and number of the interactions present in
the effective Lagrangian, at a given order in the momentum expansion.
The resulting EFT allows one to predict the low--energy amplitudes, 
except for the values of the effective couplings, which do not get fixed
by symmetry considerations. Those couplings encode the information
on higher scales, which survives at low energies. They can be fixed
experimentally, or through a matching calculation if an underlying
more fundamental EFT is known.

We have seen three important EFTs, which are associated with three
phenomenologically relevant symmetries: Chiral Perturbation Theory, 
Heavy Quark Effective Theory and the Electroweak Chiral Effective
Theory. Using quite similar tools, these three EFTs allow us to
successfully analyze three different energy regimes: 
the light quark dynamics
below 1~GeV, the physics of bottom and charm quarks, and the
electroweak symmetry breaking scale.

There are of course many more interesting applications of EFT which
are useful for phenomenology. The basic formalism that we
have discussed can be adapted to very different situations,
to obtain the most important information on the physical system being
analyzed.

The fundamental search for {\it the theory of everything} will
continue being our ultimate dream for many years.
In the meanwhile, EFT allows us to understand the main features of
the physics at a given scale.
Moreover, even if {\it the theory of everything} is found at some point,
EFT will still provide a simpler (but less fundamental) description
of nature.

%
%

\section*{Acknowledgements}

I would like to thank the organizers for the charming atmosphere of
this school, and the students for their many interesting questions
and comments.
I'm also grateful to V.~Gim\'enez, J.~Portol\'es and A.~Santamar\'{\i}a
for their critical reading of the manuscript.
This work has been supported in part by 
the EEC--TMR Program ---Contract No. ERBFMRX-CT98-0169---
and by CICYT (Spain) under grant No.~AEN-96-1718.

%
%


%
%

\end{document}